\documentclass[onecolumn, 11pt, usenames, dvipsnames]{article}

\usepackage[utf8]{inputenc}
\usepackage{lmodern}

\usepackage[top=1in, bottom=1in, left=1in, right=1in]{geometry}
\usepackage{authblk}

\usepackage[english]{babel}
\usepackage[abbreviations,british]{foreign}

\usepackage[numbers, sort&compress]{natbib}

\usepackage[]{color}
\usepackage[]{xcolor}

\usepackage[]{graphicx}
\usepackage[]{subfigure}
\usepackage[]{tcolorbox}

\usepackage[]{mathtools}
\usepackage[]{amssymb, amsmath, amsfonts}
\usepackage[]{yhmath}
\usepackage[]{stmaryrd}
\usepackage[]{nicefrac}
\usepackage[]{bm}

\usepackage[]{siunitx}
\usepackage[]{physics}

\usepackage[]{hyperref}
\usepackage[noabbrev]{cleveref}

\usepackage[]{tikz}
\usetikzlibrary{arrows}
\usetikzlibrary{shapes}
\usetikzlibrary{positioning}
\usetikzlibrary{tikzmark}
\usetikzlibrary{patterns}
\usetikzlibrary{backgrounds}

\usepackage[]{lipsum}
\usepackage[]{xparse}

\DeclareMathOperator*{\minimize}{minimize~}

\ExplSyntaxOn
\NewDocumentCommand{\colvec}{O{1}m}
{
  \scalebox{#1}{$\generalvec{#2}{\\}$}
}
\NewDocumentCommand{\rowvec}{O{1}m}
{
  \scalebox{#1}{$\generalvec{#2}{&}$}
}
\NewDocumentCommand{\generalvec}{mm}
{
  \clist_set:Nn \l_tmpa_clist { #1 }
  \begin{bmatrix}
    \clist_use:Nn \l_tmpa_clist { #2 }
  \end{bmatrix}
}
\ExplSyntaxOff

\title{Krylov methods for large-scale dynamical systems: Application in fluid dynamics}
\author[1]{Ricardo S.\ Frantz}
\author[1]{Jean-Christophe Loiseau}
\author[1]{Jean-Christophe Robinet}

\affil[1]{Arts et Métiers Institute of Technology, CNAM, DynFluid, HESAM Université, F-75013 Paris, France}

\date{\today}

\begin{document}
\maketitle

\begin{abstract}
In fluid dynamics, predicting and characterizing bifurcations, from the onset of unsteadiness to the transition to turbulence, is of critical importance for both academic and industrial applications.
Different tools from dynamical systems theory can be used for this purpose.
In this review, we present a concise theoretical and numerical framework focusing on practical aspects of the computation and stability analyses of steady and time-periodic solutions, with emphasis on high-dimensional systems such as those arising from the spatial discretization of the Navier-Stokes equations.
Using a matrix-free approach based on Krylov methods, we extend the capabilities of the open-source high-performance spectral element-based time-stepper \texttt{Nek5000}.
The numerical methods discussed are implemented in \texttt{nekStab}, an open-source and user-friendly add-on toolbox dedicated to the study of stability properties of flows in complex three-dimensional geometries.
The performance and accuracy of the methods are illustrated and examined using standard benchmarks from the fluid mechanics literature.
Thanks to its flexibility and domain-agnostic nature, the methodology presented in this work can be applied to develop similar toolboxes for other solvers, most importantly outside the field of fluid mechanics.
\end{abstract}

\maketitle 

\section{Introduction}\label{sec: introduction}

The transition to turbulence is a long-standing problem in fluid dynamics, pioneered by Osborne Reynolds~\cite{osbourne1883experimental}.
As mathematical tools have developed, a dynamical systems point of view has led to a better understanding of this phenomenon.
Before the advent of computers, theoretical analyses had to rely on simplifying assumptions, the most important ones being the \emph{parallel flow assumption} and that of \emph{infinitesimal perturbations} forming what we now know today as \emph{local stability theory}.
For simple shear flows, these assumptions lead to the famous Orr-Sommerfeld-Squire equations
\begin{equation}
\begin{aligned}
    \left[ \left( \dfrac{\partial}{\partial t} \right) \nabla^2 - U^{\prime\prime} \dfrac{\partial}{\partial x} - \dfrac{1}{Re}\nabla^4 \right] v & = 0, \\
    \left[ \dfrac{\partial}{\partial t} + U \dfrac{\partial}{\partial x} - \dfrac{1}{Re}\nabla^2 \right] \eta & = - U^{\prime} \dfrac{\partial v}{\partial z},
    \label{eq: orr-sommerfeld-squire eqs}
\end{aligned}
\end{equation}
where $U(y)$ is the \emph{base flow} velocity profile, $v(x, y, z, t)$ is the wall-normal velocity component of the perturbation, and $\eta(x, y, z, t)$ its wall-normal vorticity component.
Despite their simplicity, these assumptions led to important theorems in hydrodynamic stability theory: Rayleigh's inflection point criterion~\cite{rayleigh1879stability}, Fj{\o}rtoft~\cite{fjortoft1950application} or Squire's theorem~\cite{squire1933stability}.
They also led to a better understanding of non-normality and to the development of nonmodal stability analysis~\cite{reddy1993pseudospectra,amr:schmid:2014}.
Although $U(y)$ is formally a steady solution of the Navier-Stokes equations, a tremendous amount of understanding has been gained by replacing it with simple approximations such as the Bachelor vortex sheet model or the piecewise linear approximation of the Blasius boundary layer profile in the seminal work of Tollmien and Schlichting~\cite{tollmien1930entstehung,schlichting1933enstehung}.
Using a normal mode ansatz, the velocity fluctuation can be decomposed using a Fourier expansion
\begin{equation}
v(x, y, t) = \hat{v}(y) \exp \left( i (\alpha x - \omega t) \right),
\end{equation}
and similarly for the vorticity fluctuation.
Determining the stability of the system then amounts to solving a generalized eigenvalue problem.
Depending on the assumptions about the wavenumber $\alpha$ and the frequency $\omega$, these stability analyses fall into two categories
\begin{itemize}
    \item Temporal stability: defined by $\alpha \in \mathbb{R}$ and $\omega \in \mathbb{C}$.
In this context, one aims to determine whether a fluctuation grows over time at a given streamwise position. 
    \item Spatial stability: defined by $\alpha \in \mathbb{C}$ and $\omega \in \mathbb{R}$.
In this context, one investigates whether forcing at a particular frequency causes the perturbation to grow in space while being advected by the flow.
\end{itemize}
An important milestone was achieved by Huerre \& Monkewitz~\cite{huerre1985absolute, huerre1990local} by letting both $\alpha$ and $\omega$ be complex numbers.
Depending on subtle properties of the dispersion relation in the complex plane, they introduced the dichotomy between \emph{absolute} and \emph{convective instabilities}, thus establishing the first connection between the local stability properties of the flow and its spatiotemporal evolution.
An instability is classified as \emph{absolute} if the perturbation grows in place (\ie{} its group velocity is zero), otherwise as \emph{convective} if it grows as it propagates before leaving the region of interest.
For more details, see \cite{huerre1985absolute,cossu1998convective}.
This connection between \emph{local} and \emph{global} properties of flow was refined by Monkewitz, Huerre and Chomaz~\cite{monkewitz1993global} using the WKBJ formalism and led to the \emph{weakly non-parallel flow assumption} which is, however, still very limiting and hardly applicable to flow configurations of practical interest where separation is important.
For a good numerical overview of local stability theory, interested readers are referred to the book by Schmid \& Henningson~\cite{book:schmid:2000}.

This leap in understanding the nature of instabilities allowed for better explanations and expanded the limiting dichotomy of open and closed flows, as the dynamics themselves could now be categorized as \emph{noise amplifiers} or \emph{oscillators} (more details given in ~\cite{huerre1990local,cossu1998convective}).
This distinction is of great importance, for example, for the selection and design of flow control strategies or the placement of sensors and actuators (see the review by Schmid and Sipp~\cite{schmid2016linear} and also Cossu~\cite{cossu2014introduction}).
Flow configurations of a noise-amplifying nature are much more difficult to control and predict because the dynamics are sensitive to both the amplitude and the spectral content of the incoming disturbance (\textit{viz.} vibrations, acoustics, or turbulence intensity).
In such flows, the incoming disturbances can excite otherwise stable modes and begin to extract energy from the base flow while they are transported downstream.
Natural oscillators, on the other hand, are characterized by the presence of a dominant unstable structure (\ie{} a \emph{global} instability), which locates the physical mechanism that extracts energy from the base flow in space.
When the instability is suppressed, the flow becomes stable and returns to the laminar state. 
More complex flow configurations (such as the jet in crossflow) may exhibit convective noise amplifier behavior or self-sustaining oscillatory behavior, depending on the combination of control parameters \cite{megerian2007transverse}.

Problems such as the one presented in \cref{eq: orr-sommerfeld-squire eqs} can be analyzed theoretically, but in practice, it is common to discretize the resulting equations in the wall-normal direction using spectral methods such as Chebyshev polynomials.
A famous example is the work of Orszag~\cite{orszag1971accurate} on the temporal stability of the canonical plane Poiseuille flow. 
At about the same time as \cite{huerre1985absolute,chomaz1988bifurcations,huerre1990local,monkewitz1993global}, computers and numerical methods began to reach sufficient maturity that the (weakly non-)parallel flow assumption could be relaxed and the spectral decomposition of the Navier-Stokes operator linearized in the vicinity of a truly two-dimensional base flow started to be computed. In fact, the computation of 2D baseflows in hydrodynamics began in the 1970s with direct methods ~\cite{benjamin1979convergence,winters1979finite,meyer1980computations,schreiber1983driven} before other, more efficient numerical approaches became popular and years later enabled the computation of 3D baseflows with increasing computer power.

In the mid-1980s, Jackson~\cite{jackson1987finite} and Zebib~\cite{jem:zebib:1987} obtained 2D steady states and computed stability analysis on the flow past a circular cylinder using full-matrices (Jackson used iterative methods to approximate the leading eigenvalues).
At the same time, the first use of a time-dependent numerical solver and Krylov methods for fluid dynamics seems to point to the work of Erikson \& Rizzi~\cite{eriksson_computer-aided_1985}, which coincides with the eigenvalue calculations of Tuckerman~\cite{tuckerman1985formation} and Marcus \& Tuckerman~\cite{marcus1987simulation_a,marcus1987simulation_b} on the flow between concentric rotating spheres.
Similar numerical methods have been explored by Goldhirsch \etal~\cite{goldhirsch1987efficient}, Christodoulou \& Scriven~\cite{christodoulou1988finding}, Tuckerman~\cite{tuckerman1989steady}, and Edwards \etal~\cite{jcp:edwards:1994}.
In these works, the prohibitive memory requirements of a matrix-forming approach are replaced by methods that use more accessible computational resources and are now commonly referred to as the \emph{matrix-free/time-stepping approach}.
These advances in algorithms eventually enabled the computation of 3D eigenmodes evolving on 2D solutions in the 1990s. This began with the work of Natarajan \& Acrivos~\cite{jfm:natarajan:1993} and Ramanan \& Homsy~\cite{ramanan1994linear} on the lid-driven cavity flow and later the analysis of Barkley \& Tuckerman~\cite{barkley1999stability} on the perturbed plane Couette flow.

To distinguish these analyses, which solve the linearized 2D equations, from a \emph{local stability} framework relying on a (weakly-) parallel flow approximation, Theofilis \etal~\cite{ptra:theofilis:2000} has called them \emph{(bi-)global stability analyses}.
Since then, the linear stability of numerous two-dimensional flow configurations have been studied: the lid-driven cavity flow~\cite{pof:albensoeder:2001,theofilis2004viscous}, the backward-facing step~\cite{jfm:lanzerstorfer:2012}, or the two-dimensional flow past a bump~\cite{jfm:ehrenstein:2005,jfm:ehrenstein:2008}, to name just a few.
Because of its importance for the development of local stability theory, the two-dimensional boundary layer flow has also been the focus of many investigations~\cite{aiaa:bagheri:2009,alizard2007spatially}.

Although matrix-free methods were already available, it is important to mention that several of the previously mentioned papers still considered the explicit construction of the linearized Navier-Stokes operator in combination with standard algebraic solvers to compute its eigenpairs (other examples include \cite{theofilis2004viscous,gelfgat2007stability,gulberg2016flow}).
For a comprehensive review of research contributions up to the early 2010s, with particular emphasis on this matrix-forming approach, see Theofilis~\cite{theofilis2011global} and Juniper \etal~\cite{juniper2014modal}.

Yet, over the past decade, the time-stepping framework has become increasingly popular and enabled the investigation of the stability properties of fully three-dimensional flows.
A large body of works has focused on two configurations, namely the jet in crossflow~\cite{jfm:bagheri:2009,jfm:ilak:2012,peplinski2014stability} or the boundary layer flow past three-dimensional roughness elements~\cite{jfm:loiseau:2014,pof:citro:2015,jfm:kurz:2016,jfm:brynjell:2017,bucci2018jfm,bucci2021influence,ma_mahesh_2022,wu_romer_axtmann_rist_2022}.
The stability of lid-driven and shear-driven three-dimensional cavities with spanwise end-walls has also been investigated in~\cite{Feldman_PoF_2010,Kuhlmann_PoF_2014,loiseau2016intermittency,jfm:liu:2016,jfm:picella:2018,gelfgat2019linear}.
The same methodology has also been employed to compute the leading optimal perturbation \cite{amr:schmid:2014} in magnetohydrodynamic flows \cite{zikanov2014laminar}, or best exemplified by \cite{jfm:blackburn:2008,jfm:blackburn:2008b} on backward facing steps and stenotic pipe flows.
It has also been used to solve high-dimensional Ricatti equations for linear optimal control in \cite{tcfd:semeraro:2018}, or to study the stability properties of flow governed by the compressible Navier-Stokes equations with or without shocks~\cite{sansica2020laminar,pof:fabre:2008,fabre2018practical}.
These include modal and non-modal stability of compressible boundary layers~\cite{jfm:robinet:2007,jfm:guiho:2016,hildebrand2018simulation,bugeat20193d,bugeat2022low}, cavities~\cite{bres2008three,theofilis2003algorithm,yamouni2013interaction,sun2017spanwise}, wavepackets in jets~\cite{nichols2011global,beneddine2015global,prf:semeraro:2017}, transonic buffet~\cite{jcp:crouch:2007,jfm:crouch:2009,timme2016towards,prf:paladini:2019,jfm:crouch:2019,jfm:paladini:2019}, including the flow past the NASA Common Research wing Model~\cite{timme2016towards,timme_2020}, wakes~\cite{meliga2010effect} and bluff bodies~\cite{mack2011globala,mack2011globalb,jfm:sansica:2018,sansica2020laminar}.

The matrix-free approach has been the key to efficient computation and (Floquet) stability analysis of time-periodic solutions, beginning with Schatz, Barkley \& Swinney~\cite{schatz1995instability}, followed by the secondary instability of the flow past a circular cylinder with the canonical work of Barkley \& Henderson~\cite{jfm:barkley:1996} and later by the study of Blackburn, Marques \& Lopez~\cite{blackburn2005symmetry}.
Other work includes the backward-facing step flow~\cite{barkley2002three}, the study of the stability properties of a pulsatile stenotic pipe flow~\cite{jfm:blackburn:2008b}, the flip-flop instability in the wake of two side-by-side cylinders~\cite{jfm:carini:2014}, the secondary bifurcation in a shear-driven cavity flow~\cite{bengana2019bifurcation}, or the study of the vortex pairing mechanism in a harmonically forced axisymmetric jet~\cite{jfm:leopold:2019}.
In parallel with these developments in the hydrodynamic stability community, similar numerical methods and tools have been explored in the community looking at turbulence from the point of view of a dynamical system.
Matrix-free methods and clever exploration of flow symmetries can significantly reduce computational costs and allow the calculation of \emph{exact coherent states}\footnote{Exact coherent states can be fixed points in the original reference frame or in a co-moving reference frame (in which case they are called \emph{traveling waves}) or true periodic states~\cite{tuckerman1988global,dijkstra2014numerical}.} in the turbulent basin of attraction.
These exact coherent states include relative periodic orbits~\cite{jfm:kreilos:2013,crm:rawat:2014}, chaotic saddles~\cite{kawahara2001periodic,jfm:wedin:2004,jfm:willis:2013,paranjape2020oblique} or edge states~\cite{jpsj:itano:2001,prl:skufca:2006,canton2020critical}. Interested readers may refer to the specialized codes \textit{Channelflow}~\cite{channelflow} or \textit{Openpipeflow}~\cite{openpipeflow} and the corpus of related works.
It is also noteworthy that in the special case of weakly non-parallel shear flows, the development of \emph{parabolized stability equations}~\cite{herbert1997parabolized,li1995mathematical} paved the way for the estimation of the neutral stability curves of the Blasius boundary layer~\cite{herbert1991boundary,bertolotti1992linear,pof:liu:2008}
and its secondary instabilities~\cite{herbert1987floquet,herbert1988secondary}, which may involve curvature effects inducing centripetal Görtler vortices~\cite{malik1999secondary,ren2015secondary}.

Building on the framework of Orszag and aiming at the geometric flexibility of the finite element method, Patera~\cite{patera1984spectral} and Maday \& Patera~\cite{maday1989spectral} laid the foundation of the spectral element method (SEM). 
SEM has since become a popular discretization strategy in computational fluid dynamics (CFD).
Within the incompressible hydrodynamic stability community, the spectral element solver \texttt{Nek5000}~\cite{fischer2008nek5000} has established itself as one of the leading high-performance open-source CFD codes.
Most of the aforementioned three-dimensional stability analyses heavily relied on \texttt{Nek5000}.
Except for the KTH Framework\footnote{Freely available at \url{https://github.com/KTH-Nek5000/KTH_Framework}. Apart from some linear stability capabilities, this toolbox also provides additional capabilities such as full restarts, turbulence statistics, additional boundary conditions, or incorporating user variables and various volume forces.}, relatively few toolboxes have been developed for \texttt{Nek5000} despite its large user base.
Even then, the capabilities of this toolbox (as far as linear stability is concerned) are limited to simple fixed-point calculations using the selective frequency damping approach~\cite{pof:akervik:2006}, while the leading eigenpairs of the linearized Navier-Stokes operator are calculated using \texttt{PARPACK}~\cite{lehoucq1998arpack,maschho1996portable}, at the expense of introducing new dependencies for the code.
Linear stability analysis capabilities are also available to \texttt{Nek5000}'s brethren, including \texttt{Nektar++}~\cite{cantwell2015nektar++,book:kaniadakis:2005}, which can handle a variety of mesh types, and \texttt{Semtex}~\cite{blackburn2019semtex} designed for spanwise periodic flows, with examples given by~\cite{sherwin2005three,elston2006primary,mao2011transient,blackburn2013lower,albrecht2015triadic}.
The finite element code \texttt{FreeFem++}\cite{hecht2012new} can also be used to extract the Jacobian matrix directly with examples including~\cite{marquet2008amplifier,marquet2008sensitivity,yamouni2013interaction,citro2014three,citro2015linear,kaiser2018stability}.
The solver capabilities can be extended using the \texttt{StabFem}~\cite{fabre2018practical} MATLAB suite for compressible flows.
Finally, the \texttt{FEniCS}~\cite{alnaes2015fenics} project is also another consolidated alternative with a Python interface.
Recently several other spectral-based Python projects for solving partial differential equations became available.
These include \texttt{SpectralDNS}~\cite{mortensen2016high,mortensen2018shenfun}, \texttt{FluidSim}~\cite{augier2018fluiddyn,mohanan2018fluidsim}, \texttt{Dedalus}~\cite{burns2020dedalus} and \texttt{Coral}~\cite{miquel2021coral}.
The finite volume framework \texttt{BROADCAST}~\cite{poulain2022broadcast} has become available for the study of 2D curvilinear structured grids with support for nonlinear and linear calculations of compressible flows.

\bigskip

The aim of the present work is by no means to be an exhaustive review, but rather a comprehensive introduction to the Krylov methods underlying the most recent works on the stability of very large-scale dynamical systems such as fully three-dimensional flows.
For that purpose, this manuscript is organized as follows: first, brief overviews of the theoretical framework and numerical methods are given in \cref{sec: theoretical framework} and \cref{sec: numerical methods}, respectively.
A collection of examples illustrating the use of these techniques in fluid dynamics is provided in~\cref{sec: examples}, while some particular theoretical and practical points are discussed in \cref{sec: discussion}.
Finally, conclusions and perspectives are given in \cref{sec: conclusion}.

\section{Theoretical framework}
\label{sec: theoretical framework}

We focus on the analysis of the stability properties of high-dimensional nonlinear dynamical systems, typically arising from the discretization of partial differential equations such as the incompressible Navier-Stokes equations.
Once discretized, the governing equations are generally expressed as a nonlinear system of first-order ordinary differential equations
\[
\dfrac{d X_j}{dt} = F_j \left( \left\{ X_i ; i = 1, \cdots, n \right\}, t \right),
\]
where $n$ is the \emph{dimension} of the discretized system.
Using the notation $\mathbf{X}$ and $\mathbf{F}$ for the sets $\left\{ X_j ; j = 1, \cdots, n \right\}$ and $\left\{ F_j ; j = 1, \cdots, n \right\}$, this system can be written as
\begin{equation}
  \dfrac{d\mathbf{X}}{dt} = \mathbf{F}(\mathbf{X}, t),
  \label{eq: nonlinear dynamical system}
\end{equation}
where $\mathbf{X} \in \mathbb{R}^n$ is the \emph{state vector} of the system and $t$ is a continuous variable that denotes time. 
If the system is supplemented with constraints (\eg{} the divergence-free constraint in incompressible fluid dynamics), $\mathbf{F}$ is then understood as the restriction of dynamics in the feasible set of such constraints.
One can also consider the equivalent discrete-time system
\begin{equation}
  \mathbf{X}_{k+1} = \Phi_\tau(\mathbf{X}_k),
  \label{eq: discrete-time nonlinear dynamical system}
\end{equation}
where $\Phi_\tau(\mathbf{X})$ is the forward map defined as
\begin{equation}
  \Phi_\tau(\mathbf{X}) = \int_{0}^{\tau} \mathbf{F}(\mathbf{X} (t), t)\ \mathrm{d}t.
  \label{eq: forward map}
\end{equation}
Such a discrete-time system may result from the time discretization of the governing equations (with $\tau$ the sampling period) or in the study of periodic orbits (with $\tau$ the period of the orbit).
In practice (and due to discretization errors), an approximation to the action of an operator is made by computing many (\ie{} $\tau/\Delta t$) small time-steps (\ie{} $\Delta t \ll 1$), each consisting of a rational or polynomial approximation to the operator.
The term \textit{timestepper}, coined in 2000 by Tuckerman \& Barkley~\cite{tuckerman2000bifurcation}, refers to the adaptation of a time integration code to perform bifurcation analyses.
In what follows, we consider an exact operator notation as a shorthand notation for the timestepper approximation of operators (linear or nonlinear).

In the following sections, we present a definition of \emph{fixed points}, \emph{periodic orbits}, and \emph{linear stability analyses}.
These are the fundamental concepts required to characterize the properties of the system under investigation.
In particular, we focus on the \emph{modal} (\ie{} asymptotic) and \emph{non-modal} (\ie{} finite-time) stability analysis, which are the classical and a more modern approach that has become increasingly popular in fluid dynamics in recent decades.

We note that part of the community has also shifted its attention to \emph{nonlinear optimal perturbations}, which is beyond the scope of this work. Interested readers are referred to the work of \cite{rpp:kerswell:2014} and the references therein.

\subsection{Fixed points and periodic orbits}

Nonlinear dynamical systems, such as \cref{eq: nonlinear dynamical system} can admit different solutions or attractors that form the backbone of their phase space: \emph{fixed points} (steady dynamics), \emph{periodic orbits} (periodic dynamics), \emph{tori} (quasi-periodic dynamics) or \emph{strange attractors} (chaotic dynamics).
Hereafter, our attention will be solely focused on fixed points and periodic orbits.

\subsubsection{Fixed points}

\begin{figure}\centering
\includegraphics[width=0.48\columnwidth]{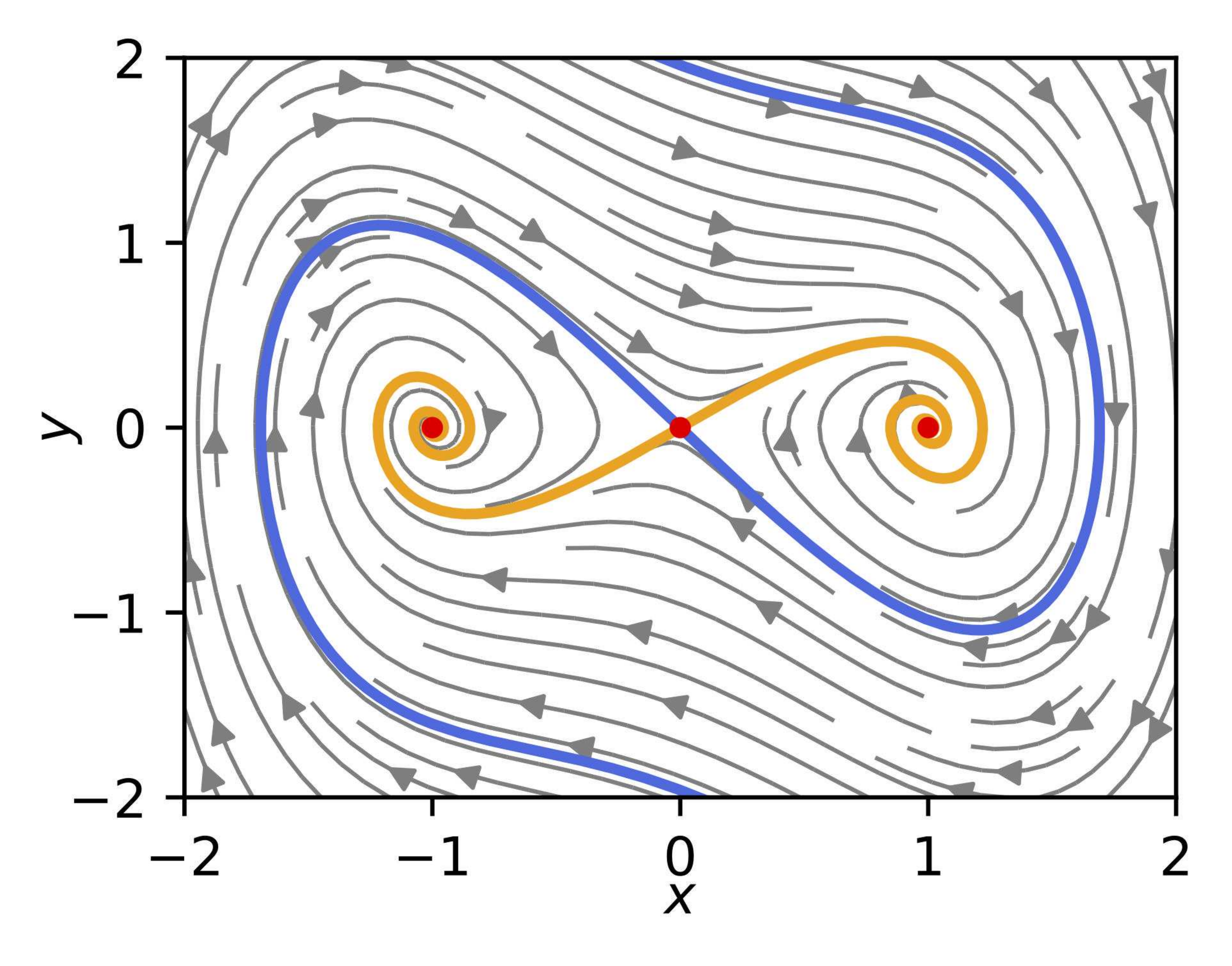}
\caption{Phase portrait of the unforced Duffing oscillator. The red dots denote the three fixed points admitted by the system. The blue (resp. orange) thick line depicts the stable (resp. unstable) manifold of the saddle point located at the origin. Grey lines highlight a few trajectories exhibited for different initial conditions.}
\label{fig: duffing oscillator}
\end{figure}

For a continuous-time dynamical system described by \cref{eq: nonlinear dynamical system}, the fixed points $\mathbf{X}^*$ are particular equilibrium solutions satisfying
\begin{equation}
  \mathbf{F}(\mathbf{X}^*) = 0.
  \label{eq: continuous-time fixed point}
\end{equation}
Similarly, for a discrete-time system described by \cref{eq: discrete-time nonlinear dynamical system}, fixed points are solutions to
\begin{equation}
  \Phi_{\tau}(\mathbf{X}^*) = \mathbf{X}^* \quad \forall \tau.
  \label{eq: discrete-time fixed point}
\end{equation}
These particular solutions are thus characterized by the absence of dynamics: the system is in a steady state.
Since we are dealing with nonlinear equations, both \cref{eq: nonlinear dynamical system} and \cref{eq: discrete-time nonlinear dynamical system} can have a multitude of fixed points.
This is illustrated by a dynamical system as simple as the \emph{Duffing oscillator}
\begin{equation}
\left\{
\begin{aligned}
  \dot{x} & = y, \\
  \dot{y} & = -\dfrac{1}{2}y + x - x^3.
\end{aligned}
\right.
\end{equation}
Because of the stabilizing cubic term in the $y$-equation, the Duffing oscillator admits three fixed points
\begin{itemize}
\item a saddle at the origin $\mathbf{X}^* = (0, 0)$,
\item two linearly stable spirals located at $\mathbf{X}^* = (\pm 1, 0)$.
\end{itemize}
These different fixed points, along with typical trajectories, are depicted in \cref{fig: duffing oscillator}.
A multiplicity of fixed points also occurs in nonlinear dynamical systems as complex as the Navier-Stokes equations, see for instance \cite{ijnmf:guevel:2018}.
The decision of which of these fixed points is the relevant one from a physical point of view depends on the problem and is left to the judgment of the user.
The computation of equilibria is a cornerstone for all analyses described in this work.
Numerical methods to solve \cref{eq: continuous-time fixed point} and \cref{eq: discrete-time fixed point} are discussed further in \cref{sec: numerical methods}.

\subsubsection{Periodic orbits}

The second type of equilibria of interest to us is \emph{periodic orbits}.
Such solutions are characterized by dynamics repeating themselves after a given period $\tau^*$, \ie{}
\begin{equation}
\mathbf{X}(t + \tau^*) = \mathbf{X}(t).
\end{equation}
They can also be understood as fixed points of the forward map $\Phi_{\tau}$ for $\tau = \tau^*$, \ie{}
\begin{equation}
  \mathbf{X}^* = \Phi_{\tau}(\mathbf{X}^*) \quad \textrm{for } \tau = \tau^*.
  \label{eq: periodic orbit}
\end{equation}
In practice, $\tau^*$ is often unknown, and therefore one must solve simultaneously for a point $\mathbf{X}^*$ on the orbit and the period $\tau^*$.
Moreover, any point on the orbit satisfies the equation above so that \cref{eq: periodic orbit} admits an infinite number of solutions.
To close the system, a \emph{phase condition} often needs to be included.
A canonical example of a periodic orbit in fluid dynamics is the periodic vortex shedding in the wake of a two-dimensional cylinder at a low Reynolds number.
As for fixed points, a nonlinear dynamical system may admit multiple periodic orbits, each with its own period $\tau^*$.
This is particularly true when the system evolves on a strange attractor (chaotic dynamics) on which an infinite number of unstable periodic orbits (UPO) coexist with arbitrary periods.
These findings go back to Kawahara \& Kida~\cite{kawahara2001periodic}, who pioneered the extraction of periodic orbits in a fully three-dimensional plane Couette flow using a Newton root-finding technique.
The discovery of a large number of UPOs buried in turbulent attractors for various flow configurations supports the view that turbulence is a very high-dimensional dynamical system whose trajectories repeatedly visit unstable exact coherent structures~\cite{duguet2008transition,kawahara2012significance}.
Since the discovery of recurrent spatiotemporal structures concealed in a turbulent attractor, UPOs have proven to be favorable kernels~\cite{ding2016estimating} for predicting turbulence statistics due to their harmonic temporal structure~\cite{cvitanovic_2013}.
These invariant solutions can be naturally sustained by the flow and have been found to be energetically relevant for predicting or even reconstructing turbulence statistics (if enough of them are found)~\cite{chandler2013invariant,pof:lucas:2015}.
Despite growing interest since the first UPOs were obtained, the methods used to obtain them involve brute force, although attempts are being made to change this~\cite{page2020searching}.
\bigskip

\begin{figure}\centering
\includegraphics[width=.48\columnwidth]{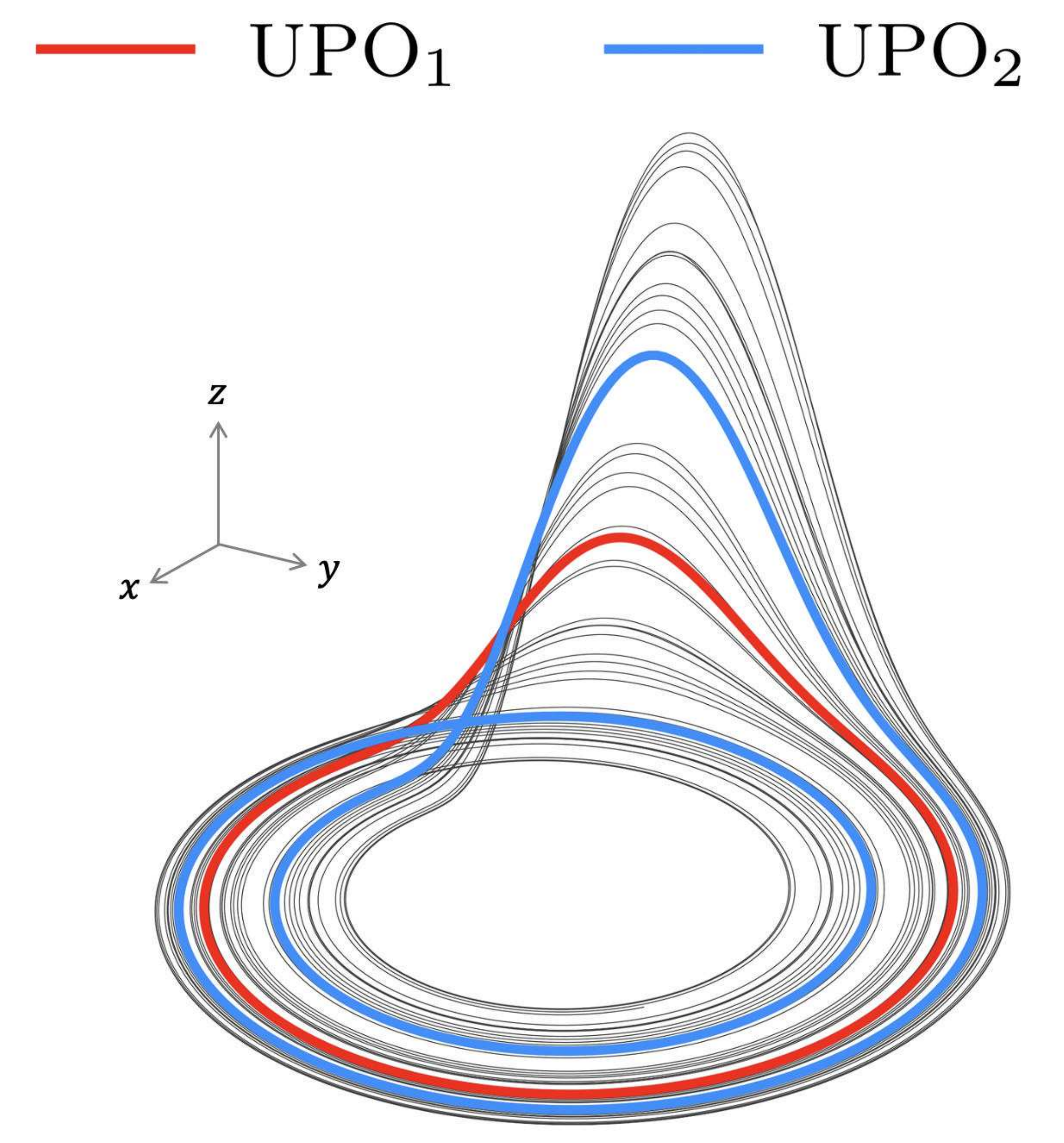}
\caption{Strange attractor for the R\"ossler system with parameters $a = 0.1$, $b = 0.1$ and $c = 14$. Colored lines depict a $\tau_1$ (red) and $\tau_2$ (blue) unstable periodic orbit.}
\label{fig: rossler system}
\end{figure}

Considering the R\"ossler system~\cite{pla:rossler:1976}
\begin{equation}
\left\{
\begin{aligned}
  \dot{x} & = -y -z, \\
  \dot{y} & = x + ay, \\
  \dot{z} & = b + z(x - c),
\end{aligned}
\right.
\end{equation}
for $a = 0.1$, $b = 0.1$, and $c = 14$, \cref{fig: rossler system} shows its famous strange attractor as well as two UPOs embedded in this attractor.
These were obtained using a simple shooting Newton method and continuation initialized with their stable counterparts at lower values of the control parameter $c$.

\subsection{Modal and non-modal linear stability}

To avoid repetition, we restrict ourselves in this section to the modal and non-modal stability of fixed points, although these concepts can easily be applied to periodic orbits by replacing the Jacobian matrix with the monodromy matrix. 
The linearly stable or unstable nature of the fixed point $\mathbf{X}^*$ is characterized by the fate of infinitesimal perturbations.
If perturbations eventually decay, the equilibrium is considered stable; otherwise, it is considered unstable.
Note, however, that an infinite time horizon is allowed for the return to equilibrium.
Thus, a fixed point can be classified as stable even if a small perturbation transiently departs very far from it before returning toward it only asymptotically with $t \to \infty$.
This distinction between asymptotic and finite-time evolution gives rise to the concepts of \emph{modal} and \emph{non-modal} stability.
The reader will find a detailed discussion of these concepts in fluid dynamics in \cite{schmid2007nonmodal,amr:schmid:2014}.

The dynamics of a perturbation $\mathbf{x} = \mathbf{X} - \mathbf{X}^*$ is governed by
\begin{equation}
\dot{\mathbf{x}} = \mathbf{F}\left( \mathbf{X}^* + \mathbf{x} \right).
\end{equation}
Assuming $\mathbf{x}$ is infinitesimally small, $\mathbf{F}(\mathbf{X})$ can be approximated by its first-order Taylor expansion around $\mathbf{X} = \mathbf{X}^*$, leading to
\begin{equation}
\dot{\mathbf{x}} = \mathbf{Lx},
\end{equation}
where
\begin{equation}
\mathbf{L} = \dfrac{\partial \mathbf{F}}{\partial \mathbf{X}},
\end{equation}
is the $n \times n$ Jacobian of $\mathbf{F}$.
Starting from an initial condition $\mathbf{x}_0$, the perturbation at time $\tau$ is given by
\begin{equation}
\mathbf{x}(\tau) = \exp(\tau \mathbf{L}) \mathbf{x}_0.
\end{equation}
The operator $\mathbf{M}_{\tau} = \exp(\tau \mathbf{L})$ is the \emph{exponential propagator} of the linearized system and corresponds to the Jacobian of the forward map $\Phi_\tau$ linearized in the vicinity of the fixed point $\mathbf{X}^*$.
For periodic orbits, this linearized propagator is defined as
\begin{equation}
\mathbf{M}_\tau = \int_0^\tau \mathbf{L}(t) \ \mathrm{d}t,
\end{equation}
with $\mathbf{L}(t + \tau) = \mathbf{L}(t)$ and is known as the \emph{monodromy matrix}.
We again use the notation of the exact operator as shorthand for its discrete counterpart obtained by integration over many small timesteps.
System operators can be linearized exactly by using a timestepper code to solve the set of linearized equations derived analytically, or by using automatic differentiation tools. 
Such an exact linear operator is in itself a discrete approximation, again requiring temporal integration over many small timesteps.
Alternatively, the action of the linearized operator can be emulated using less accurate finite-difference approximation~\cite{beneddine2015global,mathias2022optimal}.
While the computational cost remains the same for a first-order approximation, it increases with the order of the finite-difference scheme considered.

\subsubsection{Modal stability}\label{subsubsec: theoretical -- modal stability}
Introducing the (Euclidean) 2-norm of $\mathbf{x}(\tau)$
and the eigendecomposition
\begin{equation}
\mathbf{L} = \mathbf{V} \boldsymbol{\Lambda} \mathbf{V}^{-1},
\end{equation}
one can easily show
\begin{equation}
  \exp(2\tau \lambda_r) \leq \| \exp(\tau \mathbf{L}) \|_2^2 \leq \kappa(\mathbf{V}) \exp(2\tau \lambda_r),
  \label{eq: inequality}
\end{equation}
where $\lambda_r = \Re(\lambda_1)$ is the real part of the leading eigenvalue (\ie{} with greatest real part) of $\mathbf{L}$ and $\kappa(\mathbf{V}) = \| \mathbf{V} \|_2 \| \mathbf{V}^{-1} \|_2$ is the condition number of the matrix of eigenvectors $\mathbf{V}$ (with $\kappa \geq 1$).
Asymptotic stability is characterized by
\begin{equation}
\lim_{\tau \to \infty} \| \exp(\tau \mathbf{L}) \mathbf{x}_0 \|_2 = 0 \quad \forall \ \mathbf{x}_0,
\end{equation}
so a sufficient condition is that all eigenvalues of $\mathbf{L}$ have a negative real part (equivalent to all eigenvalues of $\mathbf{M}_\tau$ being inside the unit circle).
The perturbation decaying the slowest is given by the eigenvector $\mathbf{v}_1$ associated with the least stable eigenvalue $\lambda_1$.
Hereafter, a fixed point $\mathbf{X}^*$ is classified as follows
\begin{itemize}
\item if $\Re(\lambda_1) > 0$, the dynamics of simulations initialized with an initial condition non-orthogonal to the leading eigenvector (\ie{} $\mathbf{x}_0 \not\perp \mathbf{v}_1$) will grow exponentially rapidly.
The fixed point $\mathbf{X}^*$ is deemed \emph{linearly unstable}.

\item if $\Re(\lambda_1) < 0$, the dynamics of simulations initialized with any initial condition will eventually decay exponentially fast. 
The fixed point $\mathbf{X}^*$ is thus \emph{linearly stable}.
\end{itemize}
The case $\Re(\lambda_1) = 0$ is special and corresponds to a \emph{non-hyperbolic} fixed point.
Its stability cannot be determined by the eigenvalues of $\mathbf{L}$ alone and one has to resort to \emph{weakly nonlinear analysis} or \emph{center manifold reduction}.
Interested readers can refer to \cite{book:wiggins:2006, jfm:sipp:2007, book:manneville:2010, jfm:carini:2015} for more details.

\subsubsection{Non-modal stability}

The upper bound in \cref{eq: inequality} involves the condition number $\kappa(\mathbf{V})$ of the matrix of eigenvectors.
This leads to the classification of linear systems into two distinct sets with fundamentally different finite-time stability properties.
Systems for which $\kappa(\mathbf{V}) = 1$ are called \emph{normal} operators.
In this case, the eigenvectors of $\mathbf{L}$ form an orthonormal set such that
\begin{equation}
\bm{V}^{-1} = \mathbf{V}^H,
\end{equation}
where superscript $H$ denotes the Hermitian, \ie{} complex-conjugate transpose operation.
The finite-time and asymptotic stability properties of the system are identical, and the dynamics cannot exhibit transient growth: analyzing the spectrum of $\mathbf{L}$ is sufficient to fully characterize the system.
When $\kappa(\mathbf{V}) > 1$, the matrix $\mathbf{L}$ is said to be \emph{non-normal}.
This can be defined by introducing the adjoint operator $\mathbf{L}^{\dagger}$ satisfying
\begin{equation}
\langle \mathbf{y} \vert \mathbf{Lx} \rangle = \langle \mathbf{L}^{\dagger} \mathbf{y} \vert \mathbf{x} \rangle,
\end{equation}
where $\langle \cdot \vert \cdot \rangle$ is a suitable inner product (typically the inner product induced by the $\ell_2$ norm), along with appropriate boundary conditions.
Non-normality then corresponds to the fact that $\mathbf{L}$ and its adjoint $\mathbf{L}^{\dagger}$ do not commute, \ie{}
\begin{equation}
\mathbf{L}^{\dagger} \mathbf{L} \neq \mathbf{LL}^{\dagger}.
\end{equation}
Its eigenvectors no longer form an orthonormal basis for $\mathbb{R}^n$ and the dynamics may exhibit transient growth.
From a physical point of view, transient growth can be understood as a constructive interference involving almost colinear eigenvectors.
The larger the non-normality of $\mathbf{L}$, the larger the maximum transient growth with perturbations being (possibly) amplified by several orders of magnitude before the exponential decay eventually takes over (assuming all the eigenvalues have negative real parts).
In the most extreme scenario, this non-normality is characterized by $\mathbf{L}$ admitting a non-diagonalizable Jordan block leading to algebraic growth. More details on adjoint operators are provided in \cite{trefethen1997numerical,hill1995adjoint,arfm:luchini:2014}.

Given this observation, one can now ask a more subtle question about the stability of the fixed point $\mathbf{X}^*$, namely
\begin{quote}
  \emph{How far from the fixed point $\mathbf{X}^*$ can an arbitrary perturbation $\mathbf{x}_0$ go (or equivalently to what extent can it be amplified) at a finite time $\tau$?}
\end{quote}
The answer to this question can be obtained by solving the following optimization problem
\begin{equation}
  \mathcal{G}(\tau) = \max_{\mathbf{x}_0} \dfrac{\| \exp(\tau \mathbf{L}) \mathbf{x}_0 \|_2^2}{\| \mathbf{x}_0 \|_2^2},
\end{equation}
where $\| \exp(\tau \mathbf{L}) \|_2^2$ is the vector-induced matrix norm optimizing over all possible initial conditions $\mathbf{x}_0$ and $\mathcal{G}(\tau)$ is the maximal amplification gain of the perturbation at time $\tau$.
Introducing the singular value decomposition of the exponential propagator $\mathbf{M}_\tau$
\begin{equation}
\mathbf{M}_\tau = \mathbf{U} \boldsymbol{\Sigma} \mathbf{V}^H,
\end{equation}
the maximum gain is simply given by
\begin{equation}
\mathcal{G}(\tau) = \sigma_1^2,
\end{equation}
where $\sigma_1$ is the largest singular value of $\mathbf{M}_\tau$.
The optimal initial condition $\mathbf{x}_0$ is then given by the first right singular vector (\ie{} $\mathbf{x}_0 = \mathbf{v}_1$) while the associated response is $\mathbf{x}(\tau) = \sigma_1 \mathbf{u}_1$, where $\mathbf{u}_1$ is the leading left singular vector.
When/if one has access to the adjoint operator, computing these quantities can also be recast as an eigenvalue problem (EVP) rather than a singular value decomposition (SVD).

In fluid dynamics, this concept of non-normality and optimal perturbations leads to a better understanding of the formation and ubiquity of velocity streaks in the transition to turbulence of wall-bounded shear flows~\cite{trefethen1993hydrodynamic,jfm:brandt:2003,ejmbf:brandt:2006,ejmbf:brandt:2014}.
It also sheds some light on the importance of shear layer instability~\cite{ijnmf:barkley:2008,jfm:blackburn:2008,pof:cantwell:2010}.
Extension to periodic orbits has been considered for instance in~\cite{jfm:blackburn:2008b,mao2011transient}.
Although not considered herein, a similar concept exists in the frequency domain, leading to a resolvent analysis (see \cite{amr:schmid:2014} for more details).

\subsubsection*{Illustration}

\begin{figure*}\centering
\includegraphics[width=.66\textwidth]{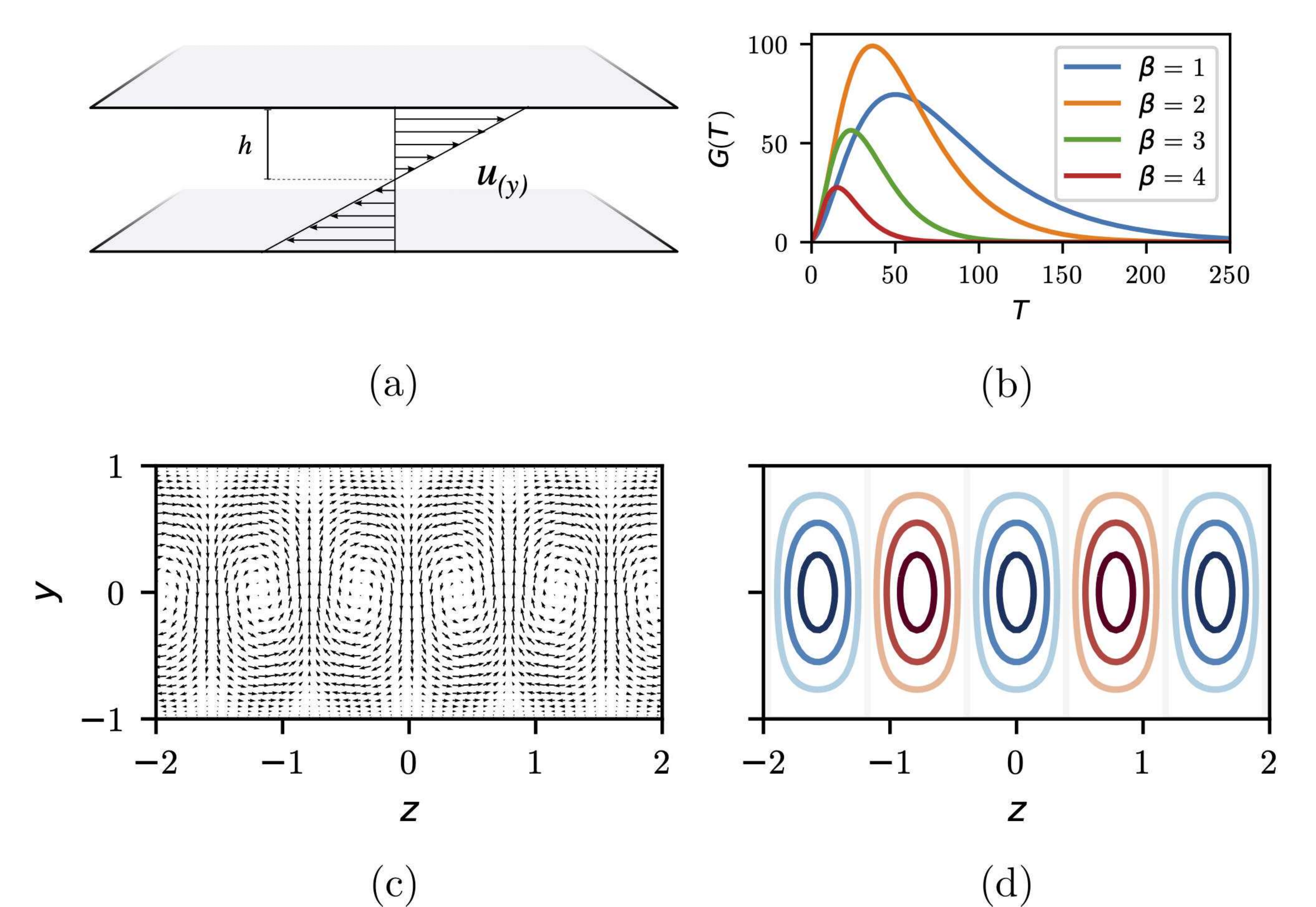}
\caption{Illustration of the optimal perturbation analysis for the plane Couette flow at $Re = 300$ for streamwise wavenumber $\alpha = 0$: (a) Schematic of the flow, (b) Optimal gain curve for different spanwise wavenumbers $\beta$, (c) Optimal perturbation in-plane velocity ($v,w$) and (d) optimal response out of plane velocity ($u$) for $\beta = 2$. The optimal perturbation consists of streamwise oriented vortices, while the corresponding response at time $T$ consists of high and low-speed streaks. Reproduced from \cite{chapter:loiseau:2019}.}
\label{fig: couette}
\end{figure*}

Let us illustrate the concepts of optimal perturbation and transient growth using a simple flow configuration.
For this purpose, consider the incompressible flow of a Newtonian fluid induced by two flat plates moving in opposite directions in the plane, as sketched in \cref{fig: couette}(a).
The resulting flow, known as \emph{plane Couette flow}, is given by
\[
U(y) = y.
\]
It is a linearly stable fixed point of the Navier-Stokes equations.
Despite its linear stability, subcritical transition to turbulence due to finite amplitude perturbations can occur at Reynolds numbers as low as $Re = 325$~\cite{mer:manneville:2016}.
Without delving too much into the mathematical and physical details of such a subcritical transition, part of the explanation can be given by linear optimal perturbation analysis.
The dynamics of an infinitesimal perturbation $\mathbf{x} = \begin{bmatrix} v & \eta \end{bmatrix}^T$, characterized by a certain wavenumber $\mathbf{k} = \alpha \mathbf{e}_x + \beta \mathbf{e}_z$, is governed by the Orr-Sommerfeld-Squire equations (matrix counterpart of \cref{eq: orr-sommerfeld-squire eqs}) written as
\begin{equation}
\dfrac{d}{dt} \begin{bmatrix} v \\ \eta \end{bmatrix}
=
\begin{bmatrix}
  \mathbf{L}_{\textrm{OS}} & 0 \\ \mathbf{C}  & \mathbf{L}_{\textrm{S}}
\end{bmatrix}
\begin{bmatrix} v \\ \eta \end{bmatrix},
\end{equation}
with $v$ the wall-normal velocity component of the perturbation and $\eta$ its wall-normal vorticity.
$\mathbf{L}_{\textrm{OS}}$ denotes the Orr-Sommerfeld operator, while $\mathbf{L}_{\textrm{S}}$ and $\mathbf{C}$ represent the Squire operator and the coupling term, respectively.
For certain pairs of wavenumbers, this Orr-Sommerfeld-Squire operator is highly non-normal.
In particular, perturbations with non-zero spanwise wavenumbers can experience large transient growth.
This is illustrated in \cref{fig: couette}(b), where the evolution of the optimal gain $\mathcal{G}(\tau)$ is shown for different spanwise wavenumbers.
The maximum amplification over all target times and wavenumber pairs is $\mathcal{G}_{\textrm{opt}} \simeq 100$.
The initial perturbation $\mathbf{x}_0$ is shown in \cref{fig: couette}(c).
This perturbation corresponds to streamwise-oriented vortices that eventually lead to streamwise velocity streaks, as shown in \cref{fig: couette}(d).
Although this perturbation eventually decays exponentially fast in a purely linear framework, it has been shown that its transient amplification even at moderately small amplitude can be sufficient to trigger the transition to turbulence when used as an initial condition in a nonlinear simulation~\cite{arfm:luchini:2014}.
For more details on subcritical transitions and the extension of optimal perturbation analysis to nonlinear operators, interested readers may refer to \cite{amr:schmid:2014} and \cite{rpp:kerswell:2014}.

\subsection{Bifurcation analysis}

\begin{figure*}\centering
\includegraphics[width=0.75\textwidth]{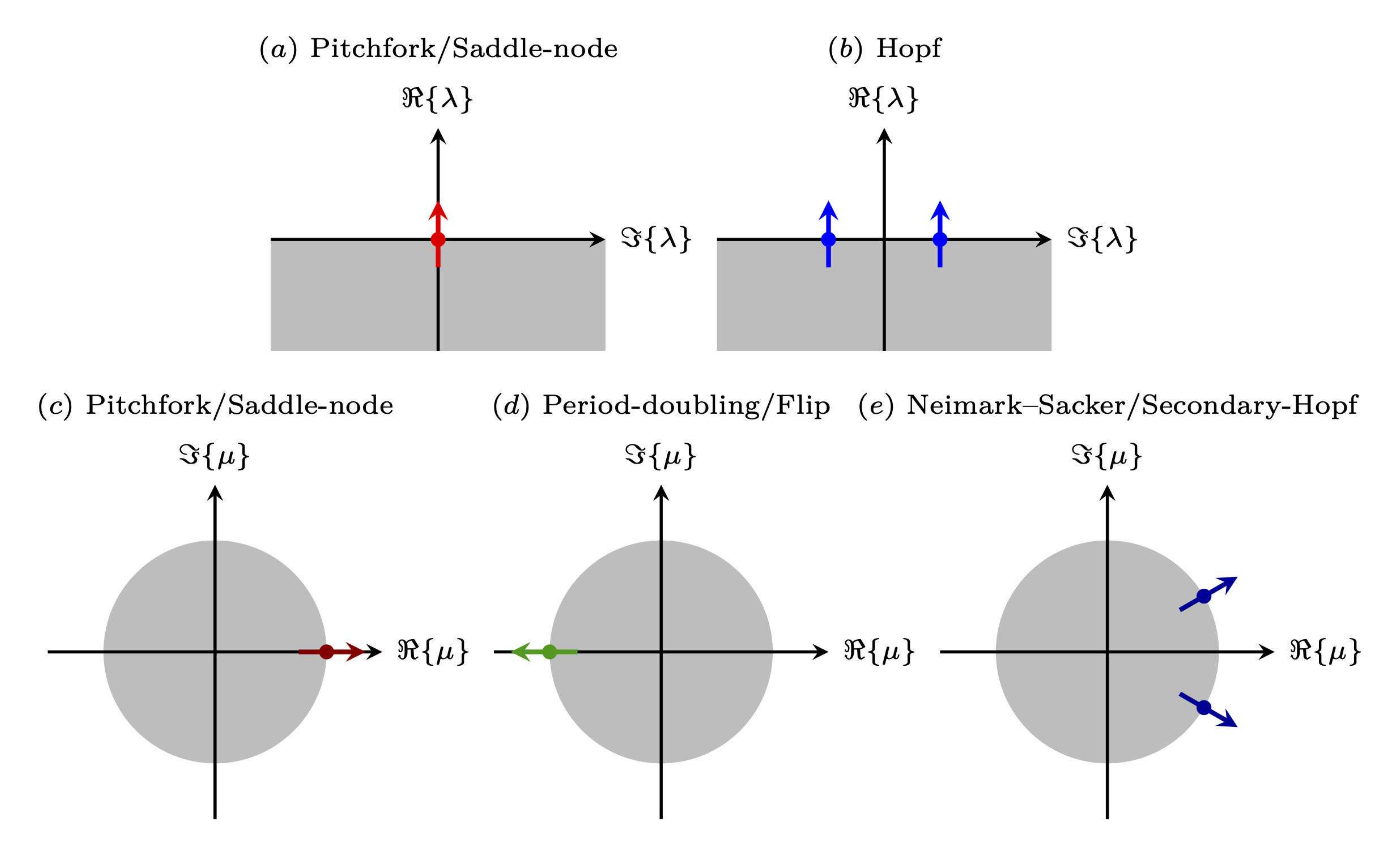}
\caption{Eigenvalue patterns associated with the standard bifurcations encountered for fixed points ($a$ and $b$), and for limit cycles ($c$ to $e$). In each case, the shaded region indicates the stable part of the spectrum (\ie{} the lower complex half-plane for fixed-point stability, and the unit disk for periodic orbits).}
\label{fig: bifurcation analysis}
\end{figure*}

The eigenvalue analysis of the Jacobian matrix $\mathbf{L}$ or monodromy matrix $\mathbf{M}_\tau$ plays a key role in determining the type of bifurcation that occurs when a control parameter (\eg{} the Reynolds number) is varied.
In the remainder of this section, a brief overview of the standard bifurcations and their correspondence to eigenvalues is given for completeness (see \cref{fig: bifurcation analysis} for a schematic representation).

\subsubsection{Bifurcations of fixed points}

The asymptotic stability properties of fixed points are related to the eigenvalues of the linearized operator $\mathbf{L}$.
Bifurcation analysis is concerned with how these properties evolve as the parameters of the system are varied.
For simplicity, we will only consider situations where a single parameter is varied.
The value of the control parameter at which a change in stability occurs is a \emph{bifurcation point}.
The bifurcations most commonly encountered in mechanics are the pitchfork, saddle-node, and Hopf bifurcations.
Each leads to a qualitatively different behavior before and after the bifurcation point.
Pitchfork and Hopf bifurcations come in two flavors, namely \emph{subcritical} and \emph{supercritical} (depending on whether the solutions created at the bifurcation point are themselves stable or not).

\paragraph{Pitchfork bifurcation}

This type of bifurcation is often encountered in systems with symmetries.
The canonical example in mechanics is that of a flexible beam on top of which a static load is applied.
Below a critical load, the beam remains upright.
As the load increases, the beam suddenly buckles to the left or right.
The upright position becomes linearly unstable, and two new stable equilibria are created as the bifurcation point is crossed.
Mathematically, the pitchfork bifurcation can be distilled into the following normal form
\begin{equation}
\dot{x} = \lambda x \pm x^3,
\end{equation}
with the sign in front of the cubic term determining whether the bifurcation is super(-) or sub(+) critical.
From a linear stability point of view, the linearized operator has a purely real eigenvalue going from being stable (\ie{} $\lambda < 0$) to unstable (\ie{} $\lambda > 0$) as we cross the bifurcation point.
This is a necessary, albeit insufficient, condition to conclude that the bifurcation is a pitchfork 
Moreover, nonlinear analyses (or simulations) are required in order to conclude whether it is supercritical or subcritical.
Examples from fluid dynamics include the flow in a sudden expansion channel~\cite{pof:drikakis:1997,jfm:lanzerstorfer:2012}, the flow past a sphere~\cite{jfm:natarajan:1993,pof:fabre:2008,jfm:sansica:2018}, the three-dimensional cavity flow~\cite{pof:albensoeder:2001,jfm:liu:2016,jfm:picella:2018} or the Rayleigh-Bénard convection between two infinite plates or inside an annular loop~\cite{tcfd:loiseau:2020}.

\paragraph{Hopf bifurcation}

The second type of bifurcation commonly encountered is the \emph{Andronov-Poincaré-Hopf} bifurcation (or simply Hopf bifurcation).
Below the bifurcation point, the system admits a single fixed point and asymptotically stationary dynamics.
As the bifurcation point is crossed, the fixed point changes stability, and a limit cycle associated with periodic dynamics is created in its vicinity.
The Hopf bifurcation can be distilled into the following normal form \cite{dijkstra2014numerical}, written as
\begin{equation}
\left\{
\begin{aligned}
  \dot{r} & = \sigma r \pm r^3, \\
  \dot{\varphi} & = \omega + \alpha r^2,
\end{aligned}
\right.
\end{equation}
where $r$ is the amplitude of the oscillations, $\varphi$ their phase, and $\omega$ the frequency of the oscillation.
The linearized system has a complex-conjugate pair of eigenvalues that change from stable (\ie{} $\sigma < 0$) to unstable (\ie{} $\sigma > 0$) when crossing the bifurcation point.
The imaginary part ($\omega$) of this complex-conjugate pair of eigenvalues then dictates the oscillation frequency.
Nonlinear analysis (or simulation) is required to determine whether it is super- or subcritical.
Examples from fluid dynamics include the broad class of so-called \emph{flow oscillators} such as the two-dimensional cylinder flow~\cite{jem:zebib:1987,jfm:noack:1994,pof:zhang:1995,arfm:williamson:1996,jfm:barkley:1996,noack2003hierarchy,epl:barkley:2006,jfm:sipp:2007,jfm:giannetti:2007,jfm:bagheri:2013,prl:mantic:2014,jfm:meliga:2016}, the lid-driven and shear-driven cavity flows~\cite{jfm:rowley:2002,jfm:sipp:2007,jfm:barbagallo:2009,yamouni2013interaction,jfm:meliga:2017,arxiv:callaham:2021}, the jet in crossflow~\cite{jfm:bagheri:2009,jfm:ilak:2012,jfm:chauvat:2020} or the roughness-induced boundary layer flow~\cite{jfm:loiseau:2014,pof:citro:2015,bucci2018jfm}.

\subsubsection{Bifurcations of periodic orbits}

The linearization of the time-periodic flow map $\Phi_\tau(\mathbf{X})$ in the vicinity of the periodic orbit $\mathbf{X}^*$ leads to the \emph{monodromy matrix} (sometimes also known as the time-shift operator).
As for fixed points, their eigenvalues (known as \emph{characteristic} or \emph{Floquet multipliers}) dictate the asymptotic stability of the periodic orbit under consideration.
The stability problem describes the development of small-amplitude perturbations during one period of evolution.
If all Floquet multipliers lie inside the unit disk, the orbit is characterized as asymptotically linearly stable, otherwise as unstable.

Bifurcations occur when one of these Floquet multipliers (or a complex-conjugate pair) steps outside the unit circle when the control parameter is varied.
Physically, the moduli of such Floquet multipliers express the orbit rate of expansion (or contraction) per unit of time (\ie{} per period of oscillation).
As an aside, in cases where the limit cycle occurs as an autonomous nonlinear oscillation (not forced), the set of Floquet multipliers contains a unit mode tangent to the limit cycle which corresponds to the time derivative of the base flow.
A limit cycle in which at least one Floquet multiplier is greater than one expands and is therefore called an unstable periodic orbit (UPO).

The most common bifurcations in this context are the \emph{pitchfork bifurcation}, the \emph{period-doubling bifurcation} (also known as a \emph{flip bifurcation}) and the \emph{Neimark-Sacker bifurcation} (see \cref{fig: bifurcation analysis} for a schematic).
Once again, these are associated with qualitatively different evolutions of the dynamics below and above the bifurcation point.

\paragraph{Pitchfork bifurcation}

As for its fixed point counterpart, the pitchfork bifurcation of periodic orbits is most often encountered in systems with spatial symmetries and comes in two flavors (super- and subcritical).
A canonical example of such pitchfork bifurcations in fluid dynamics is the three-dimensionalization of the periodic vortex shedding in the wake of a circular cylinder~\cite{jfm:barkley:1996,pre:barkley:2000}.
Below the critical Reynolds number $Re_c \simeq 189$, the flow is strictly two-dimensional and exhibits the well-known time-periodic von Kàrmàn vortex street.
All Floquet multipliers lie within the unit circle.
When the Reynolds number is increased, a Floquet multiplier leaves the unit circle at $\mu = 1$ and a pitchfork bifurcation occurs\footnote{Because of the spanwise-invariance of the unstable periodic orbit, the  dominant Floquet multiplier (\ie{} with the largest magnitude) is actually a double eigenvalue with one eigenvector exhibiting a sine dependence in the spanwise-direction while the other exhibits a cosine dependence.
In this case, the bifurcation is formally known as a \emph{circle pitchfork bifurcation}.}.
Given the synchronous nature of the pitchfork bifurcation, the frequency of the vortex shedding remains unchanged, but the spatial structure of the vortices is no longer spanwise invariant: the flow becomes three-dimensional.
If one denotes by $A_n$ the amplitude of this three-dimensionalization after $n$ periods, the corresponding normal form is given by
\begin{equation}
A_{n+1} = \mu A_n \pm \vert A_n \vert^2 A_n,
\end{equation}
where $\mu$ is the associated Floquet multiplier and the sign of the cubic term determines whether the bifurcation is super- or subcritical.
Note that, as for fixed points, other types of bifurcations (\eg{} saddle-node) are associated with a Floquet multiplier exiting the unit circle at $\mu = 1$.

\paragraph{Period-doubling bifurcation}

The second type of bifurcation commonly encountered is the \emph{period-doubling bifurcations}.
They are also known as \emph{flip} or \emph{subharmonic} bifurcations.
In this situation, a Floquet multiplier exits the unit circle at $\mu = -1$ as the bifurcation point is crossed.
In the supercritical case, the periodic orbit that was stable below the bifurcation point becomes unstable, and a new orbit with twice the period takes its place.
Considering a discrete-time system, the most famous example of this period-doubling bifurcation is the logistic map
\begin{equation}
x_{n+1} = \mu x_n ( 1 - x_n).
\end{equation}
In a continuous-time framework, this bifurcation can be found in the famous R\"ossler system.
Figure~\ref{fig: rossler system} shows two such orbits.
The one in blue, denoted as UPO\textsubscript{1} with a period $\tau_1 \simeq 6$, loses its stability via a period-doubling bifurcation at $c = 5.376 \pm 0.001$.
Above this critical value, the periodic orbit (denoted as UPO\textsubscript{2} in \cref{fig: rossler system}) is created with a period $\tau_2 \simeq 12$.
This second orbit then loses its stability through another period-doubling bifurcation (at a critical value $c = 7.771 \pm 0.001$) and a new orbit with twice the period (\ie{} $\tau_3 \simeq 24$) is created.
Systems presenting period-doubling bifurcations often exhibit a \emph{subharmonic cascade to chaos}, a universal behavior of dynamical systems put forth by Feigenbaum and others in the late 1970s.
In the context of fluid dynamics, such a subharmonic cascade was shown to occur in a confined Rayleigh-Bénard cell as the Rayleigh number is increased in the seminal work of Libchaber \etal~\cite{jpl:libchaber:1982}.
It was also observed by Buzug \etal~\cite{pre:buzug:1993} experimentally in Taylor-Couette flow and numerically in Couette flow~\cite{kreilos2012periodic}.
In plane Couette flow, a first bifurcation leads to the formation of a spatially evolving periodic orbit (also called \emph{relative periodic orbits} or \emph{traveling waves}) at $Re=236.1$~\cite{lustro2019onset}.
Above $Re \approx 240.40$, the system undergoes a period-doubling cascade.
At $Re = 240.46$ this cascade leads to a chaotic attractor with exponentially diverging trajectories that sporadically visit the various previously created UPOs.
More recently, this subharmonic cascade has also been observed numerically in rotating plane Couette flow under certain conditions~\cite{daly2014secondary}.
Period-doubling bifurcations are also observed in harmonically forced shear layers and jets where they give rise to \emph{vortex pairing}, see \cite{jfm:leopold:2019} for an example.

\paragraph{Neimark-Sacker bifurcation}

The last type of bifurcation we will consider is the \emph{Neimark-Sacker bifurcation} named after the works of Neimark~\cite{neimark1959some} in 1959 and Sacker~\cite{sacker1964invariant} in 1964 (for an overview of their work see the book by Arnold~\cite{arnold2012geometrical}).
This bifurcation is the equivalent of the Hopf bifurcation of a fixed point for periodic orbits and is therefore sometimes called a \emph{secondary Hopf bifurcation}.
From the point of view of linear stability, it is associated with a complex conjugate pair of Floquet multipliers leaving the unit disk at an angle that is neither 0 nor $\pi$, according to $\mu = e^{\pm i \omega}$.
If the new frequency is rationally related to that of the periodic orbit, the dynamics immediately after the bifurcation remains periodic, but with a different period. 
Rationally related frequencies always form a set of measure zero in the set of possible imaginary parts of the Floquet exponent.
If the new frequency is irrationally related to that of the periodic orbit, the dynamics becomes quasiperiodic immediately after the bifurcation.
In this case, the corresponding phase space object changes from a simple periodic orbit (below the bifurcation point) to a torus (above the bifurcation point)~\cite[see fig.~13]{arxiv:callaham:2021} in which the fundamental frequency and the smaller new frequency map the large and small circles of the torus, respectively.
The dynamics following a secondary Hopf bifurcation can be much more complicated, exhibiting phenomena such as frequency locking, high-order synchronization (Arnold's tongues), and devil's staircase, which are beyond the scope of this review. Interested readers are referred to the book by Pikovsky, Rosenblum \& Kurths~\cite{pikovsky2001universal}.

To help reveal the structure of the attractor, one can plot the intersections of the trajectories with a plane, called a Poincaré section, which intersects the attractor.
As the trajectories cross the plane in different positions, a Poincaré section on a torus-shaped object will continuously cover a circle~\cite{berge1987order} (provided the temporal signal is long enough).
The Fourier spectrum of a periodic system that becomes quasiperiodic is characterized by the appearance of a new peak at a frequency $f_2$ that is incommensurate with the fundamental frequency $f_1$.
Due to nonlinear interactions between these two frequencies, other peaks may appear in the spectra, which can be easily explained by the linear combinations $| n_1 f_1 \pm n_2 f_2 | $, where $n_1$ and $n_2$ are integers.
In fluid dynamics, such a bifurcation has been shown to occur in the wake of two side-by-side cylinders~\cite{jfm:carini:2014}, with the corresponding instability known as the \emph{flip-flop} instability, as well as in a two-dimensional shear-driven cavity~\cite{jfm:leclerc:2019,arxiv:callaham:2021}.

\paragraph{Beyond quasiperiodic dynamics}

Over the years, the transition from periodic and quasi-periodic dynamics to chaos has been studied in detail~\cite{gollub1980many,swinney1983observations}.
When the control parameter is varied, steady, periodic, and quasi-periodic systems can undergo a progressive loss of stability and the appearance of chaos by following characteristic paths known as \emph{routes to chaos}~\cite{argyris1993routes}, with the most relevant ones being:
\begin{itemize}
\item The Feigenbaum path: The system undergoes a cascade of period-doubling bifurcations;
\item The Ruelle-Takens-Newhouse route: gradual generation of incommensurable frequencies by a sequence of (Hopf) bifurcations;
\item The Pomeau-Manneville scenario: the existence of an intermittent alternation of regular phases and chaotic bursts. 
\end{itemize}

These characteristic routes are reviewed in Eckmann~\cite{eckmann1981roads} and can be identified by time-series analysis of physical or numerical experiments based on classical Fourier analysis to more complex phase-portrait reconstructions~\cite{physicaD:broomhead:1986}, delayed embedding~\cite{takens:1981,grassberger2004measuring} or recurrence analysis~\cite{marwan2007recurrence} techniques.

In 1971, Ruelle and Takens~\cite{cmp:ruelle:1971} showed that when the nonlinearity (or coupling) of a quasi-periodic system increases, Hopf bifurcations lead to an increase in the dimension of the torus, eventually becoming structurally unstable and collapses into a strange attractor with non-integer fractal dimension. 
The route known as Ruelle-Takens-Newhouse (RTN)~\cite{cmp:ruelle:1971,newhouse1978occurrence} is associated with the direct appearance of chaos after the formation of a $T^3$ torus (or even $T^4$ torus in some cases).
In perhaps even rarer cases, the $T^3$ torus can undergo a rapid devil's staircase to chaos \cite{kaneko1984fates}.
Flows undergoing the RTN path include the convergent-divergent channel flow \cite{guzman1994transition}, the Taylor-Couette flow \cite{coles1965transition} or the flow in the highly curved toroidal pipe flow~\cite{canton2016modal}, or even higher dimensional tori in $T^n$ with $n\ge3$~\cite{gollub1980many,jpl:libchaber:1982,oteski2015quasiperiodic}.

The path of intermittency introduced by Pomeau and Manneville~\cite{pomeau1980intermittent} in the 1980s involves an alternation between periodic and chaotic dynamics, although all system parameters remain constant and free of noise (Manneville~\cite{book:manneville:2010}'s book gives a clear overview).
Beyond the bifurcation point, dynamical systems with intermittency exhibit bursts of irregular motion with higher amplitude amidst regular motion with lower amplitude.
The duration of the irregular bursts increases with the control parameters until the chaotic dynamics predominate.
Depending on the value of the Floquet multiplier at the bifurcation point, there are different evolutionary trajectories for the decay of the periods of laminar phases.
In the theory of intermittent transitions, Floquet stability analysis provides three classifications for intermittency: Type I is associated with a pitchfork bifurcation (the Floquet multiplier crosses the unit circle at +1 at the bifurcation point), type II with a Neimark-Sacker bifurcation (two complex conjugate eigenvalues), and type III with a period-doubling bifurcation (the Floquet multiplier crosses the unit circle at -1 at the bifurcation point). 
Note that the latter two cases require a subcritical character of the bifurcation for the appearance of intermittency~\cite{berge1987order}.

\section{Numerical methods}\label{sec: numerical methods}

Numerous tools exist to study low-dimensional dynamical systems such as the R\"ossler or Lorenz~\cite{jas:lorenz:1963} systems.
These include \verb+AUTO+~\cite{auto,auto_bis} in Fortran or \verb+pde2path+~\cite{uecker2019hopf,uecker2021numerical} and \verb+MatCont+~\cite{matcont} in MATLAB.
\verb+PyDSTool+~\cite{pydstool} offers similar capabilities in Python, while \verb+BifurcationKit.jl+~\cite{bifurcationkit} is a corresponding Julia package.
Except for \verb+BifurcationKit.jl+, most of them rely on standard numerical linear algebra techniques that do not scale well for very high-dimensional problems.
Moreover, they do not necessarily integrate easily with parallel programming, which is ubiquitous when simulating discretized partial differential equations such as Navier-Stokes.
Thus, extra care may be needed to interface with the particular data structure of the original code (see Algorithm 1).

This section provides a brief overview of the standard iterative techniques used to compute fixed points or periodic orbits of very high-dimensional dynamical systems and to study their stability properties.
These techniques rely on \emph{Krylov subspaces} and associated \emph{Krylov decompositions}~\cite{siam:stewart:2001} introduced in \cref{subsec: numerics -- krylov}.
The Newton-Krylov method for fixed-point computations and its extension to periodic orbits are discussed in \cref{subsec: numerics -- newton-krylov}.
Finally, the use of Krylov techniques to compute the leading eigenvalues or singular values of the linearized operator to characterize its modal and non-modal stability properties are discussed in \cref{subsec: numerics -- eigensolvers}.
In what follows, we will assume that a time-stepping simulation code is available to simulate the nonlinear system, \ie{} the time-stepping code returning $\mathbf{X}_{k+1} = \Phi_\tau(\mathbf{X}_k)$.
Similarly, we will assume that a linearized version of this code can be used to calculate the matrix-vector product $\mathbf{M}_\tau \mathbf{x}$ by time-marching the equations, where the operator $\mathbf{M}_\tau$ is either the (numerically approximated) \emph{exponential propagator} for fixed points or the monodromy matrix for periodic orbits.
For non-modal stability analysis, we furthermore assume that an equivalent time-stepping code is available to approximate the matrix-vector product $\mathbf{M}_\tau^{\dagger} \mathbf{x}$ where $\mathbf{M}^{\dagger}_\tau$ is the exponential propagator or monodromy matrix built using the corresponding adjoint linear operator $\mathbf{L}^{\dagger}$.
The methods advocated here and in \cite{dijkstra2014numerical,chapter:loiseau:2019} can all be easily implemented with very few modifications into existing codes.

\subsection{Krylov subspaces and the Arnoldi factorization}
\label{subsec: numerics -- krylov}

In \cite{stewart2000decompositional}, the American mathematician Gilbert W. Stewart listed six of the most important matrix decompositions.
These include the pivoted LU decomposition, the QR decomposition, the spectral (\ie{} eigenvalue) decomposition, the Schur decomposition, and the singular value decomposition (SVD).
The introduction of each of these decompositions into numerical linear algebra has revolutionized matrix computations.
Nevertheless, a seventh approximate factorization should be included in this list, namely, the \emph{Arnoldi factorization}.
Introduced by Walter Edwin Arnoldi in 1951~\cite{arnoldi1951principle} it relies on the concept of \emph{Krylov subspaces}~\cite{krylov1931numerical}, named after the Russian applied mathematician Alexei Krylov.
Today, these subspaces are the workhorses of large-scale numerical linear algebra and form the foundations of numerous iterative linear solvers such as the \emph{minimal residual method} (MINRES)~\cite{siam:paige:1975} or the \emph{generalized minimal residual method} (GMRES)~\cite{siam:saad:1986}.
The book \emph{Iterative methods for sparse linear systems} by Y. Saad~\cite{saad2003iterative} is probably the most complete reference for such techniques.
Given a matrix $\mathbf{A} \in \mathbb{R}^{n \times n}$ and a starting vector $\mathbf{x} \in \mathbb{R}^n$, a Krylov subspace of dimensions $m$ can be constructed by repeated applications of $\mathbf{A}$, leading to
\begin{equation}
\mathcal{K}_m(\mathbf{A}, \mathbf{x}) = \left\{ \mathbf{x}, \mathbf{Ax}, \mathbf{A}^2 \mathbf{x}, \cdots , \mathbf{A}^{m-1} \mathbf{x} \right\}.
\end{equation}
Introducing the matrix
\begin{equation}
\mathbf{K} = \begin{bmatrix} \mathbf{x} & \mathbf{Ax} & \mathbf{A}^2 \mathbf{x} & \cdots & \mathbf{A}^{m-1} \mathbf{x} \end{bmatrix},
\end{equation}
the above sequence can be recast as the following Krylov factorization
\begin{equation}
\mathbf{AK} = \mathbf{KC} + \mathbf{e}_{m}^T \mathbf{r},
\end{equation}
where $\mathbf{C} \in \mathbb{R}^{m \times m}$ is a companion matrix of the form
\begin{equation}
\mathbf{C}
=
\begin{bmatrix}
  0 & 0 & 0 & \cdots & 0 & c_1 \\
  1 & 0 & 0 & \cdots & 0 & c_2 \\
  0 & 1 & 0 & \cdots & 0 & c_3 \\
  \vdots & \ddots & \ddots & \ddots & \ddots & \vdots \\
  0 & 0 & \cdots & \cdots & 1 & c_m
\end{bmatrix}.
\end{equation}
The coefficients $c_i$ are computed based on a least-squares procedure such that the residual $\mathbf{r}$ is not in the span of the previous $m$ Krylov vectors, \ie{} $\mathbf{r}^T \mathbf{K} = \mathbf{0}$, and $\| \mathbf{r} \|_2$ is minimized.
If $\| \mathbf{r} \| = 0$, then the columns of $\mathbf{K}$ span an invariant subspace, and the eigenvalues of $\mathbf{C}$ are a subset of those of $\mathbf{A}$.
If the starting vector $\mathbf{x}$ is random and $m$ is large enough, $\mathbf{K}$ most likely tends towards the invariant subspace of $\mathbf{A}$ associated with the eigenvalues having the largest magnitudes.
Note however that, as $m$ increases, the last Krylov vectors become increasingly collinear by virtue of the applied power iteration and, consequently, the matrix $\mathbf{K}^T \mathbf{K}$ is increasingly ill-conditioned.
This companion-based Krylov factorization is thus of little use in practice due to its numerical instability given finite arithmetic.

A simple remedy to this numerical instability is to iteratively construct each new Krylov vector such that it is orthonormal to all previously generated vectors instead of simply applying the power iteration.
Starting from a vector $\mathbf{x}_1$ (with $\| \mathbf{x}_1 \|_2 = 1$), the Krylov basis can be iteratively constructed by the following algorithm.
\begin{tcolorbox}[width=\columnwidth, title={\bf Algorithm 1 -- Arnoldi factorization}, colback=white]
  \textbf{Given :} the $n \times n$ matrix $\mathbf{A}$, and a unit-norm vector $\mathbf{v}_1$.

  \begin{enumerate}
  \item $\mathbf{w} = \mathbf{Av}_1$ ; $\alpha_1 = \mathbf{v}_1^* \mathbf{w}$ ; \\
    $\mathbf{r} = \mathbf{w} - \mathbf{v}_1 \alpha$ ; $\mathbf{V}_1 = \left[ \mathbf{v}_1 \right]$ ; $\mathbf{H}_1 = \left[ \alpha \right]$.

    \bigskip

  \item For $j = 1, 2, \cdots, m-1$ \\
    \begin{enumerate}
    \item Add the residual from the previous iteration into the Krylov basis \\
      $\beta_j = \| \mathbf{r} \|_2$ ; $\mathbf{v}_{j+1} = \beta_j^{-1} \mathbf{r}$ ; \\
      $\mathbf{V}_{j+1} = \begin{bmatrix} \mathbf{V}_j & \mathbf{v}_{j+1} \end{bmatrix}$ ; $\widetilde{\mathbf{H}}_j = \begin{bmatrix} \mathbf{H}_j \\ \beta_j \mathbf{e}_j^* \end{bmatrix}$ ;

    \item Compute the residual associated with this new Krylov basis \\
      $\mathbf{w} = \mathbf{Av}_{j+1}$ ; $\mathbf{h} = \mathbf{V}_{j+1}^* \mathbf{w}$ ; \\
      $\mathbf{r} = \mathbf{w} - \mathbf{V}_{j+1} \mathbf{h}$ ;
    \item Update the upper Hessenberg matrix accordingly \\
      $\mathbf{H}_{j+1} = \begin{bmatrix} \widetilde{\mathbf{H}}_j & \mathbf{h} \end{bmatrix}$.
    \end{enumerate}

  \end{enumerate}
\end{tcolorbox}
After $m$ steps, this leads to the Krylov factorization
\begin{equation}
\mathbf{AV} = \mathbf{VH} + \mathbf{e}_m^T \mathbf{r},
\end{equation}
known as the \emph{Arnoldi factorization}.
In this factorization, the matrix $\mathbf{V} \in \mathbb{R}^{n \times m}$ is orthonormal (\ie{} $\mathbf{V}^T \mathbf{V} = \mathbf{I}$) and $\mathbf{H} \in \mathbb{R}^{m \times m}$ is an upper Hessenberg matrix (almost triangular matrix with zero entries below the first subdiagonal).
Once again, the residual vector $\mathbf{r}$ is the component of the $(m+1)$\textsuperscript{th} Krylov vector not in the span of $\mathbf{V}$ (\ie{} $\mathbf{r}^T \mathbf{V} = \mathbf{0}$).
Knowledge of the $n \times m$ matrix $\mathbf{V}$ and the $m \times m$ matrix $\mathbf{H}$ can then be used to approximate the dominant (\ie{} with the largest absolute value) eigenvalues and eigenvectors of $\mathbf{A}$ or to obtain a reasonable solution to
\[
\mathbf{Ax} = \mathbf{b},
\]
at a reduced computational cost compared to direct inversion using Gaussian elimination or LU techniques.
The Arnoldi factorization is at the heart of the widely used GMRES technique for solving large linear systems presented in Algorithm 3.
Other Krylov factorizations exist such as the \emph{Lanczos factorization} for Hermitian matrices (where $\mathbf{H}$ reduces to a tridiagonal matrix) or the \emph{Krylov-Schur} factorization introduced by Stewart~\cite{siam:stewart:2001} (where $\mathbf{H}$ is in Schur form), enabling simple restarting strategies for computing the dominant eigenvalues and eigenvectors of $\mathbf{A}$ when the available RAM is a limiting factor.

\subsection{Newton-Krylov method for fixed points and periodic orbits}
\label{subsec: numerics -- newton-krylov}

For low-dimensional dynamical systems, fixed point (resp. periodic orbit) computations can easily be performed using the standard Newton method already implemented in numerous languages (\eg{} \verb+scipy.optimize.fsolve+ in \verb+Python+).
These implementations (see Algorithm 2) are often quite generic and rely on direct solvers for the inversion of the Jacobian matrix $\mathbf{L}$ (resp. monodromy matrix $\mathbf{M}_{\tau}$).
Due to the sheer size of the linear systems resulting from the discretization of partial differential equations, this approach however does not scale favorably.
Coupling these generic solvers with an existing time-stepping code may moreover require extra layers of code because of the particular data structure used in the simulation and possibly its parallel computing capabilities.
Although libraries such as \verb+PETSc+~\cite{petsc-efficient,petsc-user-ref,petsc-web-page} or \verb+Trilinos+~\cite{trilinos-website} exist, these may also require extra development to interface with an existing well-established code.
They moreover add extra dependencies which might complicate the deployment of the resulting applications on a large set of computing platforms with different operating systems (\eg{} from laptops for development to supercomputing facilities for production runs).

With the goal of extending the capabilities of an existing time-stepping code with a minimum number of modifications and dependencies, \cref{subsubsec: numerics -- fixed points} (resp. \cref{subsubsec: numerics -- periodic orbits}) presents a time-stepping formulation of the Newton-Krylov algorithm to compute fixed points (resp. periodic orbits).
A similar algorithm has already been introduced by~\cite{arxiv:kelley:2004} and \cite{dijkstra2014numerical} and in \texttt{ChannelFlow}~\cite{channelflow,GibsonHalcrowCvitanovicJFM08}.
As stated previously, we will assume only that we have the nonlinear time-stepper returning
\begin{equation}
\mathbf{X}_{k+1} = \Phi_\tau(\mathbf{X}_k),
\end{equation}
and its linear counterpart
\begin{equation}
\mathbf{x}_{k+1} = \mathbf{M}_{\tau}\mathbf{x}_k,
\end{equation}
where $\mathbf{M}_\tau$ is either the exponential propagator or the monodromy matrix.
We will also assume that a routine to compute the $m$-step Arnoldi factorization (see \cref{subsec: numerics -- krylov}) has been implemented. 
Along with the inclusion of a direct eigensolver (such as links to LAPACK), this implementation is the only major development needed to extend the capabilities of an existing time-stepping code.
Benefits from this implementation outperform its development costs as it paves the way for a GMRES implementation leveraging all the utilities of the existing time-stepping code.
Once available, this $m$-step Arnoldi factorization routine can also be readily used to compute the leading eigenvalues and eigenvectors of the high-dimensional linearized operator with no extra development (see \cref{subsec: numerics -- eigensolvers}).

\subsubsection{Fixed point computation}
\label{subsubsec: numerics -- fixed points}

In a time-stepper formulation, fixed points are solutions to
\begin{equation}
\mathbf{X} = \Phi_\tau(\mathbf{X}),
\end{equation}
for arbitrary integration time $\tau$.
Alternatively, they are the roots of
\begin{equation}
  \mathcal{F}(\mathbf{X}) = \Phi_\tau(\mathbf{X}) - \mathbf{X}.
  \label{eq: fixed point equation}
\end{equation}
In fluid dynamics, numerous approaches have been proposed in the literature to compute these fixed points while circumventing the need to implement a dedicated Newton solver into an existing time-stepping code.
For example, one can cite the \emph{selective frequency damping} method (SFD) proposed by \AA{}kervik \etal~\cite{pof:akervik:2006} and its variants, or \emph{BoostConv}~\cite{jcp:citro:2017}.
Although they require relatively minor modifications to an existing simulation code, they suffer from a number of limitations, such as slow convergence, which was explored in \cite{chapter:loiseau:2019}.
Because it relies on a temporal low-pass filter, the selective frequency damping procedure is, moreover, unable to compute saddle nodes, \ie{} fixed points having at least one unstable eigendirection associated with a purely real eigenvalue.

Assuming the $m$-step Arnoldi factorization has already been implemented, it requires only a relatively modest effort to integrate it into a dedicated GMRES solver.
In doing so, the roots of \cref{eq: fixed point equation} can be easily computed using the Newton-Krylov technique.
The Jacobian of \cref{eq: fixed point equation} is given by
\begin{equation}
\mathbf{J} = \exp(\tau \mathbf{L}) - \mathbf{I},
\end{equation}
where $\mathbf{L}$ is the linearized operator around the current estimate $\mathbf{X}$.
The matrix-vector product $\mathbf{Jx}$ thus requires calling
the linearized time-stepper (\ie{} to compute $\exp(\tau \mathbf{L})\mathbf{x}$), with $\mathbf{x} \in \mathbb{R}^n$ being the Newton correction.
%
\begin{tcolorbox}[width=\columnwidth, title={\bf Algorithm 2 -- Newton-Krylov solver}, colback=white]
  \textbf{Given :} the time-stepper $\mathbf{X}_{k+1} = \Phi_{\tau}(\mathbf{X}_k)$, its linearized counterpart, and an initial guess $\mathbf{X}_0$ for the fixed point.

  \bigskip
  
  For $j = 1, 2, \cdots, m$
  \begin{enumerate}
  \item Compute the residual of the nonlinear equation $\mathbf{r} = \Phi_{\tau}(\mathbf{X}_j) - \mathbf{X}_j$

    \bigskip

  \item Check for convergence.
    If $\| \mathbf{r} \| \leq \varepsilon$, return $\mathbf{X}_j$ as the solution.

    \bigskip

  \item If not converged, compute the Newton correction by solving $\mathbf{J x} = - \mathbf{r}$

    \bigskip

  \item Update the solution as $\mathbf{X}_{j+1} = \mathbf{X}_j + \mathbf{x}$.
  \end{enumerate}
\end{tcolorbox}
\noindent
The linear system in step 3 of this iteration is typically solved using a GMRES solver (or other Krylov-based solvers such as BiCGSTAB~\cite{van1992bi} or IDR~\cite{sonneveld2009idr}) making use of the previously implemented Arnoldi factorization (the matrix to be factorized being $\mathbf{J}$).
The GMRES procedure is presented in Algorithm 3.
It should be emphasized that the linear equation in step 3 does not need to be solved with high precision at each step.
It is indeed sufficient to ensure that the Newton correction $\mathbf{x}$ reduces the norm of the residual $\mathbf{r} = \Phi_{\tau}(\mathbf{X}_{j+1}) - \mathbf{X}_j$ by one or two orders of magnitude by setting the tolerance in the iterative linear solver to \eg{} $0.01 \| \mathbf{r} \|$.
Although this may increase the number of Newton steps before convergence is reached, each iteration is computationally cheaper and faster as fewer Krylov vectors need to be generated by the iterative linear solver, thus reducing the overall time to solution.
Hereafter, such a strategy is referred to as \emph{Newton-Krylov with dynamic tolerances}.

\begin{tcolorbox}[width=\textwidth, title={Algorithm 3 -- Solving $\mathbf{Ax}=\mathbf{b}$ with GMRES},]
  Consider the linear system $\mathbf{Ax} = \mathbf{b}$ with $\mathbf{A} \in \mathbb{R}^{n \times n}$, and both $\mathbf{x}$ and $\mathbf{b} \in \mathbb{R}^n$.
  We will assume furthermore that $n \gg 1$ so that solving this system with standard direct solvers (\eg{} LU factorization) is intractable.
  One of the most famous iterative methods to solve such large-scale systems is the \textbf{generalized minimal residual method} (GMRES), introduced by Yousef Saad and Martin H. Schultz~\cite{siam:saad:1986} in 1986.
  Based on the Arnoldi factorization, it iteratively approximates the solution to $\mathbf{Ax} = \mathbf{b}$ by the vector $\mathbf{x}_k$ in a $k$-dimensional Krylov subspace with minimal residual.
  This task can be formulated as the following optimization problem
  \[
  \minimize_{\mathbf{x}_k \in \mathcal{K}} \quad \| \mathbf{b} - \mathbf{Ax}_k \|,
  \]
  where $\mathcal{K}(\mathbf{r}_0, \mathbf{A})$ is a $k$-dimensional Krylov subspace iteratively constructed using the residual $\mathbf{r}_0 = \mathbf{b} - \mathbf{Ax}_0$, with $\mathbf{x}_0$ our initial guess for the solution (often taken as the zero vector), and $\| \cdot \|$ denoting the Euclidean norm.
  Consider now the $k$-step Arnoldi factorization
  \[
  \mathbf{AV}_k = \mathbf{V}_{k+1} \widetilde{\mathbf{H}}_k,
  \]
  with $\mathbf{V}_k \in \mathbb{R}^{n \times k}$ an orthonormal matrix (\ie{} $\mathbf{V}_k^T \mathbf{V}_k = \mathbf{I}_k$), and $\widetilde{\mathbf{H}}_k$ the resulting $(k+1) \times k$ upper Hessenberg matrix.
  The unknown vector $\mathbf{x}_k$ can be expressed as
  \[
  \mathbf{x}_k = \mathbf{x}_0 + \mathbf{V}_k \mathbf{y},
  \]
  with $\mathbf{y} \in \mathbb{R}^k$ an unknown low-dimensional vector.
  The columns of $\mathbf{V}_{k+1}$ being orthonormal, introducing this expression into the GMRES objective function yields
  \[
  \begin{aligned}
    \| \mathbf{b} - \mathbf{Ax}_k \| & = \| \mathbf{b} - \mathbf{A}\left( \mathbf{x}_0 + \mathbf{V}_k \mathbf{y} \right) \| \\
    & = \| \mathbf{r}_0 - \mathbf{AV}_k \mathbf{y} \| \\
    & = \| \beta \mathbf{v}_1 - \mathbf{V}_{k+1} \widetilde{\mathbf{H}}_k \mathbf{y} \| \\
    & = \| \mathbf{V}_{k+1} \left( \beta \mathbf{e}_1 - \widetilde{\mathbf{H}}_k \mathbf{y} \right) \| \\
    & = \| \beta \mathbf{e}_1 - \widetilde{\mathbf{H}}_k \mathbf{y} \|,
  \end{aligned}
  \]
  with $\mathbf{e}_1$ the first vector in the standard basis of $\mathbb{R}^{k+1}$, and $\beta = \| \mathbf{r}_0 \|$.
  The low-dimensional vector $\mathbf{y}$ is thus a solution to the following \emph{least-squares problem}
  \[
  \minimize_{\mathbf{y}} \quad \| \beta \mathbf{e}_1 - \widetilde{\mathbf{H}}_k \mathbf{y} \|,
  \]
  whose solution is given by $\mathbf{y} = \beta \widetilde{\mathbf{H}}_k^{\dagger} \mathbf{e}_1$, where $\widetilde{\mathbf{H}}_k^{\dagger} = \left( \widetilde{\mathbf{H}}_k^T \widetilde{\mathbf{H}}_k \right)^{-1} \widetilde{\mathbf{H}}_k^T$ is the \emph{Moore-Penrose pseudoinverse}.
  If the residual is too large, a new Krylov vector is generated following the Arnoldi procedure and the iteration continues until $\| \beta \mathbf{e}_1 - \widetilde{\mathbf{H}}_k \mathbf{y} \|$ is small enough.
  Numerous variants of GMRES exist, most notably when $\mathbf{A}$ is ill-conditioned.
  For more details, interested readers are referred to the excellent book \emph{Iterative methods for sparse linear systems} by Y. Saad~\cite{saad2003iterative}.
\end{tcolorbox}

\begin{figure*}[ht]\centering
\includegraphics{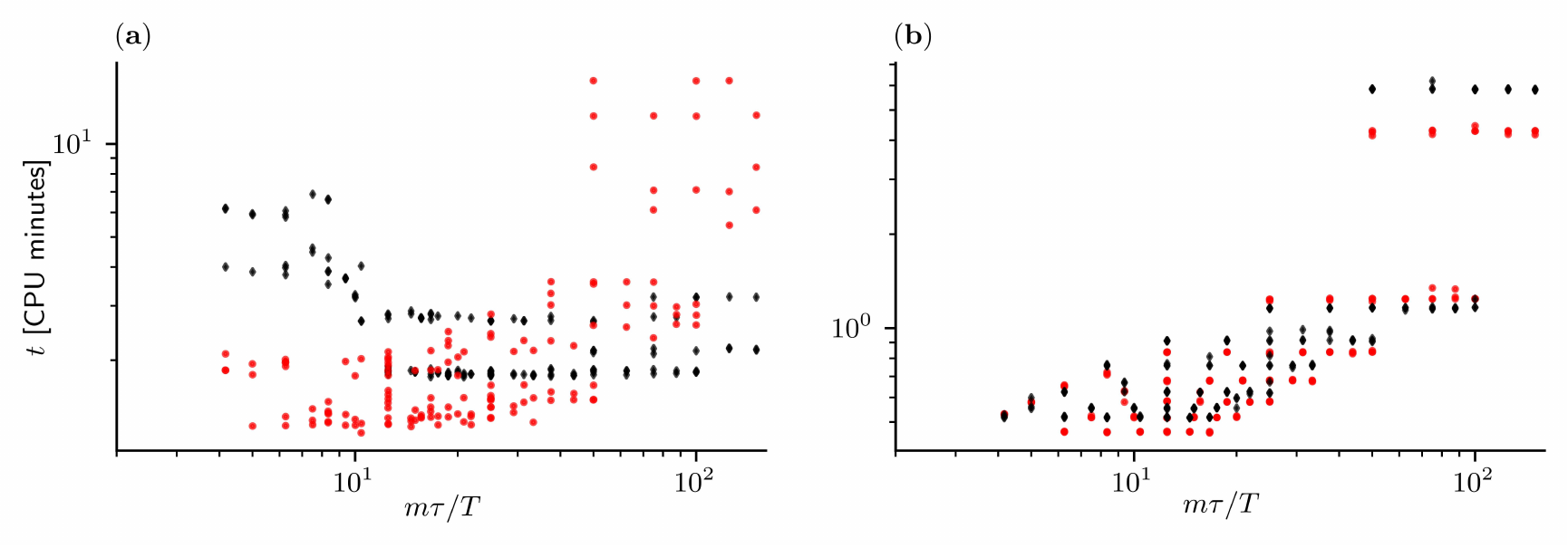}
\caption{Time to solution (in CPU minutes) versus different pairs of Krylov basis size $m$ and integration time $\tau$ for fixed point computation: ($a$) base flow of 2D flow past a circular cylinder at $Re=80$, where the initial condition is assumed to be the base flow at $Re=40$; 
($b$) base flow of the 2D open-cavity at $Re=4700$ starting from the base flow at $Re=4000$.
Black markers represent time to solution computed with fixed solver tolerances set to $10^{-10}$.
Red markers represent cases where the tolerance was tightened after each Newton step, from $10^{-5}$ at the first step to the target value $10^{-10}$ at the final step.
The parametric study is carried out using an automated Python script (found in \url{validations/newton_loop/autorun.py}) that loops over previously defined ranges of $m$ and $\tau$.}
\label{fig: numerics -- newton-krylov cylinder}
\end{figure*}

Let us now consider a critical point in the above formulation of the problem, namely its convergence and computational cost.
For a reasonable initial guess, the number of Newton iterations is expected to scale as $\mathcal{O}(k)$ where $k$ is the number of eigenvalues of $\mathbf{M}_\tau$ in the vicinity of the unit circle (see \cite{arxiv:kelley:2004} for a discussion about the convergence properties).
In typical production runs, only a handful of eigenvalues may be unstable or close to being unstable.
In our numerical experiments, the Newton solver usually converges in more or less ten iterations, irrespective of the discretization of the underlying partial differential equation.
From a computational point of view, the cost of each Newton iteration is, however, dominated by the call to the GMRES solver where each new Krylov vector is obtained from a linearized simulation over the integration time $\tau$.
This parameter $\tau$ plays a crucial role in the number of Krylov vectors that need to be generated to achieve convergence as it directly impacts the eigenvalue distribution of the exponential propagator $\mathbf{M}_\tau = \exp(\tau \mathbf{L})$ and thus of the Jacobian $\mathbf{J} = \mathbf{M}_\tau - \mathbf{I}$.
As $\tau$ increases, the gap between the leading eigenvalues and the others increases, and GMRES requires fewer iterations to converge (and thus fewer Krylov vectors must be stored in memory).
The wall-clock time of each GMRES iteration however increases.
Still, a major advantage of the time-stepper formulation is that it does not require preconditioning strategies to perform well (although it can benefit from them, as shown in \cite{dijkstra2014numerical,chapter:tuckerman:2019} and the discussion in \cref{subsec: time-stepper precond}).
A short parametric study was performed to evaluate the performance and sensitivity of the method with respect to the size $m$ of the Krylov subspace and the integration time $\tau$.
At least three runs\footnote{The tests were computed in a single node of the Jean Zay HPE SGI 8600 supercomputer with two Intel Cascade Lake 6248 processors (20 cores at 2.5 GHz) with the Intel Compiler version 2020.4
More information in \url{http://www.idris.fr/eng/jean-zay}.} are performed with $m \in (50,75,100,125,150,175,200)$ and $\tau = \left( \nicefrac{T}{12}, \nicefrac{T}{10}, \nicefrac{T}{8}, \nicefrac{T}{6}, \nicefrac{T}{4}, \nicefrac{T}{2}, T \right)$, where $T$ is the characteristic timescale of the leading eigenvalue.
This is illustrated in \cref{fig: numerics -- newton-krylov cylinder} for two different flow cases: the two-dimensional cylinder flow at $Re = 80$ and the two-dimensional open cavity flow at $Re=4700$.
For the computation of the leading eigenvalues, we can see that similar times to solution are obtained for
\[
4 < \dfrac{m\tau}{T} < 20,
\]
\ie{} the product of the dimension of the Krylov subspace and the sampling time $\tau$ is sufficient to cover between 4 and 20 periods of the instability.
For the cases of non-oscillatory instabilities, \eg{} from a pitchfork bifurcation, we suggest a sampling period of $\tau=1$ non-dimensional time unit as a starting point if no estimate of the doubling time of the instability is available.
For the cylinder flow, $m\tau=67$ using dynamic tolerances resulted in a minimum total computation time of about a minute, while $m\tau=1128$ led to a computation time of over 16 minutes.
For the open cavity 2D flow, the minimum total computation time of 30 seconds was achieved with $m\tau=10$, and the maximum time over 6 minutes was calculated with $m\tau=120$.

\subsubsection{Periodic orbit computation}
\label{subsubsec: numerics -- periodic orbits}

Periodic orbits are solutions to
\begin{equation}
\mathbf{X} = \boldsymbol{\Phi}_{\tau} (\mathbf{X}) \quad \textrm{for} \quad \tau = \tau^*,
\end{equation}
where $\tau^*$ is the period of the orbit.
They are the roots of
\begin{equation}
  \mathcal{F}(\mathbf{X}, \tau) = \boldsymbol{\Phi}_\tau(\mathbf{X}) - \mathbf{X}.
\end{equation}
The above system of equations is underdetermined: it has only $n$ equations for $n+1$ unknowns (the last unknown being the period of the orbit).
To close the system, an extra phase condition must be considered to select a particular point on the orbit (see Algorithm 4).
Various possibilities have been suggested in the literature.
For example, in \verb+AUTO+~\cite{auto,auto_bis} an integral constraint is used.
Here, a simpler condition is used.
Given an initial condition $\mathbf{X}_0 = \mathbf{X}(0)$, the phase condition is chosen as follows
\begin{equation}
\mathbf{F}(\mathbf{X}_0) \cdot \left( \mathbf{X} - \mathbf{X}_0 \right) = 0,
\end{equation}
where $\mathbf{F}(\mathbf{X}_0)$ is the time-derivative of the system evaluated at $\mathbf{X}_0$.
A solution to this bordered system
\begin{equation}
  \left\{
  \begin{aligned}
    \boldsymbol{\Phi}_\tau (\mathbf{X}) - \mathbf{X} & = 0, \\
    \mathbf{F}(\mathbf{X}_0) \cdot \left( \mathbf{X} - \mathbf{X}_0 \right) & = 0,
  \end{aligned}
  \right.
\end{equation}
can be obtained using a Newton-Krylov solver.
The Jacobian of this system is
\begin{equation}
\mathbf{J}
=
\begin{bmatrix}
  \mathbf{M}_\tau - \mathbf{I} & \mathbf{F}\left( \boldsymbol{\Phi}_\tau(\mathbf{X}_0) \right) \\
  \mathbf{F}^T(\mathbf{X}_0) & 0
\end{bmatrix}.
\end{equation}
As before, the corresponding matrix-vector product requires a single call to the linearized time-stepper (which includes many smaller time steps) to evaluate $\mathbf{M}_\tau \mathbf{x}$ (see the upper left block of the Jacobian matrix).
Evaluation of the other terms should be readily available.
\begin{tcolorbox}[width=\columnwidth, title={\bf Algorithm 4 -- Newton-Krylov solver for periodic orbits}, colback=white]
  \textbf{Given :} the time-stepper $\mathbf{X}_{k+1} = \Phi_{\tau}(\mathbf{X}_k)$, its linearized counterpart, and an initial guess $\mathbf{X}_0$ for the fixed point.

  \bigskip
  
  For $j = 1, 2, \cdots, m$
  \begin{enumerate}
  \item Compute the residual of the nonlinear equation $\mathbf{r} = \Phi_{\tau}(\mathbf{X}_j) - \mathbf{X}_j$

    \bigskip

  \item Check for convergence.
    If $\| \mathbf{r} \| \leq \varepsilon$, return $\mathbf{X}_j$ as the solution.

    \bigskip

  \item If not converged, compute the Newton correction by solving
    \[
    \begin{bmatrix}
      \mathbf{M}_{\tau} - \mathbf{I} & \mathbf{F}\left( \Phi_{\tau}(\mathbf{X}_0) \right) \\
      \mathbf{F}^T(\mathbf{X}_0) & \mathbf{0}
    \end{bmatrix}
    =
    \begin{bmatrix}
      \mathbf{x} \\ \Delta \tau
    \end{bmatrix}.
    \]

  \item Update the solution as $\mathbf{X}_{j+1} = \mathbf{X}_j + \mathbf{x}$, $\tau_{j+1} = \tau_j + \Delta \tau$.
  \end{enumerate}
\end{tcolorbox}
\noindent
When evaluating the matrix-vector product $\mathbf{M}_\tau \mathbf{x}$, both the original nonlinear system and the linearized one need to be marched forward in time.
While this increased computational cost is limited for small-scale systems, it may become quite significant for large-scale systems.
A simple strategy to alleviate this is to precompute the tentative orbit $\mathbf{X}_k(t) \textrm{ for } t \in \left[ 0, \tau_k \right]$ at the beginning of each Newton iteration and store all time steps in memory.
Then, only the linearized system needs to be marched forward in time with its coefficients updated at each time step.
If one is memory-bounded, only a limited number of time steps of the nonlinear system can be stored in memory and the intermediate steps can be reconstructed using for instance cubic spline interpolation or temporal Fourier interpolation.

\begin{figure}[ht]\centering
\includegraphics[width=.5\textwidth]{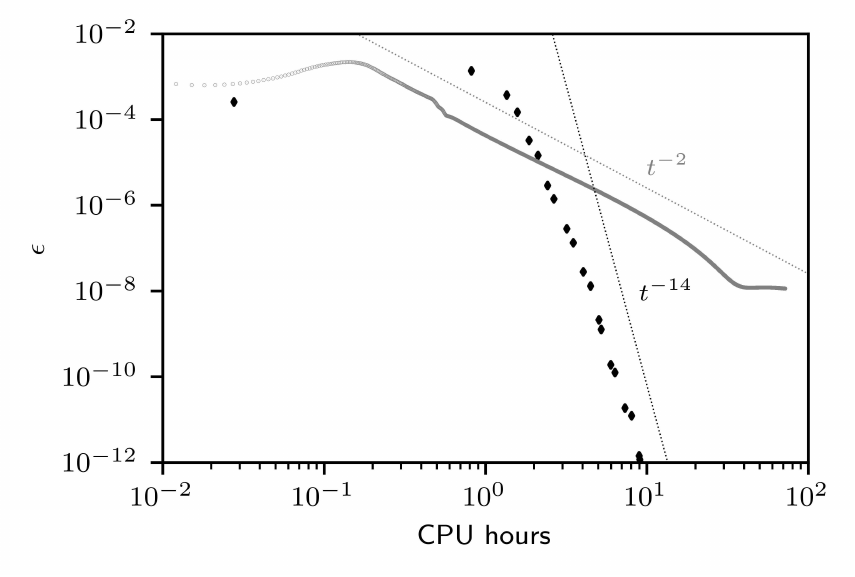}
\caption{Residual deflation as a function of the total computation time (both computed in the same hardware) for time-delayed feedback (open gray circles) and Newton GMRES (filled black diamonds) for the harmonically forced jet presented in \cref{sec:ex:hjet}. To obtain a continuous signal over time for the Newton method, we sum the number of calls to the linearized solver between each Newton iteration.
We observe decays proportional to $t^{-2}$ for the time-delayed feedback and $t^{-14}$ for the Newton.}
\label{fig:tdf_newton}
\end{figure}

The Newton-Krylov algorithm presented herein corresponds to the \emph{standard shooting} method.
It is by far the simplest method to implement for the computation of periodic orbits in large-scale systems.
Other techniques exist.
For instance, \emph{multiple shooting}~\cite{sanchez2010multiple} leverages the concept of a \emph{Poincaré section} while \emph{temporal collocation} transforms the orbit computation into a boundary value problem.
Recently, Shaabani-Ardali \etal~\cite{shaabani2017time} have also adapted ideas from feedback control with time delays to stabilize unstable periodic solutions.
In \cref{fig:tdf_newton}, we show the evolution of the residual with respect to the computational time for stabilization of a periodic base flow with an imposed frequency using the time-delayed technique and the proposed Newton GMRES.
We can observe a striking difference in the residual deflation when comparing the two techniques.

\subsection{Large-scale eigensolvers}
\label{subsec: numerics -- eigensolvers}
\begin{tcolorbox}[float*=h, width=\textwidth, title={\bf Algorithm 5 -- Solving $\mathbf{Ax} = \lambda \mathbf{x}$ with the Arnoldi factorization},]
  Consider the eigenvalue problem
  \[
  \mathbf{Ax} = \lambda \mathbf{x},
  \]
  with $\mathbf{A} \in \mathbb{R}^{n \times n}$,  $\mathbf{x} \in \mathbb{C}^{n}$ and $\lambda \in \mathbb{C}$.
  The $k$-step Arnoldi factorization reads
  \[
  \mathbf{AV}_k = \mathbf{V}_k \mathbf{H}_k + \beta \mathbf{v}_{k+1} \mathbf{e}_{k}^T,
  \]
  with $\mathbf{V}_k \in \mathbb{R}^{n \times k}$ an orthonormal matrix (\ie{} $\mathbf{V}_k^T \mathbf{V}_k = \mathbf{I}_k$), $\mathbf{H}_k$ the $k \times k$ upper Hessenberg matrix, and $\mathbf{e}_k$ the $k$\textsuperscript{th} vector in the standard basis of $\mathbb{R}^{k}$.
  Given the $i$\textsuperscript{th} eigenpair $(\lambda_i, \mathbf{y}_i)$ of the Hessenberg matrix, the $i$\textsuperscript{th} eigenpair of the original matrix $\mathbf{A}$ can be approximated as
  \[
  \lambda_i, \quad \text{and} \quad \mathbf{x}_i = \mathbf{V}_k \mathbf{y}_i,
  \]
  with the low-dimensional eigenvector $\mathbf{y}_i$ being normalized such that $\| \mathbf{y}_i \|_2 = 1$.
  The residual associated to this approximate eigenpair (also known as a \emph{Ritz eigenpair}) is given by
  \[
  \begin{aligned}
    \| \mathbf{Ax}_i - \lambda_i \mathbf{x}_i \| & = \| \mathbf{AV}_k \mathbf{y}_i - \lambda_i \mathbf{V}_k \mathbf{y}_i \| \\
    & = \| \mathbf{V}_k \left(\mathbf{H}_k -  \lambda_i \mathbf{I} \right) \mathbf{y}_i + \mathbf{v}_{k+1} \beta \mathbf{e}_k^T \mathbf{y}_i \| \\
    & = \vert \beta \vert \vert \mathbf{e}_k^T \mathbf{y}_i \vert.
  \end{aligned}
  \]
  If this residual is small enough for a sufficiently large number of Ritz eigenpairs, then the computation stops.
  Otherwise, a new Krylov vector is added to the basis $\mathbf{V}$, and the Arnoldi factorization continues until the desired number of Ritz eigenpairs have converged below the user-defined tolerance.
\end{tcolorbox}

Having computed a fixed point or a periodic orbit, one is often interested in its stability properties.
These can be its asymptotic stability (\ie{} modal stability characterized by the eigenvalues of the Jacobian matrix $\mathbf{L}$) or short-time stability (\ie{} non-modal stability characterized by the singular values of the exponential propagator $\exp(\tau \mathbf{L})$).

As discussed in \cref{subsubsec: theoretical -- modal stability}, given a fixed point $\mathbf{X}^*$, the linearized dynamics are governed by
\begin{equation}
\dfrac{d \mathbf{x}}{dt} = \mathbf{Lx},
\end{equation}
where $\mathbf{L}$ is the evolution operator linearized about $\mathbf{X}^*$.
For a periodic orbit, these dynamics are governed by
\begin{equation}
\dfrac{d \mathbf{x}}{dt} = \mathbf{L}(t) \mathbf{x},
\end{equation}
with $\mathbf{L}(t + \tau) = \mathbf{L}(t)$ and $\tau$ the period of the orbit.
In a time-stepper formulation, these continuous-time linear systems are replaced by the following discrete-time one
\begin{equation}
\mathbf{x}_{k+1} = \mathbf{M}_\tau \mathbf{x}_k,
\end{equation}
with $\mathbf{M}_\tau$ being the exponential propagator or monodromy matrix, depending on the context.
The matrix-vector product $\mathbf{M}_\tau \mathbf{x}_k$ amounts to integrating forward in time the linearized equations.
While for a fixed point, $\mathbf{M}_\tau$ and $\mathbf{L}$ have the same set of eigenvectors, their eigenvalues are related by
\begin{equation}
\lambda_i = \dfrac{\log (\mu_i)}{\tau},
\end{equation}
where $\tau$ is the sampling period.
For a periodic orbit, the eigenvalues $\mu_i$ of $\mathbf{M}_{\tau}$ are directly the Floquet multipliers needed to characterize the stability of the solution.

In both cases, the leading eigenpairs of $\mathbf{M}_\tau$ can easily be computed using the Arnoldi factorization described in \cref{subsec: numerics -- krylov} and algorithm 5.
Consider the factorization
\begin{equation}
\mathbf{M}_\tau \mathbf{V} = \mathbf{VH} + \beta \mathbf{e}^T_m \mathbf{r},
\end{equation}
with $\mathbf{V} \in \mathbb{R}^{n \times m}$ an orthonormal Krylov basis, $\mathbf{H} \in \mathbb{R}^{m \times m}$ an upper Hessenberg matrix and $\mathbf{r}$ the unit-norm residual after $m$ steps of the Arnoldi iteration.
Introducing the i\textsuperscript{th} eigenpair $(\hat{\mu}_i, \mathbf{y}_i)$ of the $m \times m$ upper Hessenberg matrix into the Arnoldi factorization leads to
\begin{equation}
\| \mathbf{M}_\tau \mathbf{Vy}_i - \mathbf{Vy}_i \hat{\mu}_i \|_2 = \vert \beta \mathbf{e}_m^T \mathbf{y}_i \vert.
\end{equation}
Hence, if the left-hand side is small enough, the pair $(\hat{\mu}_i, \mathbf{Vy}_i)$ provides a good approximation of the i\textsuperscript{th} eigenpair of the operator $\mathbf{M}_\tau$.
If one is interested in the short-time stability properties instead, the leading singular modes and associated gain can be computed using the same algorithm where $\mathbf{M}_\tau$ is replaced by $\mathbf{M}^{\dagger}_{\tau} \mathbf{M}_\tau = \exp( \tau \mathbf{L}^{\dagger} ) \exp\left( \tau \mathbf{L} \right)$.

\begin{figure*}[ht]\centering
\includegraphics[width=\textwidth]{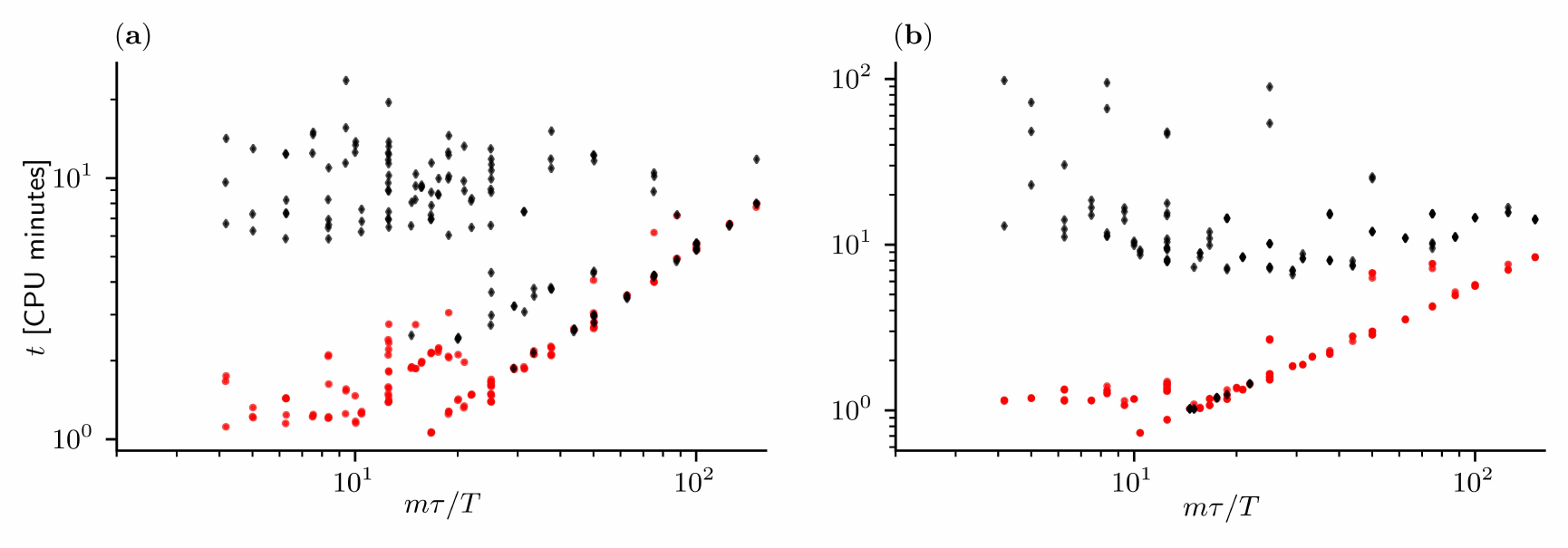}
\caption{Time to solution (in CPU minutes) versus different pairs of Krylov base size $m$ and integration time $\tau$ for the computation of eigenvalues: ($a$) 2D flow past a circular cylinder at $Re=80$ ($T=1/0.125$); ($b$) 2D open-cavity at $Re=4700$ ($T=1/1.676$). Red markers represent the time to solution for the convergence of 4 eigenmodes and black markers are for 40 eigenmodes. For each value of $m\tau$ three runs with different initial conditions are computed to  account for fluctuations of the computer. Eigenvalues with residual lower than $10^{-6}$ are considered converged. The parametric study is carried out using an automatic python script (located in \url{validations/eigen_loop/autorun.py}) looping over previously defined ranges of $m$ and $\tau$.}
\label{fig: numerics -- newton-krylov eigenvalues}
\end{figure*}

As for the Newton-Krylov solver presented in \cref{subsec: numerics -- newton-krylov}, the computational time is dominated by that of the call to the linearized solver needed to compute the matrix-vector product $\mathbf{M}_{\tau} \mathbf{x}_k$.
For fixed points, the choice of $\tau$ is also important as it plays a key role in the spectral gap between the eigenvalues of interest and the others.
This is discussed in \cref{subsec: observation eigenvalues}.
The initial vector $\mathbf{v}_1$ used to generate the Krylov subspace is also important.
Assuming random white noise distributed in the perturbation fields, the eigenpairs effectively start to converge only after a sufficiently large number of Krylov vectors have been generated such the transients are washed out of the computational domain.
The number of Krylov vectors that must be generated before this happens is, of course, dependent on the sampling period $\tau$ and the size of the computational domain under consideration.
From our experiences, assuming only one eigenvalue is unstable a good trade-off in terms of time-to-solution and computational cost is obtained when one chooses the size $m$ of the Krylov subspace and the sampling period $\tau$ such that
\[
    \dfrac{m \tau}{T} = \mathcal{O}(10),
\]
where $T$ is an \emph{a priori} estimate of the typical timescale of the instability.
If multiple eigenvalues are unstable, $T$ can be selected as the slowest time-scale.
We point out that such estimates should be regarded as indicative only, since effective computational performance depends on many other parameters not considered here (hardware, operating system, compiler, \etc).
Figure~\ref{fig: numerics -- newton-krylov eigenvalues} shows a parametric study performed to evaluate the performance and sensitivity of the method with respect to the size $m$ of the Krylov subspace and the integration time $\tau$.
At least three runs were made with $m \in (50,75,100,125,150,175,200)$ and $\tau = \left( \nicefrac{T}{12}, \nicefrac{T}{10}, \nicefrac{T}{8}, \nicefrac{T}{6}, \nicefrac{T}{4}, \nicefrac{T}{2} , T\right)$, where $T$ is the characteristic timescale of the instability.
Two cases are considered: the two-dimensional cylinder flow at $Re = 80$ and the two-dimensional open cavity flow at $Re=4700$.

\section{Examples}\label{sec: examples}

This section illustrates different applications of the use of Krylov methods to study large-scale dynamical systems.
All examples are taken from fluid dynamics applications.
Numerical simulations rely on the spectral element solver \texttt{Nek5000}~\cite{fischer2008nek5000,book:deville:2009} and the dedicated open-source toolbox \texttt{nekStab}~\url{nekstab.github.io} (see Appendix \ref{appendixA}).
It should be noted that, although we have focused our attention on a particular CFD solver, the methods presented earlier are quite general and can be relatively easily implemented in other partial differential equation solvers.
We give a brief physical description of each case, as well as details of the base flow and stability calculation, and a comparison with a reference work from the literature. 
Finally, a brief bifurcation analysis is presented for each case.
All the files needed to run these examples can be found in the \textit{nekStab/examples} folder, available in the repository \url{github.com/nekStab}.

\subsection{The flow in a two-dimensional annular thermosyphon}\label{sec:ex:thermo}
Under the influence of unstable thermal stratification, and for a range of control parameters, the two-dimensional flow in an annular thermosyphon is perhaps one of the simplest and cheapest computational test cases.
The geometry considered is the same as in \cite{tcfd:loiseau:2020}.
It consists of two concentric circular enclosures, the inner radius being $R_1$ and the outer radius $R_2$.
The ratio of the outer to inner radius is set to
\[
\dfrac{R_2}{R_1} = 2.
\]
A constant temperature $T_0$ is set at the upper walls, while the lower ones are set at a temperature $T_1 = T_0 + \Delta T$, with $\Delta T > 0$.
Hereafter, we work with the non-dimensional temperature $\vartheta(x, y)$ defined as
\[
\vartheta(x, y) = \dfrac{T(x, y) - T_0}{\Delta T},
\]
where $x$ and $y$ are the horizontal and vertical coordinates.
The origin of our reference frame is chosen to be the center of the thermosyphon.
Using this non-dimensionalization, the temperature at the lower walls is thus $\vartheta_w(y<0) = 1$, while the temperature at the upper ones is $\vartheta_w(y>0) = 0$.
Gravity acts in the vertical direction, along $-\bm{e}_y$, and is characterized by the gravitational acceleration $g$.

Assuming the working fluid is Newtonian, it is characterized by its density $\rho$, its dynamic viscosity $\mu$, its thermal expansion coefficient $\beta$, and its thermal diffusivity $\alpha$.
Using $\Delta T$ as the temperature scale, and $R_2 - R_1$ as the length scale, we can define two non-dimensional parameters, namely the Rayleigh number
\[
Ra = \dfrac{\rho g \beta \Delta T \left(R_2 - R_1\right)^3}{\mu \alpha},
\]
and the Prandtl number
\[
Pr = \dfrac{\nu}{\alpha},
\]
where $\nu = \nicefrac{\mu}{\rho}$ is the kinematic viscosity.
For this example, the Prandtl number is set to $Pr = 5$, and the Rayleigh number is varied.
For simplicity, the flow is assumed two-dimensional and incompressible, so that the effect of density variations due to temperature can be modeled using the Boussinesq approximation.
Under these assumptions, the dynamics of the flow are governed by the following Navier-Stokes equations
\[
\begin{aligned}
\dfrac{\partial \bm{u}}{\partial t} + \nabla \cdot \left( \bm{u} \otimes \bm{u} \right) & = -\nabla p + Pr \nabla^2 \bm{u} + Ra Pr \vartheta \bm{e}_y, \\
\dfrac{\partial \vartheta}{\partial t} + \left( \bm{u} \cdot \nabla \right) \vartheta & = \nabla^2 \vartheta, \\
\nabla \cdot \bm{u} & = 0,
\end{aligned}
\]
where $\bm{u}(\bm{x}, t)$ is the velocity field, $p(\bm{x},t)$ the pressure field, and $\vartheta(\bm{x}, t)$ the temperature field.
The computational domain is discretized using 32 spectral elements uniformly distributed in the azimuthal direction, and 8 elements uniformly distributed in the radial direction.
Within each element, Lagrange interpolation of order $N=7$ based on the Gauss-Lobatto-Legendre quadrature points is used in each direction, resulting in 16 384 grid points.
Temporal integration is performed using a third-order accurate scheme, and the time step has been chosen to satisfy the Courant–Friedrichs–Lewy (CFL) condition with Courant number $\mathrm{Co} < 0.5$ for all the simulations.
Despite the lack of turbulent dynamics due to the absence of the vortex stretching mechanism, the flow configuration can exhibit Lorenz-like chaotic dynamics.
This low-cost case follows the same structure as the bifurcation diagram of the Lorenz system.

\subsubsection{Pitchfork bifurcation}

\begin{figure}\centering
\includegraphics[scale=1.1]{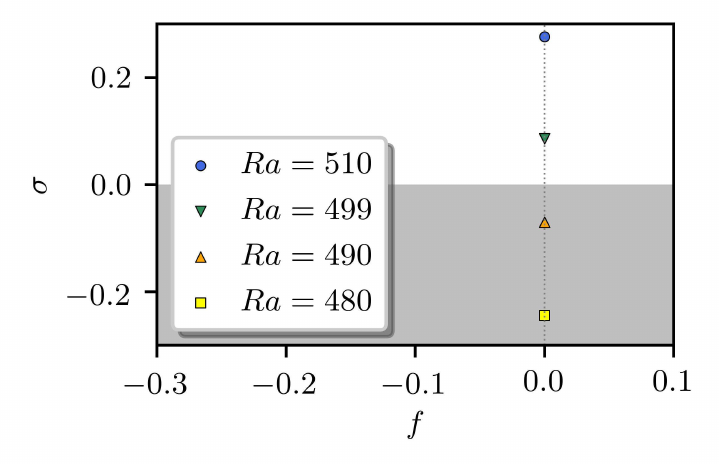}
\caption{Eigenvalues for the destabilization of the fixed point for the thermal convection on an annular thermosyphon.}
\label{fig:thermo:spectra}
\end{figure}

\begin{figure}\centering
\includegraphics[scale=1.1]{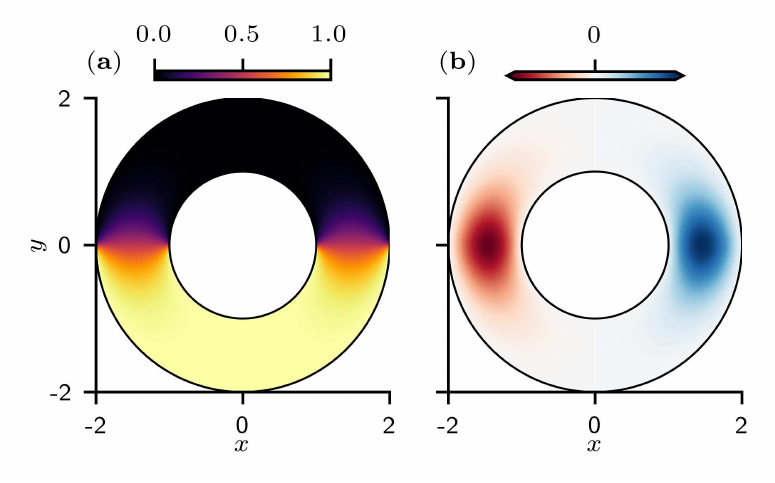}
\caption{The flow in a two-dimensional annular thermosyphon: ($a$) temperature field ($\vartheta$) of the bilateral symmetric fixed point and ($b$) real part of the temperature field ($\Re(\vartheta)$) of the unstable steady mode at $Ra\simeq499$.}
\label{fig:thermo:pitch}
\end{figure}

The first bifurcation of the flow occurs at the critical Rayleigh number $Ra_{c,1}\simeq494$.
It is associated with symmetry breaking in the temperature distribution, which leads to the emergence of a stationary convection cell.
The calculations presented were performed with $\tau=1$ (diffusive time unit) and a Krylov subspace of dimension $m=120$.
The choice of the sampling period for stationary modes is not straightforward, as a small value of $\tau$ leads to a poorly conditioned basis and the generation of spurious modes, while a large sampling period leads to an excessive computational cost.
The base flow, corresponding to pure conduction, is shown in \cref{fig:thermo:pitch}(a).
The eigenvalue spectrum in \cref{fig:thermo:spectra} shows a purely real eigenvalue stepping into the upper half-complex plane for $Ra \geq 494$.
The associated eigenvector is depicted in \cref{fig:thermo:pitch}(b) and leads to symmetry breaking.
The corresponding bifurcation is thus a pitchfork bifurcation.
Above $Ra_{c, 1}$, the conduction-dominated symmetric base flow is no longer stable and is replaced by a convection cell as shown in \cref{fig:thermo:hopf_mode}(a).
From symmetry considerations, this convection cell is equally likely to be associated with clockwise or counterclockwise flow (as shown in \cref{fig:thermo:hopf_mode}(a)).

\subsubsection{Hopf bifurcation}

\begin{figure}\centering
\includegraphics[scale=1.1]{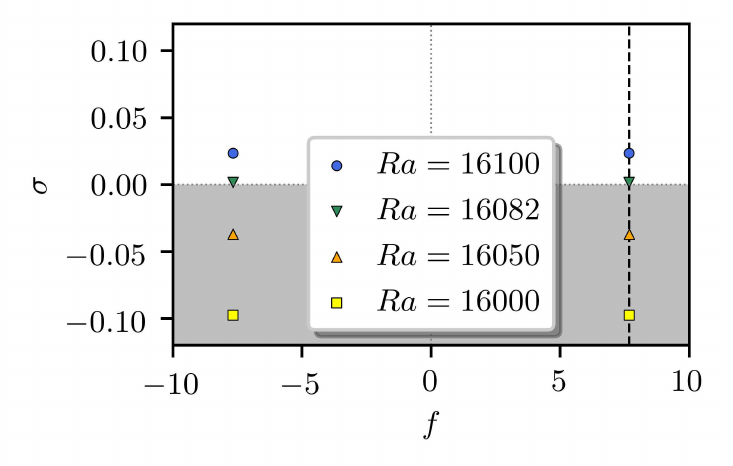}
\caption{Eigenvalues of the flow of the steady convection cell.
The dashed line represents the frequency $f=7.67$ from a DNS at $Ra=16100$.}
\label{fig:thermo:hopf}
\end{figure}

\begin{figure}\centering
\includegraphics[scale=1.1]{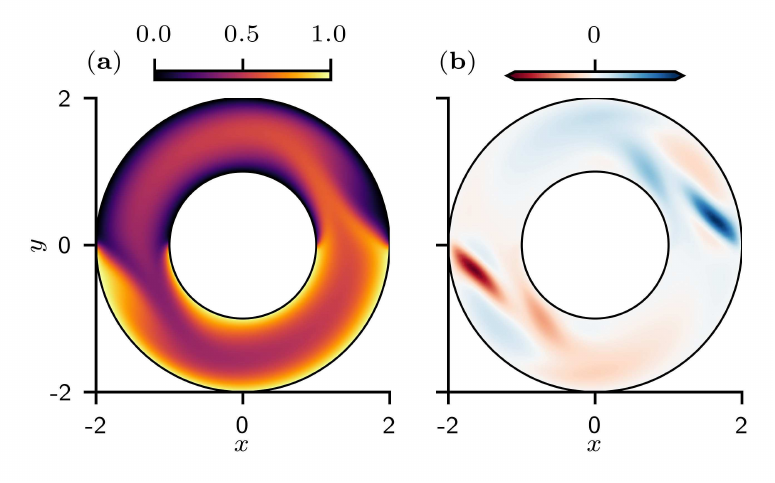}
\caption{The flow in a two-dimensional annular thermosyphon: ($a$) temperature field ($\vartheta$) of an unstable steady convection cell and ($b$) real part of the temperature field ($\Re(\vartheta)$) of the unsteady unstable mode at $Ra=16100$.}
\label{fig:thermo:hopf_mode}
\end{figure}

The new base flow is stable over a wide range of Rayleigh numbers.
It eventually becomes unstable at $Ra_{c,2} \simeq 16\ 081$ through a Hopf bifurcation.
This is indicated by a pair of complex conjugate eigenvalues of the associated linearized Navier-Stokes operator moving toward the upper half-complex plane in \cref{fig:thermo:hopf}.
The spatial structure of the leading mode is shown in \cref{fig:thermo:hopf_mode}(b).
The convection cell shown in \cref{fig:thermo:hopf_mode}(a) starts to oscillate with a characteristic Strouhal number $St=7$.
The frequency predicted by the linear analysis and the frequency measured in our DNS match very well near the bifurcation point when the nonlinear distortion of the base flow is minimal.

Eigenvalue calculations were performed with $\tau\approx0.014$ (corresponding to a standard recommendation for the sampling period of $\tau=T/8$) convective time and using a Krylov subspace of dimension $m=120$.
Base flows were calculated using the Newton-Krylov method under the same conditions. 
For comparison metrics, the calculation of the unstable base flow at $Ra=16\ 100$ starting from the base flow at $Ra=16\ 000$ took 147 seconds with Newton-Krylov compared to 1071 seconds with selective frequency damping (SFD)~\cite{pof:akervik:2006} (\ie{} one seventh of the time). 
For SFD, we considered a parameterization proposed by \cite{casacuberta2018effectivity} that leads to a more robust selection of the cutoff and gain for low-pass filtering compared to the original guidelines~\cite{pof:akervik:2006}.

\subsection{The harmonically forced jet}\label{sec:ex:hjet}

Our attention now shifts towards a time-periodic flow harmonically forced via the inflow boundary condition.
The dynamics of the flow are governed by the incompressible Navier-Stokes equations
\begin{equation}
\begin{aligned}
\dfrac{\partial \bm{u}}{\partial t} + \nabla \cdot \left( \bm{u} \otimes \bm{u} \right) & = -\nabla p + \dfrac{1}{Re} \nabla^2 \bm{u}, \\
\nabla \cdot \bm{u} & = 0.
\end{aligned}
\label{eq: Navier-Stokes eq}
\end{equation}
The Reynolds number is defined as
\[
Re = \dfrac{U_0 D}{\nu},
\]
where $U_0$ is the velocity at the jet centerline, $D$ the jet diameter, and $\nu$ the kinematic viscosity of the fluid.
For simplicity, the jet is assumed to be radially symmetric, and the axisymmetric formulation of the Navier-Stokes is considered with $\bm{x}=(z, r)$, $z$ representing the streamwise direction and $r$ the radial direction.
The time-periodic structure is \emph{forced} via a Dirichlet inflow boundary condition prescribed as
\begin{equation*}
    u(z=0,r,t)=\frac{1}{2}\left\{ 1-\tanh\left[ \frac{1}{4\theta_0 \left( r-4r^{-1}\right)}\right]\right\}(1+A\cos(\omega_f t)),
\end{equation*}
with $A=0.05$ the force amplitude, $\theta_0 = 0.025$ the initial thickness of the dimensionless shear layer, and angular frequency $\omega_f$. 
The non-dimensional frequency is the Strouhal number
\[
St= \omega_f D/(2\pi U_0).
\]
The computational domain extends from $0$ to $L_z = 40$ in the streamwise direction and $0$ to $L_r = 5$ in the radial direction.
The domain is discretized with $n_z \times n_r = 160 \times 30$ spectral elements.
Lagrange interpolants of order $N=5$ are considered, resulting in 172 800 grid points.
Based on the recent work of \cite{jfm:leopold:2019}, we reproduced a case with $St=0.6$ and investigated its modal stability properties.
Under these conditions, a forced limit cycle is formed.
For subcritical conditions (\ie{} $Re<Re_c$), vortices form periodically along the shear layer.
They are then transported in the streamwise direction before fading out because of viscous diffusion.
No pairing phenomena are identified despite the appearance of harmonics in the velocity signals.
Above the critical Reynolds number $Re_c=1371$, the vortices spontaneously start to pair, forming larger vortices.
The vortex pairing is connected to a subharmonic instability created via a period-doubling bifurcation.

\subsubsection{Period-doubling bifurcation}

\begin{figure*}[ht]
\centering
\includegraphics[width=\textwidth]{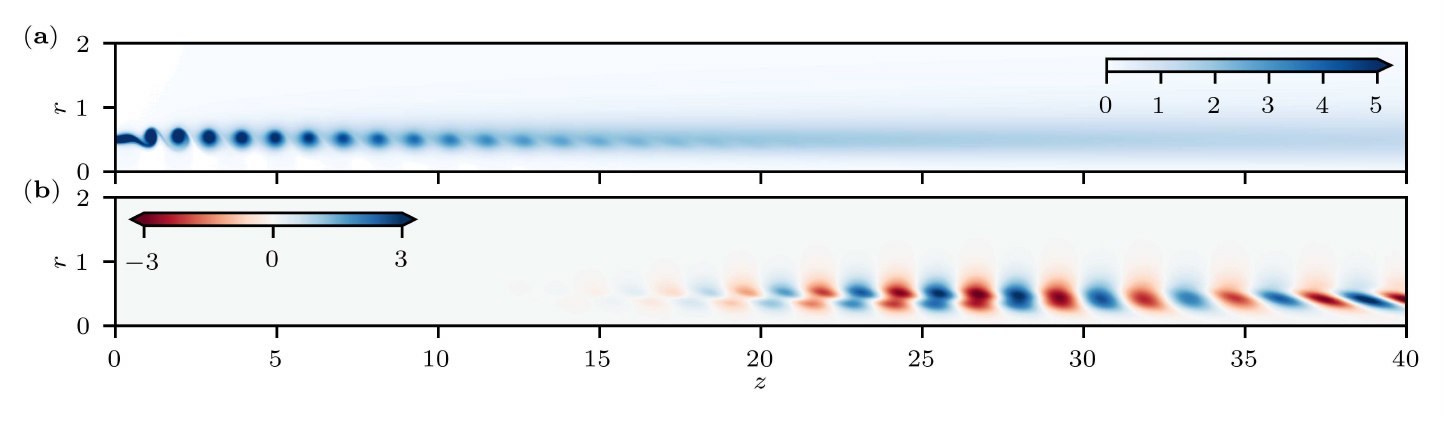}
\caption{The harmonically forced jet: ($a$) vorticity component of the stabilized unstable limit cycle at supercritical $Re=2000$ and ($b$) spatial distribution of the subharmonic Floquet mode. Inflow forcing with 5\% amplitude and $St_D=0.6$.}
\label{fig:jet:mode}
\end{figure*}

\begin{figure}
\centering
\includegraphics{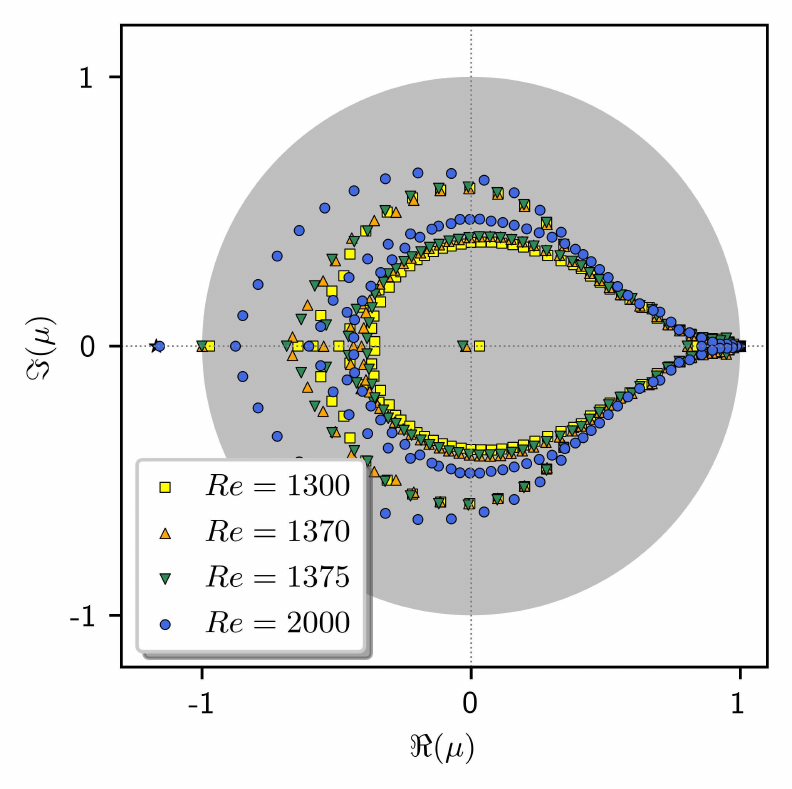}
\caption{Evolution of Floquet multipliers of the harmonic forced axisymmetric jet. The single leading unstable eigenmode is associated with the vortex pairing mechanism and a period-doubling bifurcation. The almost superposed black star represents the reference value from \cite{jfm:leopold:2019}.}
\label{fig:jet:spectra}
\end{figure}

In \cite{jfm:leopold:2019}, the time-delayed feedback technique \cite{shaabani2017time} was used to stabilize periodic orbits. 
The technique is based on the nonharmonic component filtering approach introduced by \cite{pyragas1992continuous} using an optimal filter gain derived in \cite{shaabani2017time}.
Here, the (unstable) time-periodic base flow is computed using the Newton-Krylov method with dynamic tolerances (introduced in \cref{subsubsec: numerics -- fixed points}), with $\tau=1/St$ and a Krylov subspace dimension of $m=128$ until a residual level of $10^{-11}$.
The same parameters are used for the Floquet analysis, which is sufficient to converge 20 eigenpairs to a precision of $10^{-6}$.
An unstable base flow without vortex pairing can be seen in \cref{fig:jet:mode}(a).
Figure~\ref{fig:jet:mode}(b) illustrates the dominant Floquet mode.
Figure~\ref{fig:jet:spectra} shows the spectrum of Floquet multipliers, with the leading one leaving the unit circle along $\mu = -1$, which is characteristic of a period-doubling bifurcation.
The critical number $Re_{c,1}\simeq1371.18$ is in excellent agreement with the reference value $Re_{c,1}\simeq1371$ given in \cite{jfm:leopold:2019}.
Despite the strong non-normality of the system operator, the Floquet analysis accurately predicts the leading mode responsible for the vortex pairing mechanism observed in nonlinear simulations~\cite{jfm:leopold:2019}.

\begin{figure*}\centering
\includegraphics[width=0.96\textwidth]{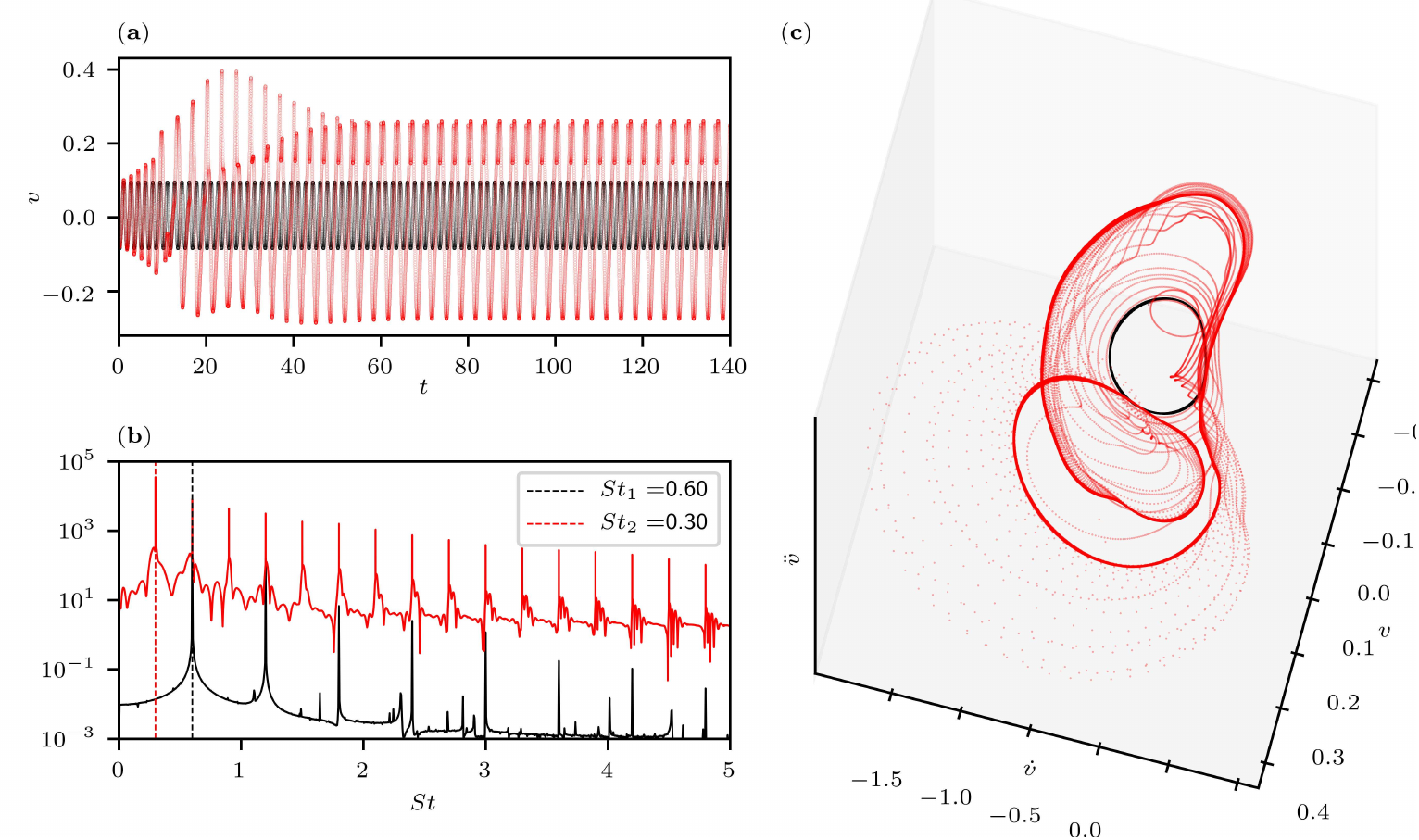}
\caption{Radial velocity signal at $x,r=(5,0.5)$ at subcritical $Re=1370$ (black) and supercritical $Re=2000$ (red): ($a$) signal evolution and ($b$) discrete Fourier transform (DFT) spectrum; ($c$) Phase portrait of the system in coordinates ($v,\dot{v},\ddot{v}$): one can observe the departure of the system trajectory from the stable limit cycle (in black), timidly exploring the phase space before settling into a period-doubled orbit (in red).}
\label{fig:jet:dft}
\end{figure*}

We verified the predictions with nonlinear simulations at subcritical $Re=1370$ and supercritical $Re=2000$, as shown in \cref{fig:jet:dft}.
At $Re=1370$ (in black), the time series of the velocity probe and its Fourier spectrum are characterized by the forced frequency $St=0.6$ and its harmonics formed by nonlinearities. 
At $Re=2000$ (in red), one can see the response of the flow in the velocity signal with the sharp increase in the subharmonic frequency at $St=0.3$, showing the growth of the secondary instability through a period-doubling bifurcation and the increase in the period of the limit cycle.

\subsection{The flow past a circular cylinder}\label{sec:ex:1cyl}

Let us now consider the example of a canonical cylinder flow assumed to be infinite in the spanwise direction.
The dynamics of the flow are governed by the Navier-Stokes equations~\eqref{eq: Navier-Stokes eq}, with the Reynolds number defined on the basis of the free-stream velocity and the cylinder's diameter.
The two-dimensional mesh considered in \cref{subsubsec: primary instability} is made of 1464 spectral elements (66 in the flow direction and 30 in the vertical), all of which are comparable to previously reported domains.
To limit the computational cost, we consider Lagrange interpolants of order $N=5$, which shows good convergence compared to $N=7$ which leads to a total of 52 704 grid points.
For the three-dimensional problem considered in \cref{subsubsec: secondary instability}, this mesh is extruded in the third direction, using 10 elements in the spanwise direction to a length $L_z=2\pi/\beta_c\simeq 3.964$.
The total number of grid points is then 3 162 240.
As for the other examples, we consider a third-order accurate temporal scheme and a time-step that satisfies $\mathrm{Co} < 0.5$ for all simulations.

\subsubsection{Primary instability and sensitivity analysis}\label{subsubsec: primary instability}
At low Reynolds numbers, the flow is two-dimensional, steady and symmetric with respect to the cross-stream direction.
According to Jackson~\cite{jackson1987finite} and more recently Kumar \& Mittal~\cite{kumar2006effect}, the flow is expected to become unstable at $Re_{c,1}= UD /\nu \approx46.6$, leading to the emergence of the well-known von Kàrmàn vortex street characterized by $St_c = 0.125$.
This primary instability is a canonical example of a supercritical Hopf-type bifurcation.
Numerous wake flows~\cite{pier2013periodic} exhibit an instability that leads to the onset of periodic vortex shedding and their characterization as \emph{flow oscillators}.

Sensitivity to base flow changes based on the linear Navier-Stokes operator was introduced by Bottaro \etal~\cite{bottaro2003effect} in a local framework and was later extended to the global framework by Marquet \etal~\cite{marquet2008sensitivity}.
Base flow sensitivity analysis has been shown to provide valuable information for shape optimization or actuator placement.
The sensitivity analysis of the complex eigenvalue $\lambda =\sigma+i\omega$, with the real part $\sigma$ being the growth rate of the eigenmode and the imaginary part $\omega$ its frequency, allows the computation of vector fields that highlight the net effects of generic small-amplitude base flow modifications $\delta \bf{U}$.
Both the sensitivity analysis for generic modifications of a base flow (\eg{} $\bf{U}$) and for a steady force (\eg{} $\bf{F}$) have been implemented in \texttt{nekStab}.

\begin{figure}\centering
\includegraphics{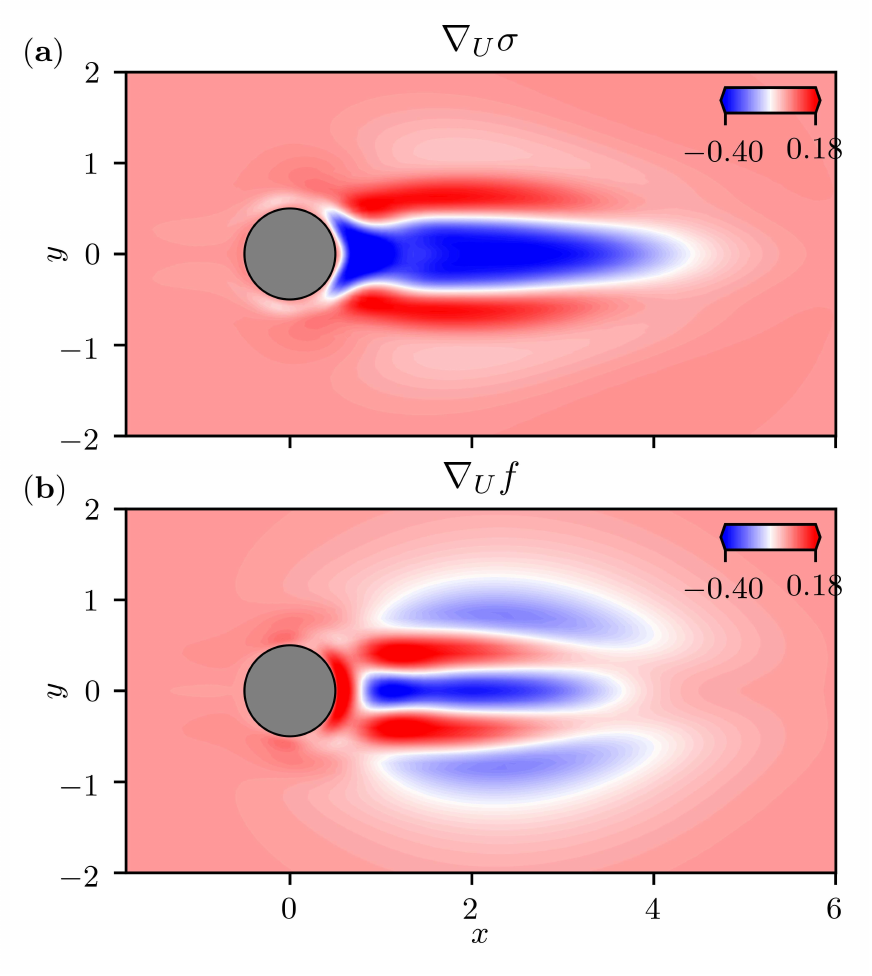}
\caption{Streamwise component of the sensitivity to base flow modifications $\nabla_{\bf{U}}\lambda$ of the leading eigenvalue $\lambda$ at Reynolds number $Re=50$. Spatial distribution of ($a$) the growth rate sensitivity $\nabla_{\bf{U}}\sigma$ and ($b$) the frequency sensitivity $\nabla_{\bf{U}}\omega$.}
\label{fig:cyl_sens_bf}
\end{figure}

Structural changes in the complex eigenvalue $\delta\lambda$ due to small-amplitude arbitrary base flow modifications $\delta \bf{U}$ can be formally related through the inner product
\begin{equation*}
    \delta\lambda = \langle \nabla_{U}\lambda | \delta \mathbf{U} \rangle.
\end{equation*}
The gradient $\nabla_{U}\lambda$ is a complex vector field that defines the sensitivity to base flow modifications, given by
\begin{equation}
    \nabla_{\bf{U}}\lambda = -(\nabla \mathbf{u})^{H}\cdot \mathbf{u}^\dagger + \nabla\mathbf{u}^\dagger \cdot \hat{\mathbf{u}}^{*},\label{eq:sensiti}
\end{equation}
with the superscript $H$ representing the conjugate transpose, $\dagger$ the adjoint, and $*$ the complex conjugate (for a complete derivation, the reader is referred to Marquet \etal~\cite{marquet2008sensitivity}).
The first term of \cref{eq:sensiti} is the sensitivity to transport modifications, while the second term is the sensitivity to production modifications.
Figure~\ref{fig:cyl_sens_bf} depicts the real and imaginary parts of $\nabla_{\bf{U}}\lambda$.
These represent the sensitivity of the growth rate of the eigenvalue to base flow modifications, and the frequency sensitivity of the eigenvalue to base flow modifications, respectively.
Generic base flow modifications in negative sensitivity zones promote stabilization of the eigenvalue (\ie{} reduction of the growth rate or frequency), whereas changes in positive sensitivity zones promote destabilization or frequency increase.

\begin{figure}\centering
\includegraphics{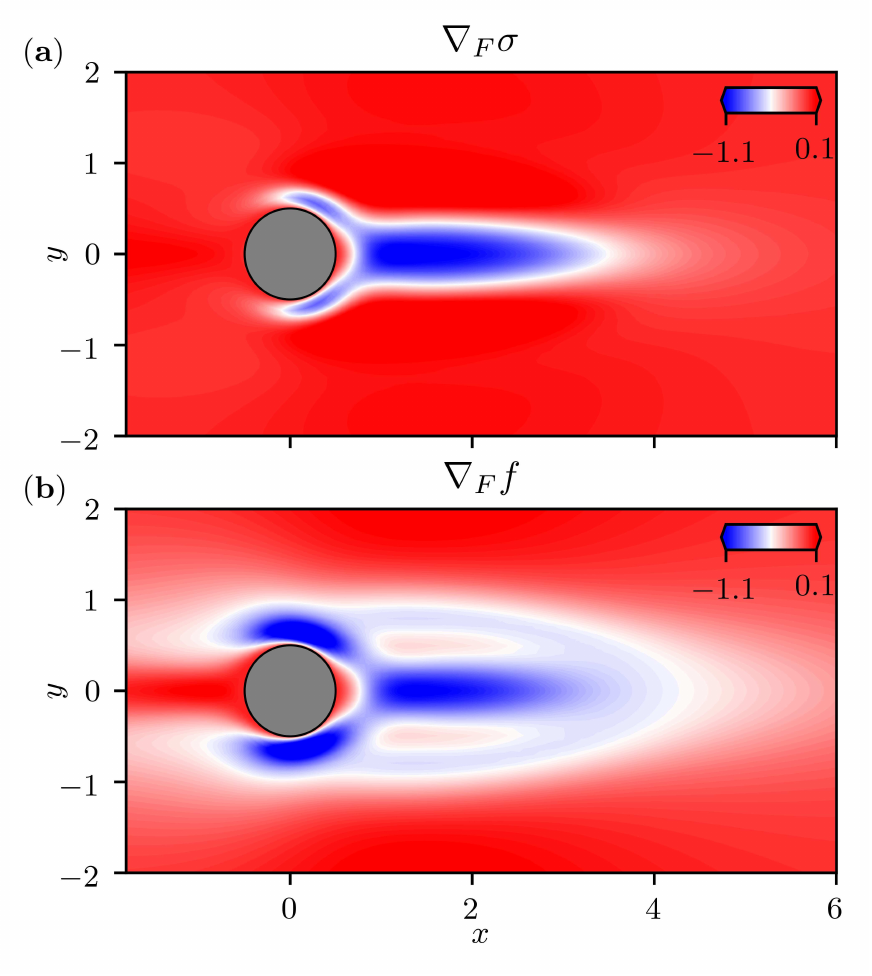}
\caption{Normalized modulus of the sensitivity to a steady force $\nabla_{\bf{F}}\lambda$ of the leading eigenvalue $\lambda$ at Reynolds number $Re=50$. 
Spatial distribution of ($a$) the growth rate sensitivity $\nabla_{\bf{F}}\sigma$ and ($b$) the frequency sensitivity $\nabla_{\bf{F}}\omega$.}
\label{fig:cyl_sens_for}
\end{figure}

Analogously, the sensitivity to a steady force can be derived by introducing an inner product in relation to a volume force $\bf{F}$ in the form
\begin{equation*}
    \delta\lambda = \langle \nabla_{\mathbf{F}}\lambda | \delta \mathbf{F} \rangle.
\end{equation*}
The complex sensitivity function $\nabla_{\mathbf{F}}\lambda$ is given by the knowledge of adjoint base flow fields in the form
\begin{equation}
    \nabla_{\mathbf{F}}\lambda = \mathbf{U}^\dagger. \label{eq:force_sensiti}
\end{equation}
This adjoint base flow is a solution to the linear system of equations
\begin{equation}
    \begin{aligned}
    -\nabla \mathbf{U}^{\dagger} \cdot \mathbf{U}_b + \left( \nabla \mathbf{U}_b \right)^T \cdot \mathbf{U}^{\dagger} - \nabla P^{\dagger} - \dfrac{1}{Re} \nabla^2 \mathbf{U}^{\dagger} & = \nabla_{\mathbf{U}} \lambda , \\
    \nabla \cdot \mathbf{U}^{\dagger} & = 0.
    \end{aligned}
    \label{eq: sensitivity linear equation}
\end{equation}
Assuming the adjoint Jacobian matrix $\mathbf{L}^{\dagger}$ has already been projected onto a divergence-free subspace, this equation can be written as
\[
\mathbf{L}^{\dagger} \mathbf{U}^{\dagger} = \nabla_{\mathbf{U}} \lambda.
\]
Note that, in a time-stepper framework, we do not have access to $\mathbf{L}^{\dagger}$, but only $\exp(\tau \mathbf{L}^{\dagger})$.
Accounting for the time derivative, the general solution of the linear system above is given by
\[
\mathbf{U}^{\dagger}(\tau) = \exp(\tau \mathbf{L}^{\dagger}) \mathbf{U}^{\dagger}(0) + \int_0^\tau \exp((\tau-t) \mathbf{L}^{\dagger}) \nabla_{\mathbf{U}}\lambda \ \mathrm{d}t.
\]
In a time-stepper framework, \cref{eq: sensitivity linear equation} is thus replaced with
\begin{equation}
\left( \mathbf{I} - \exp(\tau \mathbf{L}^{\dagger}) \right) \mathbf{U}^{\dagger} = \int_0^\tau \exp((\tau-t) \mathbf{L}^{\dagger}) \nabla_{\mathbf{U}}\lambda \ \mathrm{d}t,
\end{equation}
which is solved using the GMRES solver discussed earlier in the manuscript.
Note that the right-hand side is computed by running an adjoint simulation with the steady force $\nabla_{\mathbf{U}}\lambda$ for $\tau$ time units with a zero initial condition.
\Cref{fig:cyl_sens_for} depicts the real and imaginary parts of $\nabla_{\mathbf{F}} \lambda$.
These are the sensitivity of the growth rate of the eigenvalue to changes due to a steady force, and the associated frequency sensitivity of the eigenvalue, respectively.
These maps are consistent with those obtained in \cite{marquet2008sensitivity} and experimentally in \cite{strykowski1990formation}.
Although not reproduced here, the theoretical framework was extended by Giannetti, Camarri, and Citro~\cite{giannetti2019sensitivity} to include sensitivity analysis with respect to generic modifications and a force acting on periodic orbits.

\subsubsection{Pitchfork bifurcation}\label{subsubsec: secondary instability}

We now focus on the Floquet analysis of three-dimensional modes evolving in a two-dimensional time-periodic base flow formed in the wake of a circular cylinder.
This problem has been studied in detail by Barkley \& Henderson~\cite{jfm:barkley:1996} on the basis of a Fourier expansion in the spanwise direction, as is the gold standard for demonstrating a (secondary) pitchfork bifurcation.
In addition, this symmetry-breaking bifurcation~\cite{nayfeh2008applied} is characterized by a single real eigenvalue that becomes positive, corresponding to a single Floquet multiplier that leaves the unit cycle by $\mu=1$.
At this point, the flow experiences a steady bifurcation (synchronous with the underlying periodic orbit), thus not altering the temporal structure of the flow.
The frequency of the underlying limit cycle remains the same, but the spanwise invariance of the flow is broken, resulting in a three-dimensionalization of the flow.
Barkley \& Henderson~\cite{jfm:barkley:1996} report a secondary instability to what is called mode A occurring at $Re_{c,2} \simeq 188.5\pm 1$ with a critical wavenumber $\beta_c = 1.585$ in the spanwise direction, corresponding to a length of nearly four diameters in the spanwise direction.
Using a domain $(L_i,L_h,L_o)=-15D,\pm 22D,25D$, they report a limit cycle oscillating with $St=0.1954$ at $Re=190$, and a leading synchronous Floquet mode with $\mu= 1.034$. 
Later Giannetti, Camarri, and Luchini~\cite{giannetti2010structural}, employing a finite element code and a domain $(L_i,L_h,L_o)=-16.5D,\pm 11D,34.5D$, calculated $Re_{c,2} \simeq 189.77$ and some reference values at $Re=190$ for the limit cycle frequency $St=0.1971$ and $\mu=1.002$ for mode A. 
Recently, Giannetti, Camarri, and Citro~\cite{giannetti2019sensitivity}, using a spectral element code and a domain $(L_i,L_h,L_o)=-15D,\pm 15D,35D$, obtained $Re_{c,2} \simeq 189.71$ and reported a limit cycle with $St=0.1962$ at $Re=190$, as well as $\mu=1.009$ for mode A.

\begin{figure}\centering
\includegraphics[width=0.5\textwidth]{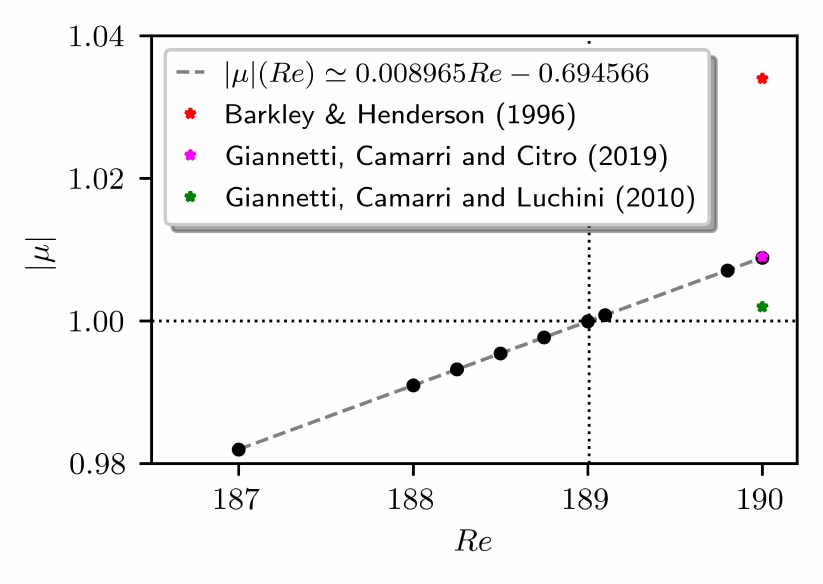}
\caption{Modulus of the dominant Floquet multiplier as a function of the Reynolds number for the flow past a circular cylinder.}
\label{fig:cyl_spec}
\end{figure}

\begin{figure*}[ht]\centering
\includegraphics[width=\textwidth]{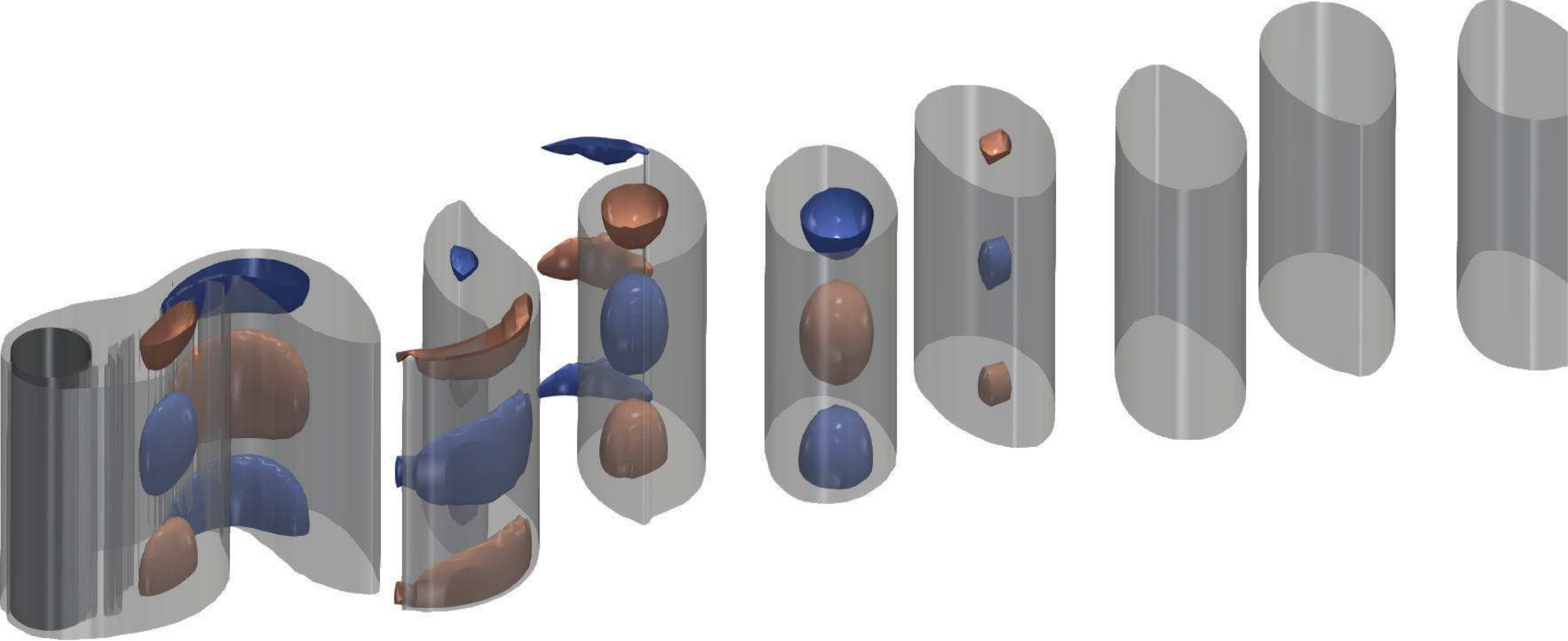}
\caption{The flow past a circular cylinder: semitransparent vorticity magnitude contours $(\omega=0.35)$ of the limit cycle at supercritical $Re=190$ superimposed with streamwise vorticity contours $(\omega_x=\pm 0.18)$ of the real part of the unstable steady mode.}
\label{fig:cyl_spec:mode}
\end{figure*}

Barkley~\cite{barkley2005confined} mentions that accurate estimates of Floquet stability analyses can be obtained in smaller domains $(L_i,L_h,L_o)=-8D,\pm 8D,25D$.
Since \texttt{nekStab} relies on the spectral element solver \texttt{Nek5000}, which does not use a Fourier decomposition in the spanwise direction, a finite-span domain is specified. A limit cycle can be computed directly in the 3D domain, or due to the geometry of the flow computational time can be saved simply by setting the spanwise velocity component to zero or by extruding a 2D solution into a 3D mesh (\eg{} using the \texttt{pymech}~\cite{pymech} package).
In our case, using the Newton-Krylov method for stabilizing UPOs, we first perform a DNS at subcritical $Re=187$ and thus transients from the initial condition are convected out of the computational domain, obtaining a 2D subcritical periodic orbit in the 3D domain.
The orbit period is measured from the DNS and used as an initial guess for the Newton GMRES algorithm, together with a snapshot of the orbit.
In this way, we can gradually stabilize the UPOs at supercritical Reynolds numbers.
A Krylov subspace of dimension $m=64$ is used, both for base flow and Floquet stability calculations.
Due to the almost degenerate nature of the eigenvalues, a Krylov-Schur iteration is used for the convergence of at least 4 direct modes.
The evolution of Floquet multipliers \textit{moduli} is shown in \cref{fig:cyl_spec}, as well as the best linear fit for their evolution along the Reynolds number. 
The fully 3D leading unstable Floquet mode at $Re=190$ is shown in \cref{fig:cyl_spec:mode} and shows excellent qualitative agreement with the unstable mode presented in Barkley \& Henderson~\cite{jfm:barkley:1996} and also in Blackburn and Lopez~\cite{blackburn2005symmetry}.
Our estimate for the critical Reynolds number $Re_{c,2}=189$ using a domain with $\beta_c=1.585$ agrees very well with the range of values reported by Barkley \& Henderson~\cite{jfm:barkley:1996} as well as Giannetti, Camarri and Luchini~\cite{giannetti2010structural}, despite our necessity to fix a spanwise wavenumber that could not precisely match the critical one reported, as well as a different mesh strategy, domain size, and polynomial order.
Specifically, our $\mu=1.012$ at $Re=190$ differs by 2.15\% with $\mu=1.034$ by Barkley~\cite{barkley2005confined} and 1\% with $\mu=1.002$ by Giannetti, Camarri, and Luchini~\cite{giannetti2010structural}.

\subsection{The flow past side-by-side circular cylinders}\label{sec:ex:2cyl}

The flow past two side-by-side cylinders is mainly governed by a Reynolds number based on the cylinder diameter and the free-stream velocity (similar to a single-cylinder wake), but with the addition of a separating distance (gap) between the surfaces.
A mesh with 5092 elements and polynomial order $N=7$ is considered, spanning from -50 to 75 in the streamwise direction and -50 to 50 in the vertical direction.
Taking into account a gap $g=0.7$, \cite{jfm:carini:2014} reports a primary Hopf bifurcation at $Re_{c,1}\simeq 55$ with a synchronized vortex shedding wake oscillating at $St_{c,1}=0.11$.
At $Re_{c,2}\simeq 61.7$, a secondary mode known as the ``flip-flop'' mode with $St_{c,2}=0.02$ becomes unstable due to a supercritical Neimark-Sacker bifurcation.
The flow is bistable: an asymmetric dual wake arises, in which much slower switching can be observed.

\subsubsection{Neimark-Sacker bifurcation}

\begin{figure}\centering
\includegraphics{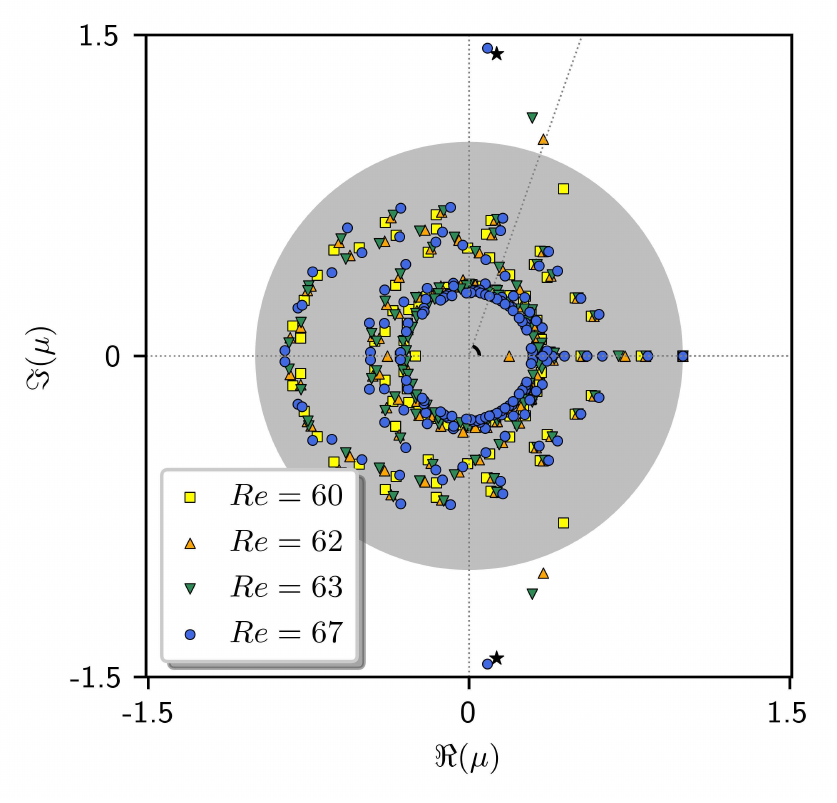}
\caption{Evolution of Floquet multipliers for the flow past two side-by-side cylinders. A pair of modes associated with the flip-flop instability leaves the unit cycle increasing its moduli (both growth rate and frequency) as a function of $Re$ in a Neimark-Sacker bifurcation. The black star represents the reference value from \cite{jfm:carini:2014} at $Re=67$.}
\label{fig:2cyl:spectra}
\end{figure}

\begin{figure}\centering
\includegraphics[scale=1.06]{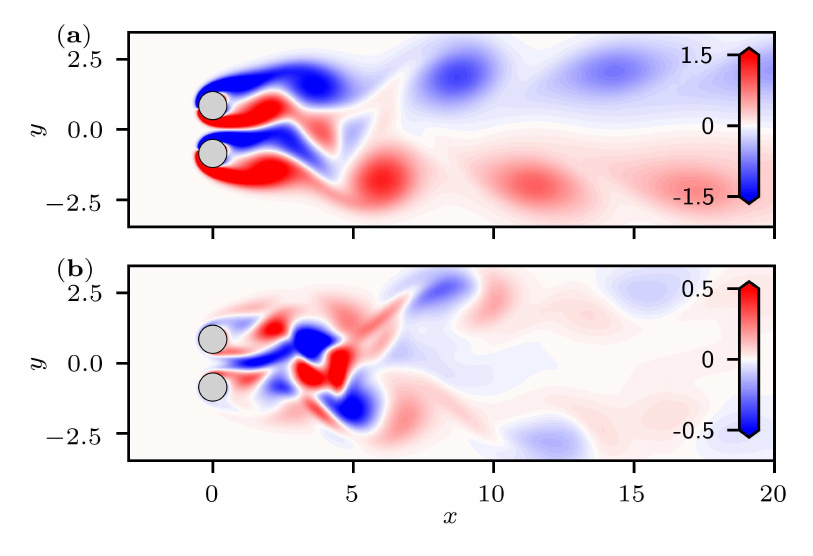}
\caption{The flow through side-by-side circular cylinders: ($a$) snapshot of the limit cycle and ($b$) Floquet ``flip-flop'' mode at $Re=67$.}
\label{fig:2cyl:mode}
\end{figure}

\begin{figure*}\centering
\includegraphics[width=\textwidth]{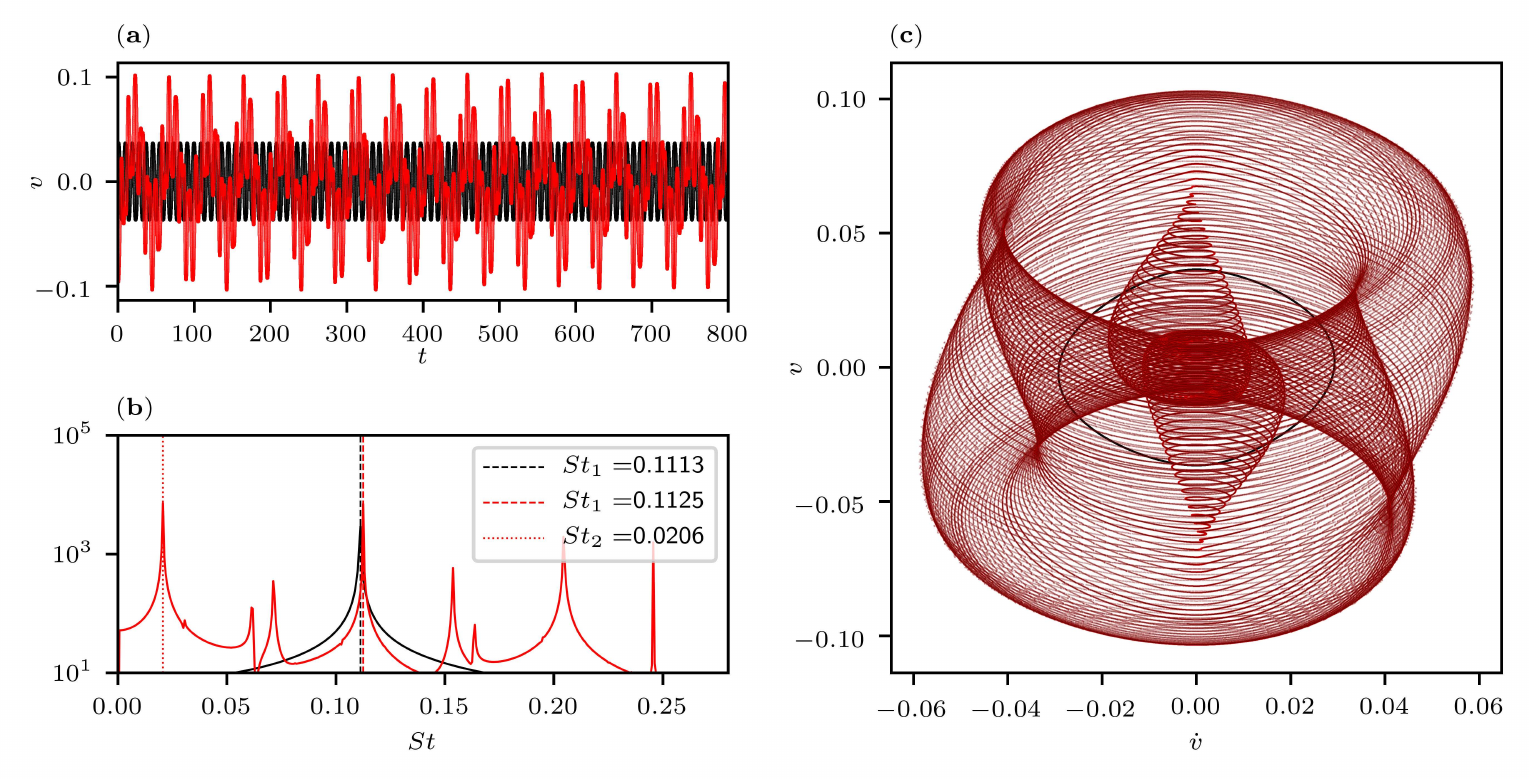}
\caption{Vertical velocity signal at $(x,y)=(5,0)$ in the flow past side-by-side cylinders with $g=0.7$ for the subcritical $Re=60$ (black) and supercritical $Re=62$ (dark red) regimes: ($a$) signal evolution and ($b$) DFT spectrum; ($c$) trajectory of the system in phase space using the coordinates (\ie{} $v,\dot{v}$) of both the limit cycle (black) and the torus (dark red).}
\label{fig:2cyl:his}
\end{figure*}

The Neimark-Sacker or secondary Hopf is characterized by the emergence of a new incommensurate frequency in the flow. 
This bifurcation differs from a pitchfork when the instability created oscillates with the same frequency of the periodic orbit and from period-doubling when the instability created is subharmonic with respect to the periodic orbit. 
Figure~\ref{fig:2cyl:spectra} shows the spectrum of Floquet multipliers for different Reynolds numbers.
The critical Reynolds number is $Re_{c,2}\simeq61.17$, which is in excellent agreement with the reference value $Re_{c,2}\simeq61.6$ reported in \cite{jfm:carini:2014}.
The mode leaves the unit disk at an angle of 71 degrees (\ie{} $\phi=\tan^{-1}(\Im( \mu)/\Re( \mu))$).
Figure~\ref{fig:2cyl:mode} shows a snapshot of the vorticity distribution of the UPO under supercritical conditions $Re=67$ and the vorticity distribution of the unstable Floquet mode associated with the flip-flop mechanism. 
Figure~\ref{fig:2cyl:his}(a,b) show the time trace and the Fourier spectrum of the velocity recorded by a probe in the wake.
The main discrete Fourier transform (DFT) spectrum peak is located at the fundamental frequency of the UPO, with another (second) peak occurring at the new incommensurable frequency.
Figure~\ref{fig:2cyl:his}(c) shows the phase-space representation of dynamics with the formation of a torus object characteristic of quasiperiodic dynamics.

\subsection{Backward-facing step}

\begin{figure}\centering
\includegraphics{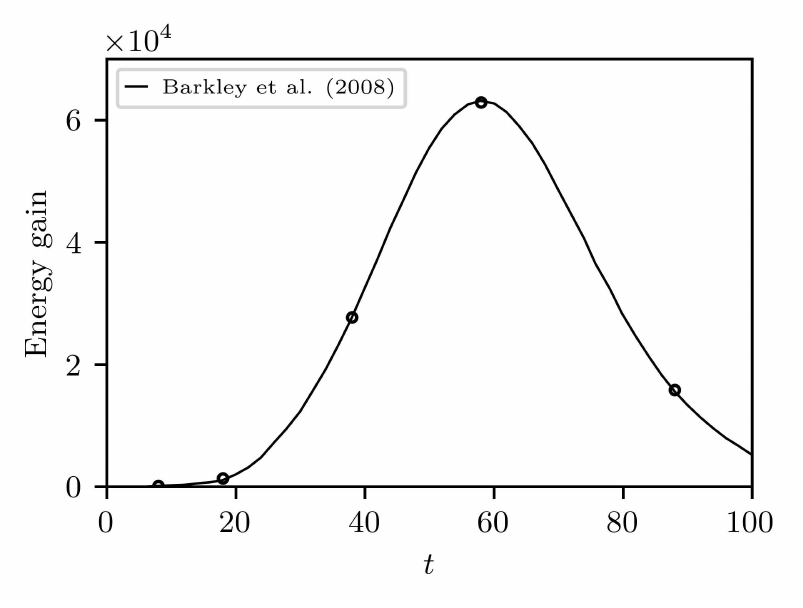}
\caption{Envelope of the optimal gain for the 2D backward-facing step problem computed for $Re=500$. The parametric study is carried out using an automatic Python script (located in \url{examples/backward_facing_step/autocomp_tg.py}) looping over a predefined range of $\tau$.}
\label{fig:backward_ts}
\end{figure}

\begin{figure*}[ht]\centering
\includegraphics[width=\textwidth]{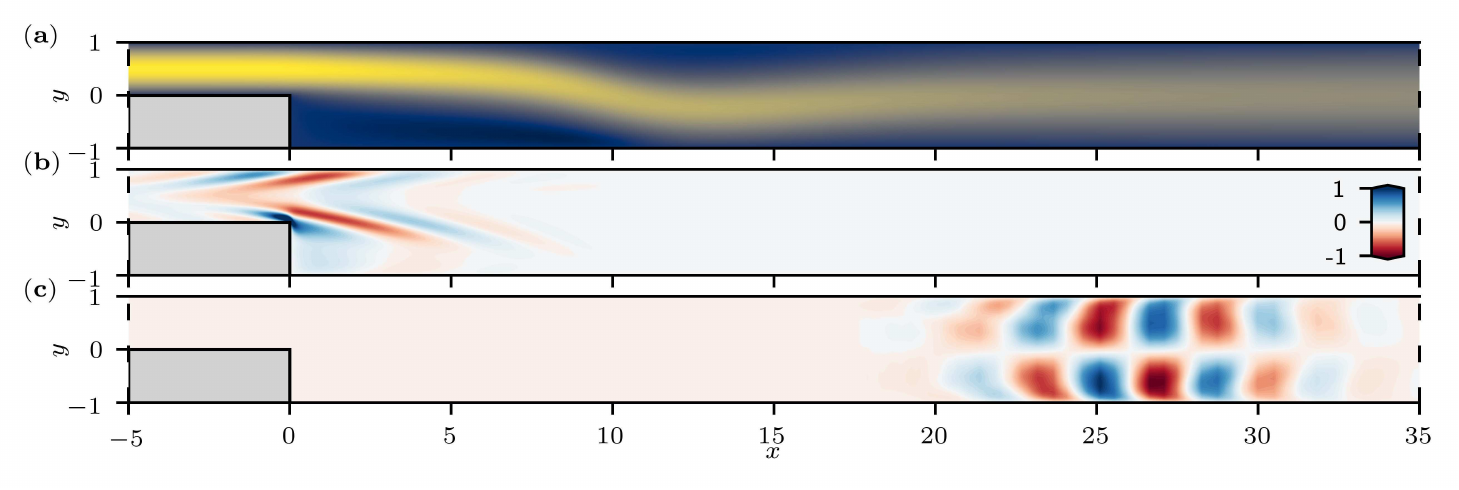}
\caption{The backward facing step problem: ($a$) fixed point at $Re=500$, ($b$) optimal disturbance, and ($c$) optimal response at the maximum amplification time horizon $\tau=58$.}
\label{fig:backward_ts:mode}
\end{figure*}

The flow past a backward-facing step is chosen as an example of a transient growth analysis.
We reproduce the analysis presented in \cite{jfm:blackburn:2008} at $Re=500$ and take the step height as the characteristic length scale.
As before, the dynamics are governed by the incompressible Navier-Stokes equations~\eqref{eq: Navier-Stokes eq}.
We considered a mesh made up of 1670 elements, each locally discretized with a polynomial order $N=5$ in both directions.
The domain spans from -10 to 50 in the streamwise direction, and from -1 to 1 in the vertical direction.
At $Re = 500$, the stationary base flow is linearly stable.
Yet, small perturbations can experience large transient growth due to the strong non-normality of the linearized Navier-Stokes operator.
The maximum gain envelope is calculated and compared with the reference in \cref{fig:backward_ts}, showing excellent agreement.
The peak is located at $\tau=58$, which corresponds to the maximum possible transient growth associated with the optimal initial condition.
Figure~\ref{fig:backward_ts:mode} shows the fixed point computed by the Newton-Krylov method, the spatial distribution of the leading eigenvector of the direct-adjoint eigenproblem corresponding to the optimal perturbation with maximum energy gain, and the optimal response at target time $\tau=58$.
Excellent agreement with reference~\cite{jfm:blackburn:2008} is obtained.

\section{Discussion}\label{sec: discussion}

In this section, we discuss various practical and theoretical aspects of using a time-stepper formulation.
This includes the interpretation of time-stepping as an effective preconditioner in \cref{subsec: time-stepper precond} and some observations on the convergence of leading eigenvalues for the linearized Navier-Stokes operator in open shear flows in \cref{subsec: observation eigenvalues}.

\subsection{Time-stepper and preconditioning}\label{subsec: time-stepper precond}

Solving a linear system forms the most computationally intensive part of Newton solvers.
Using a standard approach, one needs to solve linear systems of the form
\[
\mathbf{Lx} = \mathbf{b},
\]
where $\mathbf{L} \in \mathbb{R}^{n \times n}$ is the Jacobian matrix of the Navier-Stokes equations.
For the sake of this discussion, we will assume furthermore that $n$ is sufficiently large so that direct solvers cannot be employed.
No matter the discretization technique employed, this matrix tends to be ill-conditioned.
Consequently, directly solving the above system of equations with an iterative solver may lead to relatively poor computational performances as the convergence rate of such solvers is directly impacted by the conditioning of the matrix~\cite{saad2003iterative}.
It should be emphasized furthermore that explicitly computing this matrix-vector product might not moreover be easily accessible given a general-purpose CFD solver.
In order to overcome the first issue, practitioners typically use \emph{preconditioning}, either left preconditioning leading to a new system of the form
\[
\mathbf{PLx} = \mathbf{Pb},
\]
or right preconditioning, leading to
\[
\mathbf{LPy} = \mathbf{b}, \quad \textrm{and} \quad \mathbf{Py} = \mathbf{x}.
\]
In both cases, the matrix $\mathbf{P} \in \mathbb{R}^{n \times n}$ (known as the \emph{preconditioner}) should be a reasonably good, and more importantly cheap, approximation of the inverse of $\mathbf{L}$.
A judicious choice of $\mathbf{P}$ can lead to a substantial reduction of the number of steps needed for the iterative solver to converge.
This choice however strongly depends not only on the spectral content of $\mathbf{L}$ but also on its structure directly inherited from the discretization scheme employed.
The most standard preconditioners include Jacobi preconditioning, block LU, additive Schwarz methods, or algebraic and geometric multigrids.
In fluid dynamics, preconditioners designed specifically for fixed point computation using a Newton solver include the Stokes and Laplace preconditioning by Tuckerman and colleagues~\cite{barkley_stokes_1997,tuckerman2000bifurcation,tuckerman_stokes_2000,brynjell2017computing}, or their extensions by Gelfgat~\cite{gelfgat_acceleration_2019}.
Given a classical time-stepping solver for the incompressible Navier-Stokes equations, these are relatively easy to implement (see more discussions in \cite{brynjell2017studies}).
Yet, to the best of our knowledge, no systematic methods are however available to choose \emph{a priori} the best preconditioner.

In contrast, the linear system involved in the time-stepper formulation of the Newton-Krylov method reads
\[
\left( \exp(\tau \mathbf{L}) - \mathbf{I} \right) \mathbf{x} = \mathbf{b},
\]
where the sampling period $\tau$ plays a crucial role.
At first, not much seems to be gained from this formulation as, for a sampling period $\tau$ comparable to the discretization time step $\Delta t$, we have
\[
\begin{aligned}
\exp(\tau \mathbf{L}) - \mathbf{I} & \simeq \left( \mathbf{I} + \tau \mathbf{L} \right) - \mathbf{I} \\
& \propto \mathbf{L}.
\end{aligned}
\]
Hence, for a small sampling period, the condition number of $\exp(\tau \mathbf{L}) - \mathbf{I}$ is directly proportional to that of the Jacobian matrix $\mathbf{L}$. However, in practice, $\tau$ is of the order of a few hundred or a few thousand time-steps $\Delta t$ (\ie{} $\tau \gg \Delta t$) such that the first-order Taylor approximation of the matrix exponential does not apply.

An upper bound on the number of iterations needed to drive the residual $\| \mathbf{r}_k \|_2 / \| \mathbf{r}_0 \|_2$ below a given tolerance $\varepsilon$ can be derived analytically.
Assume for the moment that all the eigenvalues of $\mathbf{L}$ lie in the stable complex half-plane.
Furthermore, we will assume that the leading eigenvalue $\lambda_1$ satisfies
\[
\Re(\lambda_1) \leq -\delta,
\]
with $\delta > 0$.
The spectrum of the matrix $\mathbf{J} = \exp(\tau \mathbf{L}) - \mathbf{I}$ then satisfies
\[
\mathrm{spec}(\mathbf{J}) \in \mathcal{D} = \left\{ z : \vert z + 1 \vert \leq \exp(-\tau \delta) \right\}.
\]
Adapting the derivation in \cite{saad2003iterative} of the general purpose GMRES method to our particular case, it can be shown that the number $k$ of iterations needed to drive the normalized residual below a given tolerance $\varepsilon$ satisfies
\[
\varepsilon \leq \kappa(\mathbf{V}) \exp(-k \tau \delta),
\]
where $\mathbf{V}$ denotes the matrix of eigenvectors of $\mathbf{J}$ (which are identical to the eigenvectors of $\mathbf{L}$), and $\kappa(\mathbf{V}) = \| \mathbf{V} \|_2 \| \mathbf{V}^{-1} \|_2$ its condition number.
From this expression, we can then write
\[
k \leq \dfrac{1}{\tau \delta} \log \left( \dfrac{\kappa(\mathbf{V})}{\varepsilon} \right).
\]
It is well known that such upper bounds tend to be overly pessimistic, and typically are worst-case scenarios.
It nonetheless highlights the fact that increasing the sampling period $\tau$ is likely to reduce the number of GMRES iterations needed for convergence.
This is consistent with the fact that, if $\mathbf{L}$ has only stable eigenvalues, then
\[
\lim_{\tau \to \infty} \exp(\tau \mathbf{L}) - \mathbf{I} = -\mathbf{I}.
\]
Moreover, when $\exp(\tau \mathbf{L})$ has $p$ eigenvalues outside of the unit disk, it can be shown that GMRES only needs $\mathcal{O}(p)$ extra iterations to converge, and this upper bound remains unchanged.
Note however that, as we increase $\tau$, the computational cost associated to generating each Krylov vector also increases, and a trade-off needs to be found between the computational cost of generating a Krylov vector and the memory footprint of storing a large Krylov basis.
It should be noted furthermore that, while $\exp(\tau \mathbf{L}) - \mathbf{I}$ is well conditioned, this is obtained at the cost of taking many small timesteps.
In contrast, the operator in \cite{barkley_stokes_1997,tuckerman2000bifurcation,tuckerman_stokes_2000} is not as well conditioned, but consists of a single timestep.
This trade-off between fewer GMRES iterations for a well-conditioned but costly operator vs.\ more GMRES iterations for a less well-conditioned but inexpensive operator is discussed in detail in \cite{chapter:tuckerman:2019}.
From a practical point of view, using a time-stepper formulation can nonetheless be understood as an effective and easy-to-use preconditioning strategy as already pointed out by Tuckerman and collaborators~\cite{barkley_stokes_1997,tuckerman2000bifurcation,tuckerman_stokes_2000}.

\subsection{Observations about the convergence of eigenvalue computations}\label{subsec: observation eigenvalues}

Let us now discuss some practical aspects associated with eigenvalue computations.
These include some general statements about the convergence of the leading eigenvalues, the choice of the sampling period $\tau$, as well as practical considerations regarding the influence of the streamwise extent of the domain for open flows.

\subsubsection{General statements about the convergence of the Arnoldi iteration}

Suppose the matrix $\exp(\tau \mathbf{L}) \in \mathbb{R}^{n \times n}$ is diagonalizable, \ie{}
\[
\exp(\tau \mathbf{L}) = \mathbf{V} \mathbf{D} \mathbf{V}^{-1},
\]
with $D_{ii} = \mu_i$, and consider the orthonormal matrix $\mathbf{Q} \in \mathbb{R}^{n \times k}$ obtained after a $k$-step Arnoldi factorization started with the unit-norm vector $\hat{\mathbf{b}} = \frac{\mathbf{b}}{\| \mathbf{b} \|_2}$.
The vector $\tilde{\mathbf{b}}$ can be expressed as the linear combination of the eigenvectors such that
\[
\hat{\mathbf{b}} = \sum_{i=1}^n \alpha_i \mathbf{v}_i.
\]
It can be shown that the approximation of the $i$\textsuperscript{th} eigenvector $\mathbf{v}_i$ in the Krylov basis $\mathbf{Q}$ satisfies
\begin{equation}
\bigg\| \left( \mathbf{I} - \mathbf{QQ}^T \right) \mathbf{v}_i \bigg\| \leq \left( \sum_{\substack{j=1 \\ j \neq i}}^n \dfrac{\vert \alpha_j\vert}{\vert \alpha_i \vert} \right) \varepsilon_{i}^{(k)}.
\label{eq: Arnoldi equation}
\end{equation}
A detailed derivation of this statement can be found in Appendix \ref{appendixB}.
In the expression above, $\varepsilon_{i}^{(k)}$ is given by
\[
\varepsilon_{i}^{(k)} = \min_{\substack{p \in \mathbb{P}_{k-1} \\ p(\mu_i) = 1}} \quad \max(\vert p( \mu_1 ) \vert, \cdots, \vert p(\mu_n) \vert),
\]
where $\mathbb{P}_{k-1}$ denotes the set of polynomials of degree $k-1$.
Two key observations can be derived from this upper bound.

\paragraph{Fast convergence for isolated eigenvalues}

Suppose $\mu_i$ is an isolated eigenvalue such that the rest of the spectrum can be enclosed in the disk
\[
\mathcal{D} = \left\{ z : \vert z - c \vert < \rho \right\},
\]
\ie{} a disk centered in $c$ and of radius $\rho$.
In a time-stepping framework, the center $c$ typically is the origin.
Following \cite{saad2003iterative}, an upper bound for $\varepsilon_i^{(k)}$ is given by
\[
\varepsilon_i^{(k)} \leq \kappa(\mathbf{V}) \left( \dfrac{\rho}{\vert \mu_i - c \vert} \right)^k,
\]
where $\kappa(\mathbf{V}) \geq 1$ is the condition number of the matrix of eigenvectors.
Clearly, the farther away $\mu_i$ is from the rest of the spectrum, the smaller this upper bound, and hence the faster the convergence.
This observation is of utmost importance in numerous fluid dynamics applications where branches of eigenvalues are pretty common.
These branches are notoriously associated with the non-normality of the linearized Navier-Stokes operator.
Not only do they impact directly on the condition number $\kappa(\mathbf{V})$, but they are also non-isolated eigenvalues which are thus much harder to converge.

\paragraph{Influence of the starting vector}
Note that $\| \hat{\mathbf{b}} \| = 1$, and suppose the eigenvectors have been normalized such that $\| \mathbf{v}_i \| = 1$.
Then, the coefficients $\left\{ \alpha_i \right\}_{i=1, n}$ satisfy
\[
1 = \| \alpha_1 \mathbf{v}_1 + \cdots + \alpha_n \mathbf{v}_n \| \leq \vert \alpha_1 \vert + \cdots + \vert \alpha_n \vert,
\]
and let
\[
\xi_i = \dfrac{1}{\vert \alpha_i \vert} \sum_{j=1}^n \vert \alpha_j \vert - 1 \geq \dfrac{1}{\vert \alpha_i \vert} - 1.
\]
Suppose now that $\alpha_j = \delta$ for all $j \neq i$, with $\delta$ small.
Then, $\alpha_i = \mathcal{O}(1)$.
Under these conditions, the prefactor $\xi_i$ is small, and the Arnoldi iteration exhibits a fast convergence for the $i$\textsuperscript{th} eigenpair.
Alternatively, if $\alpha_i$ is very small, then $\xi_i \gg 1$, and we have slow convergence.
This observation gave rise to the rule-of-thumb that, for eigenvalue problems, the Arnoldi method will favor eigenvectors having large components in the starting vector.
Unless we have prior knowledge of the eigenvectors we wish to compute, starting the Arnoldi iteration with a genuinely random vector increases the probability that the eigenpairs of interest will converge.

\subsubsection{Influence of the domain extent for open shear flows}

\begin{figure}\centering
\includegraphics{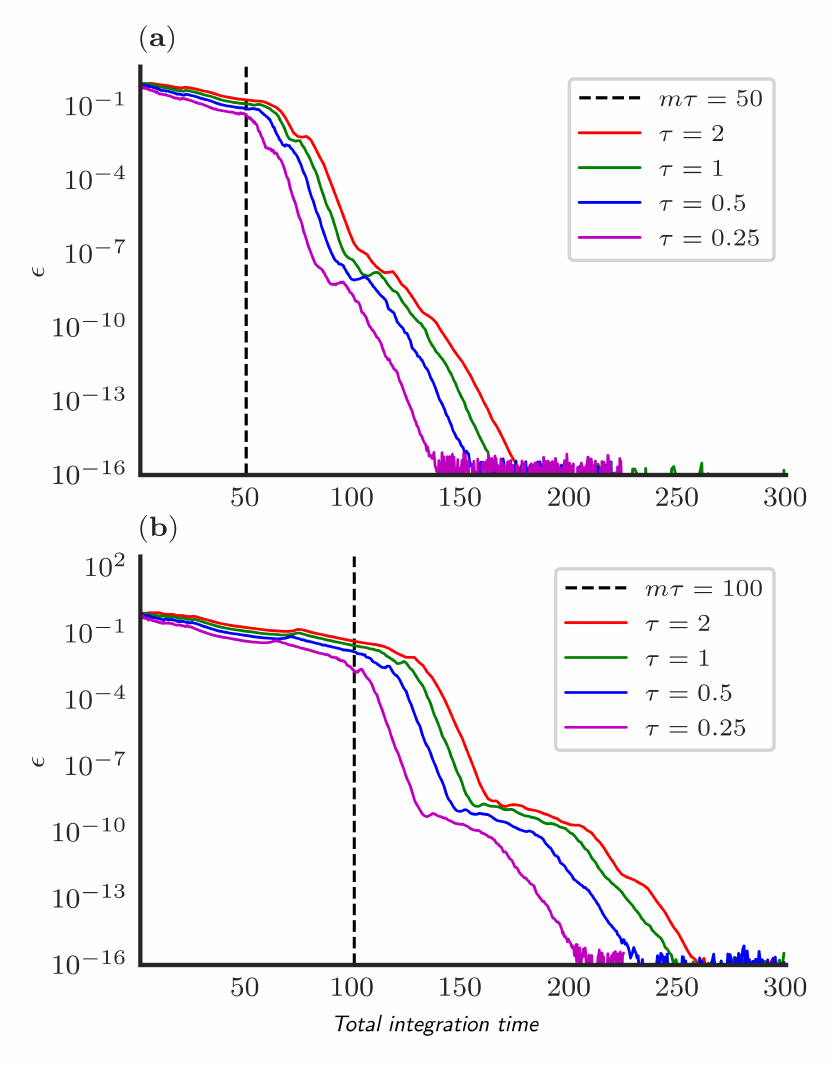}
\caption{Convergence history of the leading eigenvalue as a function of the total integration time (defined as the product of the sampling period $\tau$ and the number $m$ of Krylov iterations.) for the flow past a circular cylinder at $Re=50$ given different values of the sampling period $\tau$: ($a$) the domain extends from $-16$ to $+50$ in the streamwise direction and ($b$) the domain extends from $-16$ to $+100$ in the streamwise direction.}
\label{fig:1cyl:res_evol}
\end{figure}

We now verify the influence of the streamwise domain length on the convergence of the leading eigenpairs in open shear flows.
We performed a series of computations for the flow past a circular cylinder at $Re=50$.
In this case, the leading eigenpair corresponds to the von Kármán vortex street mode associated with a supercritical Hopf bifurcation (as previously introduced in \ref{sec:ex:1cyl}).
\Cref{fig:1cyl:res_evol} depicts the evolution of the leading eigenvalue residual as a function of the total integration time $T$ defined as the product of the number $m$ of Krylov iterations and the sampling period $\tau$, \ie{} $T = m \tau$.
Letting $L_x$ be the streamwise extent of the computational domain and $U_{\infty}$ the velocity scale, this figure highlights a rapid convergence of the leading eigenvalue once the total integration time exceeds
\[
m \tau > \dfrac{L_x}{U_{\infty}},
\]
with $U_{\infty} = 1$ in the non-dimensional case.
This is easily explained by the fact that the usual choice is to start from an initial perturbation vector consisting of noise (which aims to excite the entire computational domain while minimizing any bias).
The leading eigenvalue then starts to converge after one flow-through time.

\section{Conclusion}\label{sec: conclusion}

Transition to turbulence is a long-standing problem in fluid dynamics, for which adopting a dynamical system point of view has greatly increased our understanding.
Specialized codes such as \texttt{ChannelFlow}~\cite{channelflow,GibsonHalcrowCvitanovicJFM08}, \texttt{OpenPipeFlow}~\cite{openpipeflow} or \texttt{Semtex}~\cite{blackburn2019semtex} are equipped with the set of tools necessary to probe the phase space of the Navier-Stokes equations.
Yet, by design, these are limited to canonical configurations (\eg{} plane Poiseuille and plane Couette flow for \texttt{ChannelFlow}) with infinite spans, and no such library is available for a general-purpose CFD code.
This work therefore aims at providing a general-purpose introduction to the Krylov methods underlying numerous recent works on stability analysis of really high-dimensional systems.
These methods are implemented in \texttt{nekStab}, an open-source and user-friendly toolbox to perform large-scale bifurcation analysis in \texttt{Nek5000}.

Using a time-stepper formulation and leveraging Krylov-based techniques, \texttt{nekStab} is capable of computing fixed points and periodic orbits, as well as computing the leading eigenpairs and singular triplets of the corresponding linearized Navier-Stokes operator to characterize their stability properties.
The capabilities of \texttt{nekStab} and its underlying Krylov methods are showcased in a number of test cases available in the literature, including the canonical cylinder flow~\cite{jfm:barkley:1996}, the annular thermosyphon~\cite{tcfd:loiseau:2020}, the harmonically forced jet~\cite{jfm:leopold:2019} and the flip-flop instability in the wake of side-by-side cylinders~\cite{jfm:carini:2014}.
In all cases, excellent agreement has been obtained with the results available in the literature.

We believe that this bifurcation analysis plug-in for a code as established as \texttt{Nek5000} could have far-reaching applications, both in academia and industry, and could improve our understanding of the transition to turbulence in non-canonical configurations.
By making the code open source, we hope to foster the ideas and efforts of our community as a whole and incorporate them into the development of a high-quality tool from which we can all benefit.

\section*{Data availability}
The code and scripts presented in this review are freely available at \url{https://github.com/nekStab}.

\section*{Acknowledgment} 

This work is built on the research of numerous former Ph.D. students, including Frédéric Alizard~\cite{phd:alizard:2007}, Stefania Cherubini~\cite{phd:cherubini:2010}, Jean-Christophe Loiseau~\cite{phd:loiseau:2014}, Alessandro Bucci~\cite{phd:bucci:2017}, Mirko Farano~\cite{phd:farano:2017}, Francesco Picella~\cite{phd:picella:2019} and Ricardo A. S. Frantz~\cite{phd:frantz:2022}. 
The authors acknowledge the support of GENCI under the following projects: A0072A06362/2020, A0092A06362/2021, and A0112A06362/2022.
We are also very grateful to Laurette Tuckerman for all the valuable discussions and insights that greatly improved this manuscript.

\appendix
\section{\texttt{nekStab}, an open source toolbox for \texttt{Nek5000}}\label{appendixA}
The numerical methods introduced in this paper are implemented and validated in a toolbox for linear stability analysis of steady and periodic flows.
\texttt{nekStab} is an open-source toolbox that implements all the algorithms described in this article in less than 9000 lines of Fortran 90.
The toolbox uses the flexible and highly parallel data structure of \texttt{Nek5000}, which allows efficient computation of flows in complex geometries.
In the current implementation, when compiling \texttt{Nek5000} \texttt{nekStab} is appended as a submodule so that the core of \texttt{Nek5000} is only minimally modified. 
\texttt{Nek5000}, the workhorse of \texttt{nekStab}, is valued for its minimal dependencies and parallel performance, enabling computations on simple laptops up to high-performance computers.
Great care has been taken to limit the additional dependencies required to use \texttt{nekStab}, thanks to a non-invasive and self-contained design, no additional dependencies are introduced, and the need to modify the source code of \texttt{Nek5000} has also been eliminated.
To avoid compatibility issues and ease of use, \texttt{nekStab} interacts only with \texttt{Nek5000} through the existing user interface, without requiring any changes to the original source code while retaining its native performance.
\texttt{nekStab} directly uses \texttt{Nek5000} time-stepper and closely adheres to its data structure, but only interacts with it through the existing user interface.
Given the design strategy for \texttt{nekStab}, potential conflicts with future updates to the main solver are unlikely, as the variables are separated by subroutines crafted to interface with \texttt{Nek5000} variables.
Like \texttt{Nek5000}, our toolbox also makes use of a few subroutines from the Linear Algebra PACKage (LAPACK) provided with \texttt{nekStab} to avoid external dependencies (version 3.10 available at \url{github.com/Reference-LAPACK} included).
All results presented in this work were computed with \texttt{nekStab} using the latest version of \texttt{Nek5000} (commit \textit{7ae03b1}), available in \url{github.com/Nek5000}.
The well-commented source code is maintained online in the public Git repository at \url{github.com/nekStab}, as is the collaborative online documentation at \url{nekstab.github.io} with instructions and lightweight examples that can be quickly computed on a laptop. 
These include canonical and more complex flows, such as flow past a cylinder or adjacent cylinders, a time-periodic axisymmetric jet, open and closed cavity flow, a stratified annular thermosyphon, flow past an airfoil, a backward step, sudden channel expansion, channel flow, and a Blasius boundary layer.
The examples are compared to a reference case from the literature and aim to provide all the elements that new users might need to develop their own cases.

To the best of our knowledge, \texttt{nekStab} is the first general-purpose computational framework capable of stabilizing fully 3D unstable periodic orbits (forced or unforced) and fixed points, as well as computing direct, adjoint modes and transient growth analyses for both steady and periodic flows.
The menu of options includes a matrix-free Newton GMRES solver as well as other classical techniques such as selective frequency damping (SFD)~\cite{pof:akervik:2006}, BoostConv~\cite{jcp:citro:2017}, Time-Delayed Feedback (TDF)~\cite{pyragas1992continuous,shaabani2017time}, and Dynamic Mode Tracking (DMT)~\cite{queguineur2019dynamic}, as well as post-processing routines such as the kinetic energy budget of the leading modes based on the Reynolds-Orr decomposition~\cite{ejmbf:brandt:2006,jfm:loiseau:2014}, steady-state base flow, and sensitivity analyzes~\cite{marquet2008sensitivity}.

To facilitate adoption by the community (both academic and industrial), \texttt{nekStab} is released under the permissive BSD 3-Clause license.
All source code, examples, scripts, and tutorials can be viewed and downloaded for free at \url{github.com/nekStab} or \url{nekstab.github.io}.

We foresee more updates and future developments for \texttt{nekStab} on the horizon:
\begin{itemize} \item integration of the algorithms proposed in~\cite{tcfd:semeraro:2018} for the synthesis of linear optimal LQR controllers for large-scale systems, \item integration of the optimization algorithms in~\cite{foures2014optimal} and already implemented in \texttt{Nek5000} by M. Farano~\cite{farano2017optimal,farano2018nonlinear} for linear and nonlinear optimal perturbation analysis, \item extension of the Newton-Krylov solver with pseudo-arclength continuation to compute branches of solutions even in the presence of folding points, \item Computation of Lyapunov exponents using the algorithms presented in \cite{benettin1980lyapunov} and \cite{geist1990comparison},
\end{itemize}

\section{Proof of \cref{eq: Arnoldi equation}}\label{appendixB}

Let us prove that, given the Arnoldi factorization
\[
\mathbf{AQ} = \mathbf{QH} + \beta \mathbf{e}_{k}^T \mathbf{q}_{k+1},
\]
the approximation of the i\textsuperscript{th} eigenvector $\mathbf{v}_i$ in the Krylov basis $\mathbf{Q}$ satisfies
\begin{equation}
\bigg\| \left( \mathbf{I} - \mathbf{QQ}^T \right) \mathbf{v}_i \bigg\| \leq \left( \sum_{\substack{j=1 \\ j \neq i}}^n \dfrac{\vert \alpha_j\vert}{\vert \alpha_i \vert} \right) \varepsilon_{i}^{(k)}.
\end{equation}
Recall furthermore that the vector $\mathbf{b}$ used to seed this Krylov vector can be expressed as a linear combination of the eigenvectors, \ie{}
\begin{equation}
\mathbf{b} = \sum_{i=1}^n \alpha_i \mathbf{v}_i.
\end{equation}
The proof below follows closely the derivation given by Elias Jarlebring in his lecture notes at KTH (see \url{https://www.math.kth.se/na/SF2524/matber15/arnoldiconv.pdf}).
It proceeds in three steps.

\paragraph{Step 1} Consider an arbitrary vector $\mathbf{u} \in \mathbb{C}^n$.
Then the problem
\begin{equation}
\minimize_{\mathbf{z} \in \mathbb{C}^k} \| \mathbf{u} - \mathbf{Qz} \|_2
\end{equation}
is a least-squares minimization problem.
The matrix $\mathbf{Q}$ being orthonormal, its Moore-Penrose pseudoinverse $\mathbf{Q}^{\dagger}$ is equal to $\mathbf{Q}^T$.
Hence, the vector $\mathbf{z}$ solution to the minimization problem is simply given by $\mathbf{z} = \mathbf{Q}^T \mathbf{u}$.
It implies in particular that
\begin{equation}
\min_{\mathbf{z} \in \mathbb{C}^k} \| \mathbf{u} - \mathbf{Qz}\|_2 = \| \left( \mathbf{I} - \mathbf{QQ}^T \right) \mathbf{u} \|_2.
\end{equation}

\paragraph{Step 2} Our objective is to obtain an upper bound for the term $\| \left( \mathbf{I} - \mathbf{QQ}^T \right) \mathbf{v}_i \|_2$, where $\mathbf{v}_i$ is the i\textsuperscript{th} eigenvector of $\mathbf{A}$.
The proof is simplified by rescaling the right-hand side with $\alpha_i$, \ie{}
\begin{equation}
\begin{aligned}
  \| \left( \mathbf{I} - \mathbf{QQ}^T \right) \alpha_i \mathbf{v}_i \|_2 & = \min_{\mathbf{z} \in \mathbb{C}^k} \| \alpha_i \mathbf{v}_i - \mathbf{Qz} \|_2 \\
  & = \min_{\mathbf{y} \in \mathcal{K}_k(\mathbf{A}, \mathbf{b})} \| \alpha_i \mathbf{v}_i - \mathbf{y} \|_2.
\end{aligned}
\end{equation}
The Krylov subspace $\mathcal{K}_k(\mathbf{A}, \mathbf{b})$ can be characterized with polynomials.
The statement $\mathbf{y} \in \mathcal{K}_k(\mathbf{A}, \mathbf{b})$ is thus equivalent to the existence of a polynomial $p \in \mathbb{P}_{k-1}$ such that $\mathbf{y} = p(\mathbf{A}) \mathbf{b}$.
Hence
\begin{equation}
\| \left( \mathbf{I} - \mathbf{QQ}^T \right) \alpha_i \mathbf{v}_i \|_2 = \min_{p \in \mathbb{P}_{k-1}} \| \alpha_i \mathbf{v}_i - p(\mathbf{A}) \mathbf{b} \|_2.
\end{equation}

\paragraph{Step 3} The last step consists in inserting the expansion of $\mathbf{b}$ in terms of the eigenvectors.
This leads to
\begin{equation}
\begin{aligned}
  \| \left( \mathbf{I} - \mathbf{QQ}^T \right) \alpha_i \mathbf{v}_i \|_2 & = \min_{p \in \mathbb{P}_{k-1}} \| \alpha_i \mathbf{v}_i - p(\mathbf{A}) \sum_{j=1}^n \alpha_j \mathbf{v}_j \|_2 \\
  & = \min_{p \in \mathbb{P}_{k-1}} \| \alpha_i \mathbf{v}_i - \sum_{j=1}^n \alpha_j p(\mu_j) \mathbf{v}_j \|_2.
\end{aligned}
\end{equation}
It can then easily be shown that the expression above is bounded from above by
\begin{equation}
\| \left( \mathbf{I} - \mathbf{QQ}^T \right) \alpha_i \mathbf{v}_i \|_2 \leq \left( \sum_{\substack{j=1 \\ j \neq i}}^n \vert \alpha_j \vert \right) \cdot \epsilon_{i}^{(k)},
\end{equation}
where $\epsilon_{i}^{(k)}$ is given by
\begin{equation}
\epsilon_{i}^{(k)} = \min_{\substack{p \in \mathbb{P}_{k-1} \\ p(\mu_i) = 1}} \max_{j \neq i}( \vert p(\mu_j) \vert ).
\end{equation}
The proof of \cref{eq: Arnoldi equation} is completed by dividing this upper bound by $\vert \alpha_i \vert$.

\bibliographystyle{plainnat} 
\bibliography{biblio} 

\end{document}